%% file: thesis.tex
\documentclass[myorder,lof,lot]{puthesis}

\include{component_definitions}

\author{Leonid M. Malyshkin}
\title{Magnetized Turbulent Dynamo\\ in Protogalaxies}

\departmentprefix{Department of}
\department{Astrophysical Sciences}

\dedication{\centerline{To my mother}}

\include{component_abstract}

\include{component_acknowlegements}

\begin{document} 

\frontmatter\maketitlepage\makecopyrightpage\makededication\makeabstract\makeacknowledgements\tableofcontents\clearpage\makelof\clearpage\makelot


\clearpage\mainmatter

\include{chapter_1}

\include{chapter_2}

\include{chapter_3}

\include{chapter_4}

\include{chapter_5}

\appendix

\include{component_appendix_a}

\include{component_appendix_b}

\include{component_appendix_c}

\include{component_appendix_d}

\include{component_appendix_e}

\include{component_bibliography}

\end{document}

%% file: component_definitions.tex
\def	\be	{\begin{equation}}
\def	\ee	{\end{equation}}
\def	\beq	{\begin{eqnarray}}
\def	\eeq	{\end{eqnarray}}

\def	\simlt	{{\lower.5ex\hbox{$\;\buildrel{\mbox{\footnotesize$<$}}\over{\mbox{\footnotesize$\sim$}}\;$}}}
\def	\simgt	{{\lower.5ex\hbox{$\;\buildrel{\mbox{\footnotesize$>$}}\over{\mbox{\footnotesize$\sim$}}\;$}}}
\def	\define	{\stackrel{\rm def}{{}={}}}
\def	\etal	{{\em et~al.~}}

\def	\b		{{\hat b}}

\def	\k		{{\hat k}}
\def	\v		{{\mbox{\large$\upsilon$}}}
\def	\vT		{{\mbox{\large$\tilde\upsilon$}}}
\def	\vbf		{{\mbox{{\large\boldmath$\upsilon$}}}}

\def	\BST		{{\widetilde{B^2}}}

\def	\Bo		{{{}{}^0\!B}}
\def	\BoT		{{{}{}^0\!{\tilde B}}}
\def	\Bobf		{{\bf{}{}^0B}}
\def	\BoTbf		{{\bf{}{}^0{\tilde B}}}
\def	\bo		{{{}{}^0\b}}
\def	\bobf		{{\bf{}{}^0\b}}
\def	\vo		{{{}{}^0\v}}
\def	\vobf		{{\bf{}{}^0\vbf}}
\def	\bbo		{{{}{}^0b}}
\def	\bboT		{{{}{}^0{\tilde b}}}
\def	\MoT		{{{}{}^0\!{\cal M}}}		
\def	\Vo		{{{}{}^0V}}
\def	\BSo		{{{}{\vphantom{B^2}}^0\!B^2}}
\def	\BSoT		{{{}{\vphantom{B^2}}^0\!\BST}}

\def	\Bi		{{{}{}^1\!B}}
\def	\BiT		{{{}{}^1\!{\tilde B}}}
\def	\Bibf		{{\bf{}{}^1B}}
\def	\BiTbf		{{\bf{}{}^1{\tilde B}}}
\def	\bi		{{{}{}^1\b}}
\def	\bibf		{{\bf{}{}^1\b}}
\def	\vi		{{{}{}^1\v}}
\def	\viT		{{{}{}^1\vT}}
\def	\viTbf		{{\bf{}{}^1{\bf\tilde\vbf}}}
\def	\vibf		{{\bf{}{}^1\vbf}}
\def	\Vi		{{{}{}^1V}}
\def	\ViT		{{{}{}^1{\tilde V}}}
\def	\Vibf		{{\bf{}{}^1V}}
\def	\ViTbf		{{\bf{}{}^1{\tilde V}}}
\def	\bbi		{{{}{}^1b}}
\def	\bbiT		{{{}{}^1{\tilde b}}}
\def	\Pi		{{{}{}^1\!P}}
\def	\PiT		{{{}{}^1\!{\tilde P}}}
\def	\FiT		{{{}{}^1\!{\cal F}}}		
\def	\BSi		{{{}{\vphantom{B^2}}^1\!B^2}}

\def	\Bii		{{{}{}^2\!B}}
\def	\Biibf		{{\bf{}{}^2B}}
\def	\BiiT		{{{}{}^2\!{\tilde B}}}
\def	\bii		{{{}{}^2\b}}
\def	\biibf		{{\bf{}{}^2\b}}
\def	\vii		{{{}{}^2\v}}
\def	\viiT		{{{}{}^2\vT}}
\def	\viibf		{{\bf{}{}^2\vbf}}
\def	\Vii		{{{}{}^2V}}
\def	\ViiT		{{{}{}^2{\tilde V}}}
\def	\Viibf		{{\bf{}{}^2V}}
\def	\Pii		{{{}{}^2\!P}}
		
\def	\BSii		{{{}{\vphantom{B^2}}^2\!B^2}}
\def	\BSiiT		{{{}{\vphantom{B^2}}^2\!\BST}}

%% file: component_abstract.tex
\abstract{
The prevailing theory for the origin of cosmic magnetic fields is that they 
have been amplified from insignificant seed values to their present values 
by the turbulent dynamo inductive action driven by the plasma turbulent
motions in the protogalactic and galactic medium.
Up to now, in calculation of the turbulent dynamo, it has been customary 
to assume that there is no the back reaction of the magnetic field on the 
turbulence, as long as the magnetic energy is less than the turbulent 
kinetic energy. This assumption leads to the kinematic dynamo theory that 
has been well developed in the past. 

However, the applicability of the kinematic dynamo theory to protogalaxies 
is rather limited. The reason is that in protogalaxies the temperature is 
very high, and the viscosity is dominated by ions. As the magnetic field 
strength grows in time because of the dynamo action, the ion cyclotron time 
becomes shorter than the ion collision time, and the plasma becomes strongly 
magnetized. As a result, the ion viscosity becomes the Braginskii viscosity, 
and the magnetic field starts to strongly affect the turbulent motions on the 
viscous scales. Thus, in protogalaxies the back reaction sets in much earlier, 
at field strengths much lower than those which correspond to energy 
equipartition between the field and the turbulence, and the turbulent dynamo 
becomes what we call the magnetized turbulent dynamo.

The main purpose of this thesis is to lay the theoretical groundwork for the 
magnetized turbulent dynamo. In particular, we predict that the magnetic energy 
growth rate in the magnetized dynamo theory is up to ten time larger than that 
in the kinematic dynamo theory, and this could lead to the dynamo creation of 
cluster fields. We also briefly discuss how the Braginskii viscosity can aid 
the development of the inverse cascade of magnetic energy, which happens after 
the energy equipartition time.
}


%% file: component_acknowlegements.tex
\acknowledgements{

First, I thank my advisor, Russell Kulsrud, who was, in my opinion, the
ideal advisor. He taught me how to think about science. He always had
time to explain the finest details to me and to patiently answer even my 
most trivial questions. Most important, he was not just an advisor, but 
he was both my advisor and my colleague. On one hand, whenever I met 
problems, he led me and helped me, thus, being my advisor. On the other 
hand, he always listened to my opinion, always first let me try to 
correct my mistakes by myself, and always allowed me the freedom to direct 
my research, thus, being my colleague. This was extremely important for my 
scientific growth. I was very lucky to be a Russell's student, and I 
shall always be very grateful to him.

I am deeply indebted to Vasily S. Beskin, my master thesis advisor in
Moscow. I successfully carried out my first scientific projects with him. 
This PhD thesis and my scientific career would not be possible without 
his support and encouragement four years ago.

I would like to thank other people in Princeton, who I worked with: 
Jeremy Goodman, Gillian Knapp and Scott Tremaine. They were great 
advisors. I got priceless experience of scientific research, while 
working with them. 

I would like to thank my committee, Bruce Draine, Jeremy Goodman, Jeremiah
Ostriker and David Spergel, for stimulating discussions and for carefully 
reading the thesis. I also thank Eric Blackman, Steven Cowley and 
Alexander Schekochihin for very useful discussions and for a number of 
important comments.

My special thanks to the Department of Astrophysical Sciences at Princeton
University and to Bruce Draine for financial support of my research. I 
would also like to thank Nancy Baker for her help in completing all 
administrative procedures and forms requirements for the PhD defense, and 
Jane Holmquist for her help in finding journals and books in the university 
libraries.

There are so many friends of mine, who I am very grateful to. It would be
impossible to thank everyone who supported and helped me during the
last four years. I would particular like to mention Sergei Dobrovolsky
and Alexander Karpikov. They were always there for me when I was lonely,
tired, frustrated or joyful. Michael Way and Joseph Weingartner became
my first American friends. They really helped me with English language 
and with learning American culture. My first positive judgment about 
this country was based on their friendliness. I would also like to thank 
Marilyn Fagles, who helped me a lot with English and patiently 
answered my numerous questions about the United States during my 
first years in Princeton.

I thank my son, Alesha. I am blessed to have you and I love you. You have
been an unlimited source of joy and support for me since you were born. I
would also like to thank Marina for taking care of Alesha with so much 
love.

Finally, words can not express my gratitude and my undying love to my 
mother. She was and is the best mother in the world, and I owe everything 
to her.

}


%% file: chapter_1.tex
\chapter
[Introduction]
{Introduction}
\label{INTRODUCTION}

\section
[Galactic and Extragalactic Magnetic Fields:\\ 
Observational Results]
{Galactic and Extragalactic Magnetic Fields:\\ 
Observational Results}
\label{MAGNETIC_FIELDS}

\begin{figure}[!p]
\vspace{14.5truecm}
\includegraphics{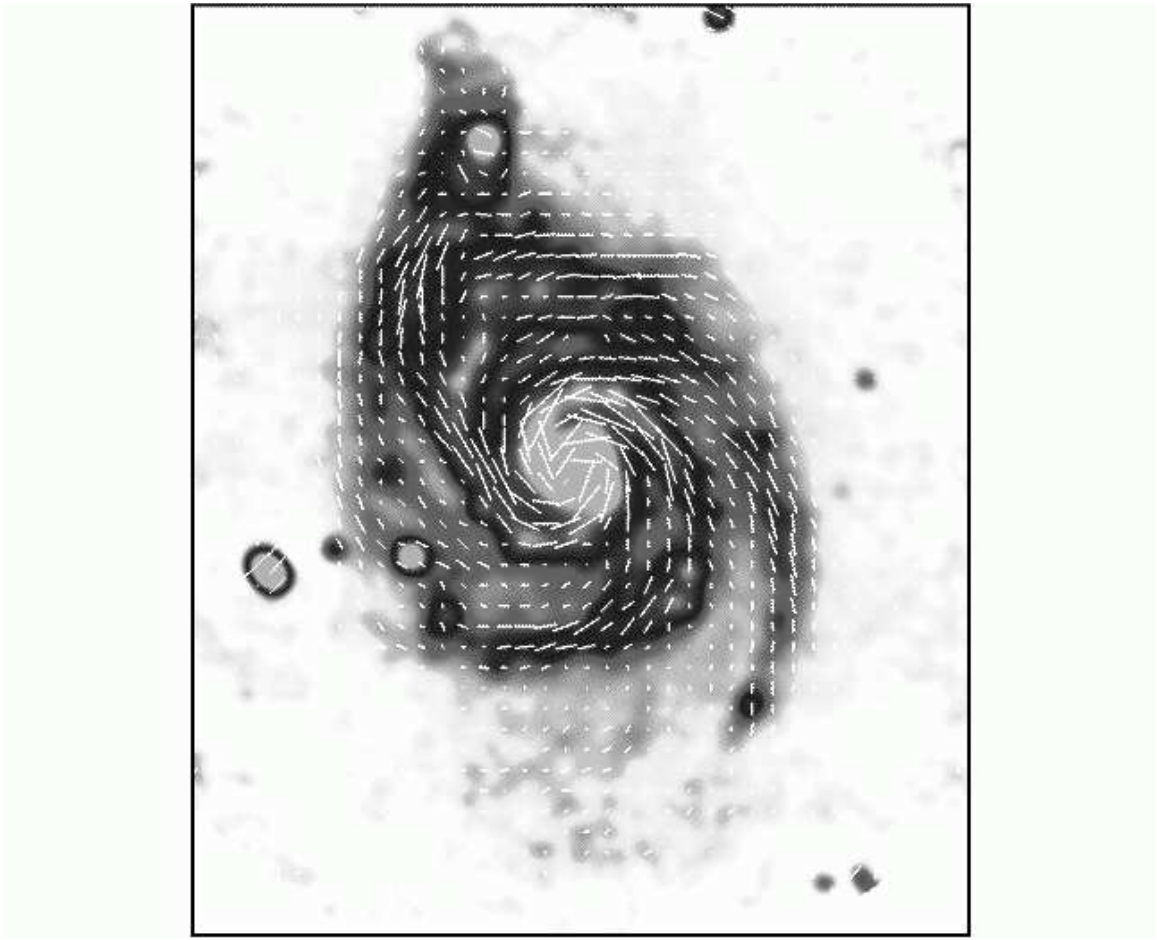}
\caption{
Magnetic field vectors (white bars) for the galaxy M51 (measured by fitting 
a model of the polarized synchrotron emission to radio observations of M51). 
The map is superimposed on a M51 image in $6\,{\rm cm}$. The magnetic field 
lines approximately follow the spiral structure. By courtesy of Beck, Horellou, 
Neininger, and the MPIfR collaboration~\cite{NHBBKK_93}.
}
\label{FIGURE_M51_GALAXY}
\end{figure}

\begin{figure}[!p]
\vspace{8.0truecm}
\includegraphics{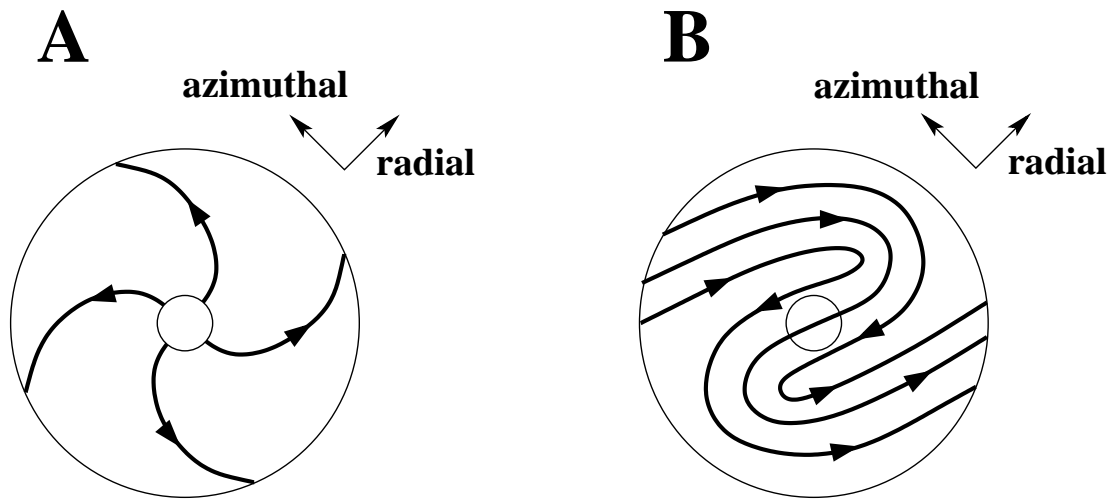}
\caption{
Two possible magnetic field configurations in disk galaxies, face-on view.
The bold lines are magnetic field lines with the field direction shown by
the arrows. 
A: axisymmetric spiral structure, no radial reversals of the field (this
structure is predicted by the galactic dynamo theory); 
B: bisymmetric spiral structure, there are radial reversals of the field 
(this structure arises in the rotating galactic disk if the field has 
primordial origin~\cite{ZH_97}).
}
\label{FIGURE_FIELD_GEOMETRY}
\end{figure}

One of the most important and challenging questions in astrophysics is the origin 
of galactic and exragalactic magnetic fields.
The existence of interstellar magnetic fields in our Galaxy was first proposed by 
Alfven in 1943~\cite{A_43}. In 1949 the polarization of starlight was observed 
by Hiltner~\cite{H_49} and independently by Hall and Mikesell~\cite{HM_49}. The 
starlight polarization was interpreted as a consequence of light scattering by 
interstellar dust grains aligned in the galactic magnetic fields. In the same year
Fermi employed galactic magnetic field for the acceleration of the cosmic rays and
for their confinement in the Galaxy~\cite{F_49}. In $60$'s the Faraday rotation
and the Zeeman effect were measured for different sources distributed over the 
sky~\cite{GW_66,V_69}. In 1970 Mathewson and Ford found a large-scale magnetic field 
in Magellanic Clouds by measuring the polarization of stars in the Magellanic 
System~\cite{MF_70}. In 1974 Manchester first correctly measured the galactic magnetic 
field in the vicinity of the Sun (within $\approx 2\,{\rm Kpc}$) by measuring the Faraday 
rotation for thirty eight nearby pulsars~\cite{M_74}. He found that the local field has 
a longitudinal component in the Galactic plane. In 1989 Rand and Kulkarni confirmed and 
significantly improved the measurements of the local magnetic field~\cite{RK_89}. They 
analyzed the Faraday rotation measures for nearly two hundred pulsars within 
$\approx 3\,{\rm Kpc}$, and they found that the uniform component of the local magnetic 
field  has a strength of $\approx 1.6\,\mu{\rm G}$ towards a galactic longitude of 
96$^{\rm o}$, with a reversal of the field at a distance about $0.6\,{\rm Kpc}$ towards 
the galactic center. The random field component was estimated by Rand and Kulkarni as 
${}\sim 5\,\mu{\rm G}$. 
Since early $90$'s a significant progress has been made in measuring magnetic fields in
our Galaxy and in other spiral galaxies, by measuring polarization of starlight, by 
measuring the Faraday rotation from pulsars and extragalactic radio sources, by detecting
synchrotron radiation from relativistic electrons, and by measuring the Zeeman splitting
of spectral lines; see recent excellent review papers~\cite{BBMSS_96,K_94,ZH_97}. 
The observations indicate that galaxies possess magnetic fields with strengths of 
several microgauss (up to several tens of microgauss). These fields have uniform 
components, with strengths comparable to those of the random components (e.~g.~in the 
vicinity of the Sun the uniform$/$random field strength ratio is about $1/2$). 
The uniform field components lie in the galactic disks and have typical correlation 
lengths from several hundreds parsecs to one kiloparsec.
Figure~\ref{FIGURE_M51_GALAXY} shows the magnetic field pattern for the galaxy M$51$. 
However, most galaxies do not show such a nice regular field pattern
as M$51$ does. As a result, the question about the prevailing geometrical structure 
of magnetic fields in spiral galaxies is still open, see 
Figure~\ref{FIGURE_FIELD_GEOMETRY}. The magnetic field in Milky Way exhibits reversals 
with radius, which suggests a bisymmetric spiral structure for it, see 
Figure~\ref{FIGURE_FIELD_GEOMETRY}(B).

It is very important to know that magnetic fields may exist in very young galaxies 
(at high redshifts)~\cite{ZH_97}. In 1992 Wolfe, Lanzetta and Oren found that the 
probability for Faraday rotation for radio QSOs (quasi-stellar objects) is 
significantly higher in damped Ly$\alpha$ systems~\cite{WLO_92}. These systems are 
associated with early forming galactic disks. Wolfe \etal estimated the magnetic field 
in two damped Ly$\alpha$ systems with $z\approx 2$ as a few microgauss. More recent 
extensive observations have confirmed this result. Thus, magnetic fields with similar 
spatial scale and strength to those in the local universe exist at redshifts above 
$z=1$, see review papers~\cite{K_94,P_94,ZH_97}. The field strength estimates all 
fall in the range of $1$ to $5\,\mu{\rm G}$, and there is an absence of observational 
evidence for cosmological evolution of magnetic field strengths! In addition,
there also exist some recent observational data which indirectly indicate the 
existence of rather strong magnetic fields in the distant past. The observations 
of~ $^6$Li and $^7$Li abundances in old metal-pure halo stars by Lemoine \etal 
indicate a massive production of lithium isotopes by cosmic rays in a very early 
phase of the Milky Way Galaxy~\cite{LSTC_97}. Without magnetic fields the cosmic 
rays would escape and would not be able to produce the lithium isotopes. Other 
indirect evidence for magnetic fields in the past is the primordial star 
formation process, which must have happened according to the observed 
metallicities of globular clusters. To form stars at early times ($z\sim 3$), 
primordial magnetic fields are believed to be required in order to remove significant 
angular momentum of self-gravitating gas by the magnetic braking effect~\cite{PS_89}.

Strong magnetic fields have been observed not only in galaxies, local and at high 
redshifts, but also in clusters of galaxies~\cite{E_99,K_94,ZH_97}. The most recent 
direct observations of the Coma Cluster by Fusco-Femiano \etal~\cite{FFFGGMMS_99} 
manifest a volume averaged intracluster magnetic field of ${}\sim 0.15\,\mu{\rm G}$ 
in the cluster. Fusco-Femiano \etal used the inverse Compton scattering of 
relativistic electrons on the cosmic microwave background photons to model the hard 
nonthermal X-ray radiation flux from the Coma Cluster. As a result, they estimated 
the electron density in the cluster. Then, they used the synchrotron radio flux to 
estimate the field strength. In 2000 Sarazin and Kempner revised the results of 
Fusco-Femiano \etal. Sarazin and Kempner used nonthermal bremsstrahlung radiation 
models for the hard X-ray emission from the Coma Cluster, and depending on the 
model they used, they estimated the field strength to be ranging from 
${}\sim 0.4\,\mu{\rm G}$ (as also implied by equipartition in the radio halo) to 
${}\sim 6\,\mu{\rm G}$ (as the typical field in individual galaxies in the 
cluster)~\cite{SK_00}.

Thus, galaxies, early galaxies (at $z\sim 2$ redshift) and galaxy clusters do 
have strong magnetic fields, with the strengths 
$0.1\,\mu{\rm G}\simlt B\simlt 10\,\mu{\rm G}$! Where did these strong fields
come from?

\section
[The Origin of Galactic and Extragalactic Magnetic Fields:\\ 
Primordial and Galactic Dynamos]
{The Origin of Galactic and Extragalactic\\
Magnetic Fields: Primordial and\\ 
Galactic Dynamos}
\label{ORIGIN_OF_FIELDS}

\begin{figure}[!p]
\vspace{12.0truecm}
\includegraphics{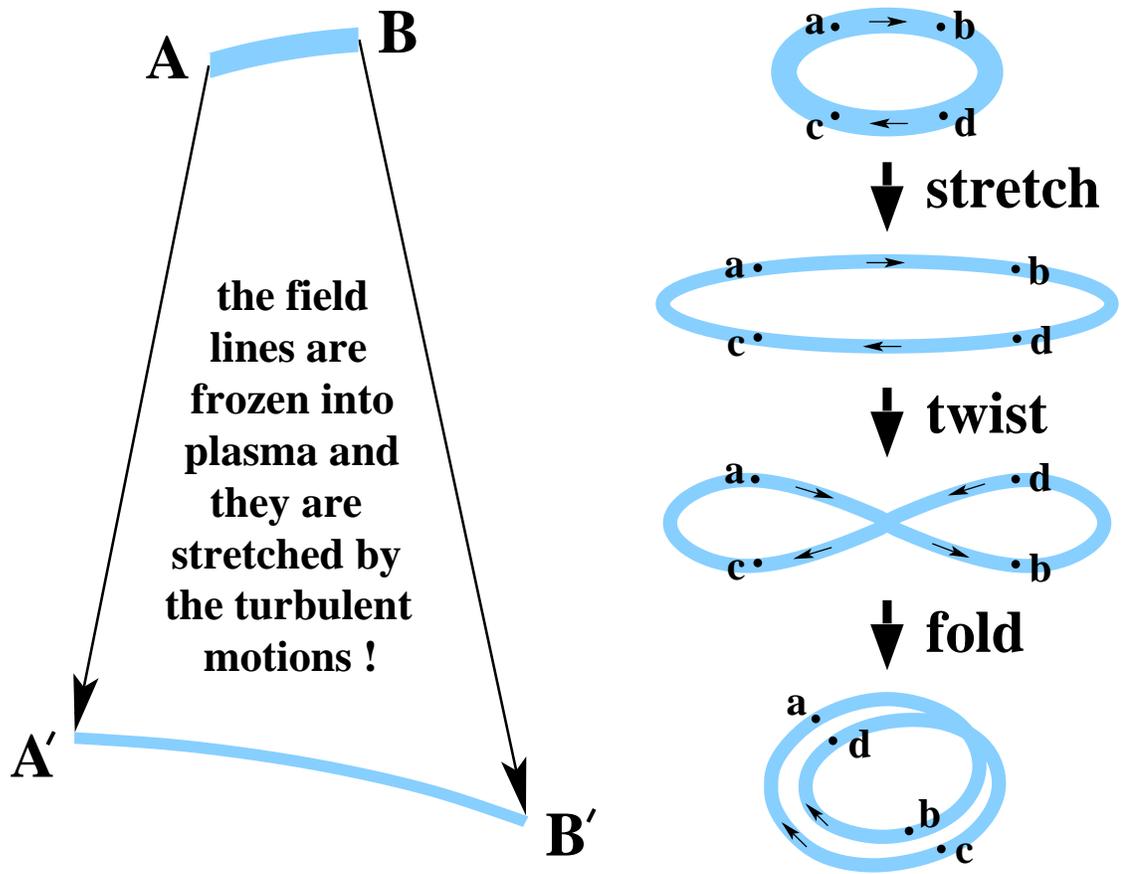}
\caption{
Left plot: stretching of magnetic field lines in a turbulent plasma because 
of the Kolmogorov-Lyapunov exponential divergence of closely neighboring 
points. Right plot: three-dimensional Zeldovich ``figure 8'' dynamo mechanism.
}
\label{FIGURE_DYNAMO}
\end{figure}

The prevailing theory for the origin of strong cosmic magnetic fields is that they 
were produced by the turbulent dynamo inductive action driven by the fluid motions 
in galactic and/or protogalactic medium.
The turbulent dynamo works as follows. The astrophysical plasmas are very hot 
and they have low densities (e.~g.~see Table~\ref{TABLE_PARAMETERS}). Therefore, the 
effect of electrical resistivity is negligible over an extremely broad range of scales. 
As a result, the magnetic field lines are frozen into plasma~\cite{LL_84,VZ_72}.
Because in a turbulent plasma the distance between any two infinitesimally neighboring 
points increases exponentially in time (the Kolmogorov-Lyapunov exponentiation), the 
magnetic field lines are stretched and the field strength grows exponentially 
fast~\cite{K_68,VZ_72}, see the left plot in Figure~\ref{FIGURE_DYNAMO}. 
There is an additional mechanism for the growth of the mean field strength in the 
case of three-dimensional turbulent dynamos, which is called the Zeldovich 
``figure 8'' mechanism~\cite{LL_84,VZ_72}. This mechanism is shown in the 
right plot of Figure~\ref{FIGURE_DYNAMO}, where the closed magnetic field lines are 
frozen into the torus. The field strength is increased exponentially fast in time 
by repetition of first, stretching the torus, second, twisting it into ``figure 8'', 
and third, folding it.

Yet full understanding of all stages of production of the strong galactic and 
extragalactic magnetic fields by the turbulent dynamo action has not been achieved. 
There are two alternative theories on how and when the fields have been 
produced.

The first theory, the galactic dynamo theory, also known as the $\alpha$--$\Omega$
dynamo theory, states that the fields have been primarily amplified in differentially 
rotating galactic disks after the galaxies had been formed. This theory considers 
the evolution of a large-scale mean magnetic field in a galactic disk. The turbulent 
dynamo action and the differential rotation of the disk are parametrized 
by transport coefficients. These transport coefficients enter the evolution 
equation for the mean field and they represent the destruction and the induction of 
the mean field~\cite{BBMSS_96,P_70,P_71,P_79,V_70,VR_72a,VR_72b,VZ_72}. 
The typical time scale of the mean field growth in the galactic disk is approximately 
equal to the disk rotation period, ${}\sim 300$~million years. The galactic dynamo 
theory is a beautiful theory, but it involves several crucial unsolved 
problems~\cite{CV_91,GD_94,K_99,K_00,RD_89,ZH_97,VC_92}. First, in the 
$\alpha$--$\Omega$ theory it seems to be extremely difficult to expel a fraction of 
the magnetic flux from the galactic disk in order to produce the net magnetic 
flux~\cite{K_99,RK_00}. The second problem is that the effect of small-scale 
fields on the amplification of the mean magnetic field remains unclear. These 
small-scale fields are amplified on the smallest turbulent eddy turnover time 
scale, much faster than the mean field is being amplified (see 
Figure~\ref{FIGURE_FIELD_SCALES}). Thus, the small-scale fields must saturate 
and possibly significantly alter the $\alpha$--$\Omega$ dynamo action well
before the mean field is amplified appreciably~\cite{K_99,KA_92}.
The third problem is that the galactic dynamo theory requires rather large seed fields
in the distant past, in order to successfully build up the fields observed at the 
present time. The Biermann battery itself and other similar effects do not seem to 
be able to provide the required seed fields~\cite{BPSS_94,K_99}. Finally, as we wrote 
in Section~\ref{MAGNETIC_FIELDS}, the observational data indicate that galaxy clusters
and galaxies at low and high redshifts possess magnetic fields of up to several 
microgauss. It is hard to explain these strong fields by the galactic dynamo 
theory~\cite{ZH_97}. On one hand, the galactic dynamo just did not have enough time 
to build microgauss fields in early galaxies at, say, redshift $z\sim 2$. On the 
other hand, the fields in clusters of galaxies are several orders of magnitude 
larger than the fields that would be obtained from galactic ejecta alone. Thus, 
sufficiently strong initial magnetic fields must be generated during the 
pregalactic era!

In this thesis we accept the second theory for the origin of cosmic 
magnetic fields, {\it the primordial dynamo theory}, which states that the galactic 
and extragalactic magnetic fields have primarily been produced in protogalaxies, 
i.~e.~before the galaxies were formed~\cite{K_99,K_00,KA_92,PS_89}. Of course, these 
fields were subsequently modified in the rotating galactic disks after the galaxies 
were formed~\cite{BPSS_94}.

\begin{table}[p]
\begin{tabular}{||l|c|c|c||}
\hline
\hline
Parameter				&Notation			
	&Value$^{*\,}$ 			&Scaling$^{*\,}$\\
\hline
\multicolumn{4}{||c||}{Physical Quantities\vphantom{\Large !}} \\
\hline
{\bf total mass, g}			&{\boldmath$M$}
	&{\boldmath$2\times 10^{45}$} 	&{\boldmath$\bf\sim 10^{12}$$\:{\rm M}_\odot$}\\
\hline
{\bf total{\boldmath$/$}baryon mass ratio}   &{\boldmath$\xi$}
	&$\bf 10$ 			&\\
\hline
temperature, K				&$T\sim T_i\sim T_e$		
	&$2\times10^6$ 			&$ML^{-1}$\\
\hline
ion \& $e^-$ density, cm$^{-3}$ 	&$n$				
	&$5\times10^{-4}$ 		&$\xi^{-1}ML^{-3}$\\ 
\hline
neutral density, cm$^{-3}$		&				
	&$0$ 				&\\ 
\hline
ion thermal speed, cm$/$s	 	&$V_T$				
	&$2\times10^7$	 		&$M^{1/2}L^{-1/2}$\\ 
\hline
ion viscosity, cm$^2/$s			&$\nu\equiv\nu_i\sim V_T^2t_i$	
	&$5\times10^{26}$ 		&$\xi M^{3/2}L^{1/2}$\\
\hline
$e^-$ thermal speed, cm$/$s	 	&$(m_i/m_e)^{1/2}V_T$  		
	&$9\times10^8$			&$M^{1/2}L^{-1/2}$\\
\hline
$e^-$ viscosity, cm$^2/$s 		&$\sim(m_e/m_i)^{1/2}\nu_i$  
	&$10^{25}$			&$\xi M^{3/2}L^{1/2}$\\
\hline
neutral viscosity, cm$^2/$s		&				
	&$0$ 				&\\
\hline
magnetic diffusivity, cm$^2/$s 	 	&$\eta_s$			
	&$8\times 10^4$	 		&$M^{-3/2}L^{3/2}$\\
\hline
smallest eddy speed, cm$/$s		&$V_\nu\sim R^{-1/4}V_T$		
	&$2\times 10^6$			&$\xi^{1/4}M^{3/4}L^{-1/2}$\\
\hline
\multicolumn{4}{||c||}{Dimensionless numbers\vphantom{\Large !}} \\
\hline
hydrodynamic Reynolds			&$R\sim V_TL/\nu$		
	&$3\times10^4$			&$\xi^{-1} M^{-1}$\\
\hline
magnetic Reynolds			&$R_m\sim V_TL/\eta_s$		
	&$2\times 10^{26}$		&$M^2L^{-1}$\\
\hline
Prandtl number				&$Pr\sim\nu/\eta_s$		
	&$6\times10^{21}$		&$\xi M^3L^{-1}$\\
\hline
field$/$smallest eddy energy  		&$B^2/4\pi m_p n V_\nu^2$ 	
	&$\!3\times10^{13}B^2\!$	&$\xi^{1/2}M^{-5/2}L^4B^2\!$\\
\hline
\multicolumn{4}{||c||}{Length Scales\vphantom{\Large !}} \\
\hline
{\bf system size, cm}			&{\boldmath$L$}				
	&{\boldmath$6\times10^{23}$}	&{\boldmath$\sim 0.2$}~{\bf Mpc}\\
\hline
viscous cutoff scale, cm    		&$2\pi k_\nu^{-1}\sim R^{-3/4}L$	
	&$3\times10^{20}$		&$\xi^{3/4}M^{3/4}L$\\
\hline
ion mean free path, cm  		&$\lambda_i=V_Tt_i\sim R^{-1}L$  
	&$7\times10^{19}$  		&$\xi ML$\\
\hline
ion gyroradius, cm			&$r_i=V_T/\omega_i$		
	&$2\times10^3/B$  		&$M^{1/2}L^{-1/2}B^{-1}$\\
\hline
$e^-$ mean free path, cm		&$\sim(m_e/m_i)^{1/2}\lambda_i$  
	&$2\times10^{18}$  		&$\xi ML$\\
\hline
$e^-$ gyroradius, cm			&$\sim(m_e/m_i)^{1/2}r_i$  	
	&$50/B$				&$M^{1/2}L^{-1/2}B^{-1}$\\
\hline
resistive cutoff scale, cm 		
	&$\!\!2\pi k_{\eta_s}^{-1}\!\!\sim\!Pr^{-1/2\,}\!R^{-3/4\,}\!L\!\!$
	&$4\times10^{9}$ 		&$\xi^{1/4}M^{-3/4}L^{3/2}$\\
\hline
Debye length, cm			&$(k_{\rm B}T/2\pi ne^2)^{1/2}$		
	&$6\times 10^5$			&$\xi^{1/2} L$\\
\hline
\multicolumn{4}{||c||}{Time Scales\vphantom{\Large !}} \\
\hline
gravitational collapse time, s		&$\sim L/V_T$		
	&$3\times 10^{16}$		&$M^{-1/2}L^{3/2}$\\
\hline
largest eddy turnover time, s		&$\sim L/V_T$	
	&$3\times10^{16}$		&$M^{-1/2}L^{3/2}$\\
\hline
$\!$smallest eddy turnover time, s$\!\!$ 	&$\sim R^{-1/2}L/V_T$
	&$2\times 10^{14}$		&$\xi^{1/2}L^{3/2}$\\
\hline
ion collision time, s			&$t_i$  	
	&$3\times10^{12}$ 		&$\xi M^{1/2}L^{3/2}$\\
\hline
ion cyclotron frequency, s$^{-1}$	&$\omega_i=eB/m_i c$
	&$9\times 10^3B$  		&$B$\\
\hline
$e^-$ collision time, s		 	&$\sim(m_e/m_i)^{1/2}t_i$  
	&$7\times10^{10}$ 		&$\xi M^{1/2}L^{3/2}$\\
\hline
$e^-$ cyclotron frequency, s$^{-1}$ 	&$(m_i/m_e)\omega_i$ 
	&$2\times10^7B$  		&$B$\\
\hline
\hline
\end{tabular}
\caption{
Physical parameters in a protogalaxy calculated for a fully ionized hydrogen plasma, 
assuming definite values for the parameters printed in boldface.~$^{*\,}$The field $B$ 
in column ``Value'' is expressed in gauss, the Coulomb logarithm is taken to be $30$.
}
\label{TABLE_PARAMETERS}
\end{table}

\begin{figure}[!p]
\vspace{11.5truecm}
\includegraphics{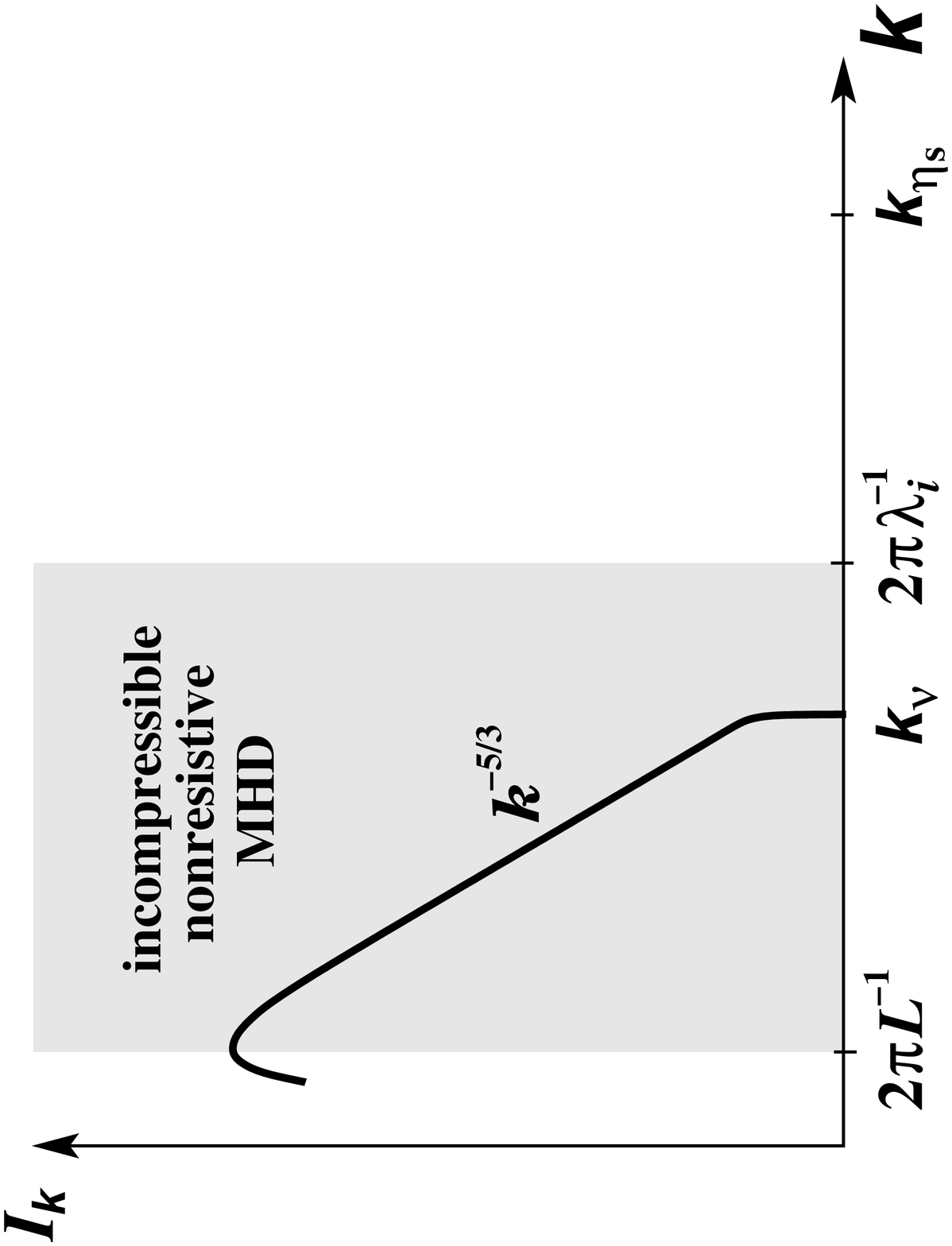}
\caption{
The Kolmogorov spectrum of kinetic energy, $I_k$, is shown by the thick solid 
line. The $x$-axis has the wave number $k$ plotted on it, and demonstrates the 
hierarchy of scales in a protogalaxy: $L$ is the system size, $2\pi k_\nu^{-1}$ 
is the viscous cutoff scale, $\lambda_i$ is the ion mean free path, and 
$2\pi k_{\eta_s}^{-1}$ is the resistive scale. Note that 
$L \gg 2\pi k_\nu^{-1} \gg \lambda_i \gg 2\pi k_{\eta_s}^{-1}$. The 
equations for a nonresistive incompressible MHD turbulence are valid only inside 
the range of scales shown by the shaded area (see also footnote~\ref{COMMENT_MFP} 
on page~\pageref{COMMENT_MFP}).
}
\label{FIGURE_LENGTH_SCALES}
\end{figure}

In order to understand how the magnetic fields can be built up in protogalaxies,
let us first discuss the physical conditions that were present there. The typical
values of physical parameters in a protogalaxy are given in
Table~\ref{TABLE_PARAMETERS}, and the hierarchy of scales is displayed on the
$x$-axis of Figure~\ref{FIGURE_LENGTH_SCALES}. The most important facts are
the following.
\begin{itemize}
\item
First, the gas is very hot in a protogalaxy, so it is fully ionized. Therefore,
the viscosity is dominated by ions, not by neutrals. In addition, the Spitzer
resistivity~\cite{S_62} is tiny, and thus, the resistive cutoff scale for the 
magnetic field, $2\pi k_{\eta_s}^{-1}$, is extremely small compared to the viscous
cutoff scale for the turbulent velocities, $2\pi k_\nu^{-1}$. As a result, we can
neglect resistivity as long as we consider scales $\gg 2\pi k_{\eta_s}^{-1}$, see
Figure~\ref{FIGURE_LENGTH_SCALES}.
\item
Second, the Reynolds number is very large. Therefore, the turbulence develops over
a very broad range of scales, starting at the largest scale, $L$, and ending at
the viscous cutoff scale, $2\pi k_\nu^{-1}$ (see Figure~\ref{FIGURE_LENGTH_SCALES}). 
The velocities at the largest scale are of the order of the sound speed, which is
approximately equal to the thermal speed. All turbulent velocities at smaller
scales are smaller. As a result, we can treat the plasma as incompressible on
scales $2\pi k^{-1}\ll L$.
\item
Third, the ion mean free path $\lambda_i$ is considerably shorter than the
viscous cutoff scale for the turbulent velocities, $2\pi k_\nu^{-1}$. The Debye 
length and the resistivity scale are even much shorter than $\lambda_i$. Thus, we 
can use the single-fluid magnetohydrodynamic (MHD) equations for description of
a nonresistive incompressible plasma on scales
$\lambda_i\simlt 2\pi k^{-1}\simlt L$.~\footnote{
We should consider scales larger than $\lambda_i$ in order to use MHD description
for the plasma. However, for estimation purposes, one can use MHD equations even 
for smaller scales, $2\pi k^{-1}\ll\lambda_i$, if one reduces the molecular 
viscosity by the ratio of the scale to the ion mean free path, i.~e.~by factor 
$1/(k\lambda_i)\ll 1$.
\label{COMMENT_MFP}
}
\item
Finally, it is very important that the cyclotron period of ions in the 
magnetic field is shorter than the ion collision time, $\omega_i^{-1}\ll t_i$, 
provided that the field strength is larger than ${}\sim 10^{-16}\,{\rm G}$ 
(see Table~\ref{TABLE_PARAMETERS}). On the other hand, the energy of the magnetic 
field becomes comparable to the kinetic energy of the smallest turbulent eddies
(which are on on the viscous cutoff scale) if the field exceeds 
${}\sim 10^{-7}\,{\rm G}$. As a result, there is a very broad range of magnetic 
field strength values at which the magnetic pressure and tension are still negligible,
while the presence of the field is already important. This is because the plasma is 
strongly magnetized, and the magnetic field controls the microscopic motions of ions, 
so that the plasma viscosity is different from that in a field-free plasma.
\end{itemize}

\begin{figure}[!p]
\vspace{17.5truecm}
\includegraphics{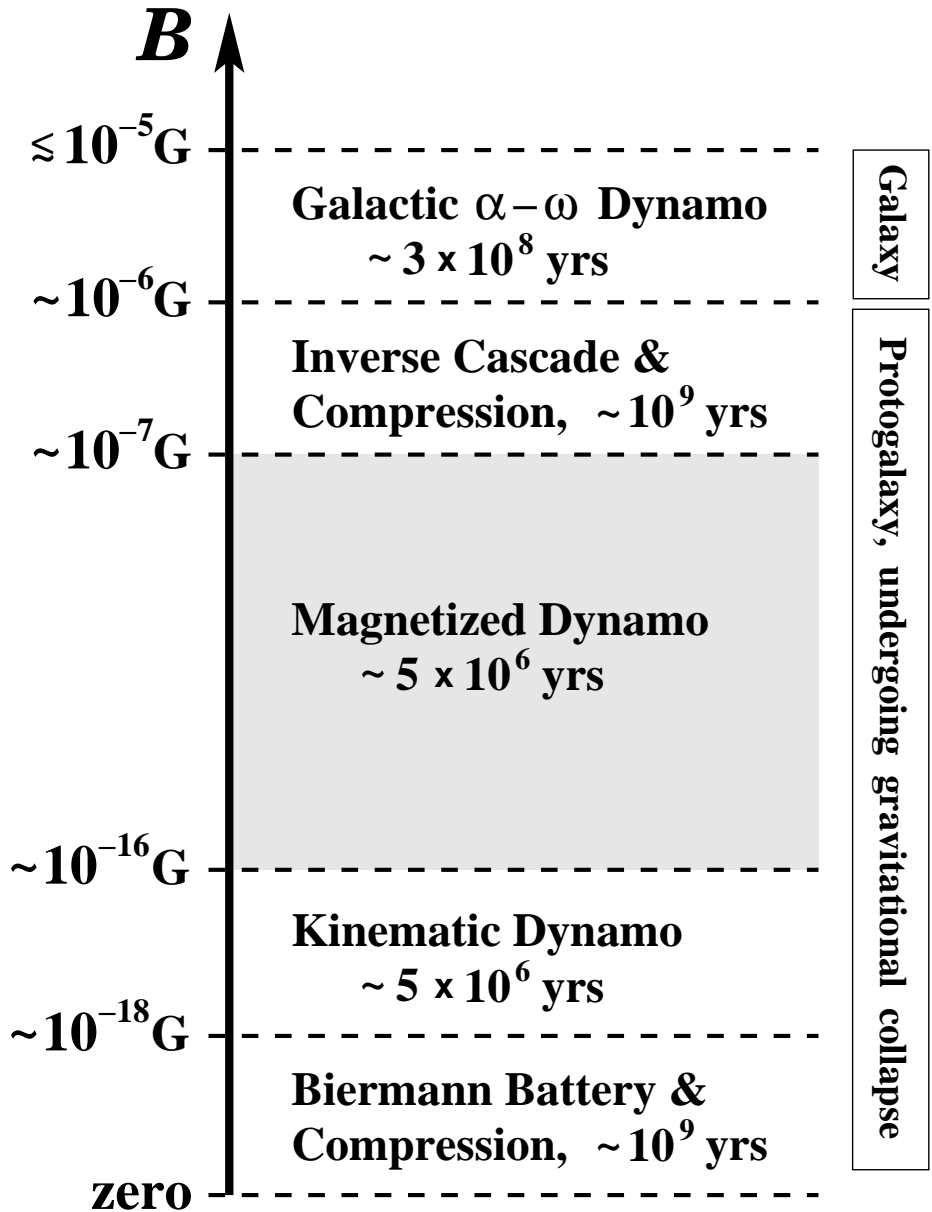}
\caption{
There are five major stages of the production of strong galactic and 
extragalactic magnetic fields, as the field strength grows from zero up to 
${}\simlt 10^{-5}\,{\rm gauss}$. The results of this thesis apply to the 
magnetized turbulent dynamo stage in protogalaxies, shaded on the plot. This 
stage makes the largest contribution to the built up of the magnetic fields.
}
\label{FIGURE_FIELD_SCALES}
\end{figure}

The primordial dynamo theory is as follows.
We believe that there are five major stages of the production of the strong 
magnetic fields. Four of them happen in protogalaxies, and one happens in 
galaxies, see Figure~\ref{FIGURE_FIELD_SCALES}. These stages are the following:
\begin{enumerate}
\item
During the first stage, in a protogalaxy undergoing gravitational collapse, the 
Biermann battery~\cite{B_50} builds a seed magnetic field linearly in time. The
Biermann battery requires the pressure to be non-barotropic, which is achieved 
behind hydrodynamic shocks driven by the gravitational instability~\cite{KCOR_97}.
The resulting seed field is of the order of 
$10^{-21}$--$10^{-19}\,{\rm G}$ at the largest scales~\cite{DW_00,KCOR_97,PS_89}. 
The Biermann battery inductive action is proportional to the vorticity of the 
turbulence, which is $R^{1/2}$ times larger at the viscous scales compared to 
the vorticity at the largest scales ($R\gg 1$ is the Reynolds number). 
Therefore, the seed field at the smallest scales of the turbulence (the viscosity 
scales), is about ${}\sim 10^{-18}\,{\rm G}$. The time scale of the Biermann 
battery action is approximately equal to the free-fall time in a protogalaxy, 
${}\sim 1$~billion years.
\item
During the second stage, when the plasma is unmagnetized and $\omega_i^{-1}\gg t_i$, 
the seed field is exponentially amplified by {\it the kinematic turbulent dynamo} 
inductive action. The kinematic dynamo builds the field up to approximately 
$10^{-16}\,{\rm G}$, when the plasma becomes magnetized, $\omega_i^{-1}\ll t_i$, 
and the field becomes strong enough to affect microscopic processes in plasma, such 
as the viscosity. The time scale of the kinematic dynamo is very short, it is of 
the order of the smallest eddy turnover time, ${}\sim 5$~million years. The main 
assumption of the kinematic dynamo theory and, basically, the strict definition of 
it, are that the growing magnetic field stays so weak, that it does not affect the 
fluid motions, i.~e.~that there is no {\it the back reaction} of the field on the 
turbulence. The kinematic dynamo has been intensively studied during the last several 
decades (see the references given in Section~\ref{KINEMATIC_DYNAMO}). Up to now it 
has been customary to assume that there is no the back reaction, and the main 
assumption of the kinematic dynamo is valid, as long as the magnetic energy is less 
than the turbulent kinetic energy, so that the Lorentz forces are small. However, 
the plasma becomes strongly magnetized and the back reaction sets in much earlier, 
at field strengths much lower than those which correspond to the energy equipartition 
between the field and the turbulence (see Figure~\ref{FIGURE_FIELD_SCALES}). Thus, 
the applicability of the kinematic dynamo theory to protogalaxies is rather limited. 
In Section~\ref{KINEMATIC_DYNAMO} we discuss the main results obtained for the 
kinematic dynamo theory. The results we obtain in this thesis reduce to those of 
the kinematic dynamo theory in the limit of very weak field strength (i.~e.~when 
the plasma is unmagnetized, and there is no the back reaction).
\item
The third stage starts when the field grows above ${}\sim 10^{-16}\,{\rm G}$, the 
ion cyclotron time in the magnetic field becomes shorter than the ion collision 
time, and the plasma becomes strongly magnetized. As a result, the magnetic field 
starts to strongly affect the dynamics of the turbulent motions on the viscous
scales~\footnote{
Note that the turbulent motions on the viscous scales are the most important 
in the dynamo theory. This is because the magnetic field is primarily amplified 
by the turbulent eddies on these scales. These eddies are the smallest ones,
they have the shortest turnover times and produce the largest velocity 
shearings~\cite{KA_92}.
\label{VISCOUS_IMPORTANT_SCALES}
},
by completely changing the viscosity (see Section~\ref{MAGNETIZED_DYNAMO}), despite 
the fact that the field energy is still very small compared to the kinetic energy
of the turbulence! We call this stage as {\it the magnetized turbulent dynamo} 
(see Figure~\ref{FIGURE_FIELD_SCALES}). In previous theories this stage has not 
been recognized. The main goal of this thesis is to construct a theoretical model 
for it. The time scale of the magnetized dynamo is roughly about the same as the 
time scale of the kinematic dynamo, ${}\sim 5$~million years.
\item
So far the field scale is of the order of the viscous scale or less, and the 
magnetic field is incoherent in space. The fourth stage starts when the magnetic 
field strength grows up to ${}\sim 10^{-7}\,{\rm G}$. The field energy becomes 
comparable to the kinetic energy of the smallest turbulent eddies (the energy 
equipartition), and the Lorentz forces become dynamically important in the plasma. 
During this stage the turbulent motions are dissipated by the growing field, the 
turbulent energy spectrum becomes truncated at larger and larger scales, and the 
turbulent energy is eventually transfered into a large-scale strong magnetic field 
with its energy comparable to the kinetic energy of the fluid motions on the largest 
scales in the protogalaxy (which is approximately the same as the thermal energy). 
This process is called the inverse cascade~\cite{B_01,BBMSS_96,KA_92,PFL_76,V_82}.
We discuss it only qualitatively in Chapter~\ref{DISCUSSION}. The time scale 
available for the inverse cascade process is of the order of the largest eddy 
turnover time, ${}\sim 1$~billion years. The inverse cascade may not have time
to amplify the field up to microgauss values, which are observed in galaxies.
A crucial question is how far it goes. This thesis addresses this question and 
is concerned with the rate of the magnetic field built up by the magnetized 
dynamo.
\item
Finally, the fifth stage is the galactic dynamo, which happens in the 
differentially rotating galactic disc after the galaxy is formed (see the 
discussion of the galactic dynamo above and Figure~\ref{FIGURE_FIELD_SCALES}). 
This process modifies the strong field that was built up in the protogalaxy in 
advance~\cite{HK_97}. The galactic dynamo may also further amplify the mean magnetic 
field in the disk by a factor of three or so, on a time scale of the order of the 
rotation time in the galaxy, ${}\sim 300$~million years. The galactic dynamo theory 
is beyond the scope of this thesis.
\end{enumerate}

As we discussed in the beginning of this section, the galactic mean field dynamo 
theory seems to be unable, by itself, to build magnetic fields from the seed
fields of ${}\sim 10^{-18}$~gauss, which are provided by the Biermann battery action, 
up to ${}\simgt 10^{-6}$~gauss, which exist in the present Universe. The observational 
evidences and the theoretical problems of the galactic dynamo theory, discussed above, 
persuade us to accept the primordial dynamo theory, which states that the observed 
cosmic fields were primarily built up in protogalaxies. The primordial dynamo 
theory seems to be free of the shortcomings of the galactic dynamo theory. The 
turbulent dynamo in protogalaxies does seem to be able to amplify the seed field up 
to ${}\sim 10^{-7}\,{\rm G}$ in a time interval ${}\simlt 10^7$ years, which is 
negligible in comparison to the Hubble time. 
However, up to now, there was one major deficiency in the primordial dynamo theory. 
It is well known that the turbulent dynamo builds the magnetic fields first on 
small subviscous scales~\cite{KA_92}, while the observed cosmic fields have rather 
large correlation lengths. Therefore, the magnetic field lines must be unwrapped 
on the small scales by the Lorentz tension forces, while the field energy is 
transfered and amplified on larger scales during the inverse cascade stage. Up to 
now, there was a strong argument against the possibility of such unwrapping. This 
argument is as the following. Both theoretical calculations and numerical 
simulations~\cite{C_99,SCMM_01,SMOC_01} show that the magnetic field, built up by 
the standard dynamo action, has folding structure with the wave number perpendicular 
to the field lines much larger than the wave number parallel to the field lines, 
as demonstrated in Figure~\ref{FIGURE_FOLDING}. Therefore, the Lorentz tension 
forces, which are proportional to the parallel wave number, are not very large. 
As a result, the standard isotropic viscous forces, calculated for an unmagnetized
or partially ionized plasma, can balance the Lorentz tension 
forces~\footnote{
The magnetic pressure is balanced by the hydrodynamic pressure in an incompressible 
plasma.
},
and block the unwrapping of the field lines (see Chapter~\ref{DISCUSSION}). At the 
same time, the perpendicular wave number can grow in time, and the perpendicular 
scale of the field lines can unrestrictively 
decrease~\footnote{
Until it reaches the scales at which resistivity becomes important, see
Figure~\ref{FIGURE_LENGTH_SCALES} on page~\pageref{FIGURE_LENGTH_SCALES}.
}.

This anti-unwrapping argument~\cite{C_99} applies only if the viscous forces are 
isotropic. However, in the case of the magnetized turbulent dynamo in a fully ionized 
plasma, the magnetic field controls the microscopic motions of particles on the viscous 
scales. The transport of ion momentum across the field lines is inhibited, and the 
viscosity stress is given by the Braginskii tensor [see equation~(\ref{PI}) and the 
discussion before it]. As a result, in the magnetized dynamo theory the viscous forces 
do not prevent the unwrapping of the magnetic field lines by the Lorentz tension forces 
on the small scales and do allow the field to become converted into a large-scale field 
during the inverse cascade stage in a protogalaxy (see Chapter~\ref{DISCUSSION} for more 
discussion). On the other hand, the theory of turbulent dynamos with the Braginskii 
viscosity has not been worked out. It turns out that the dynamo theory is qualitatively 
modified by including the Braginskii viscosity. The main purpose of this thesis is to 
lay the theoretical groundwork for the magnetized turbulent dynamos.

\section
[The Thesis Preview]
{The Thesis Preview}
\label{THESIS_PREVIEW}

Let us briefly outline the thesis content. 

In Chapter~\ref{DYNAMOS} we give the main equations and the main results of the 
kinematic dynamo theory (in Section~\ref{KINEMATIC_DYNAMO}), and we formulate the basic 
equations of the magnetized dynamo theory (in Section~\ref{MAGNETIZED_DYNAMO}). 

In Chapter~\ref{VELOCITIES} we calculate the statistics of turbulent velocities, $\bf V$, 
in strongly magnetized plasmas. In Section~\ref{EXPANSION} we write down the quasilinear 
expansion in time of the MHD equations for both the velocities and the magnetic field, 
similar to the expansion used by Kulsrud and Anderson~\cite{A_92,KA_92}. In 
section~\ref{LAPLACE_TRANSFORM} we make use of the Laplace transformation in time to find 
the turbulent velocities. We assume that the tensor $b_{\alpha\beta}$, which is the product 
of two unit vectors in the direction of the magnetic field, see equation~(\ref{bb}), can 
be taken to be constant in space in the beginning of the expansion in time, 
$b_{\alpha\beta}=\b_\alpha\b_\beta={\rm const}$ at zero time. 
This is our first hypothesis. It basically relies on our assumption that in the case 
of the magnetized turbulent dynamo the magnetic field has a folding structure 
similar to the one that exists in the case of the kinematic turbulent dynamo 
(see Figure~\ref{FIGURE_FOLDING} and the discussion in Section~\ref{LAPLACE_TRANSFORM}).
We find that there are velocity modes which are not damped by the Braginskii viscous 
forces. These undamped velocity modes grow unrestrictively in time unless we incorporate 
the non-linear inertial terms of the MHD equations into our quasilinear expansion. 
In Section~\ref{DUMPING} we argue that the non-linear damping of the growing velocity 
modes may be included into our theory by allowing for rotation of velocity vectors 
relative to the magnetic field unit 
vectors~\footnote{
This rotation is essentially due to Coriolis forces that make the velocity rotate 
differently than the magnetic field direction.
}.
This is our second hypothesis. 
In section~\ref{FOURIER_TRANSFORM} we again solve for the turbulent velocities by now
including our effective rotational damping into the MHD equations and by making use of 
the Fourier transformation in time. As a result, we obtain the correlation functions 
for the turbulent velocities in a strongly magnetized plasma, similar to those given 
by equations~(\ref{U_STATISTICS}) and ~(\ref{U_U_STATISTICS}) for the Kolmogorov
velocities in the kinematic dynamo case. We find that, contrary to the Kolmogorov
velocities, the turbulent velocities in the magnetized plasma are strongly anisotropic, 
as one might expect, because the magnetic field sets ``a preferred axis in space''.

In Chapter~\ref{MAGNETIC_SPECTRUM} we use the correlation functions for the 
turbulent velocities found in Chapter~\ref{VELOCITIES} to calculate the energy 
spectrum of random magnetic fields in the magnetized plasma.
We start with calculations of the total magnetic energy growth rate in 
Section~\ref{MAGNETIC_ENERGY}. In Section~\ref{MODE_COUPLING_EQUATION} we derive 
the mode coupling integro-differential equation for the evolution of the magnetic 
energy spectrum. Our mode coupling equation is a more general version of
equation~(\ref{MODE_COUPLING_KA}) of Kulsrud and Anderson obtained in the kinematic 
dynamo theory~\cite{KA_92}. In Section~\ref{SMALL_SCALES} we consider the magnetic 
energy spectrum on small subviscous scales. On these scales our mode coupling equation 
greatly simplifies and becomes a homogeneous differential equation, similar to the
equation~(\ref{SMALL_SCALES_MODE_COUPLING_KA}) of Kulsrud and Anderson. We find
the Green's function solution of our differential equation.

Finally, in Chapter~\ref{DISCUSSION} we give our conclusions. We discuss the 
possibilities of further research on the magnetized turbulent dynamos.
We also discuss the peculiarities of the inverse cascade in a strongly magnetized 
turbulent plasma.

%% file: chapter_2.tex
\chapter
[Basic Dynamo Equations]
{Basic Dynamo Equations}
\label{DYNAMOS}

\section
[The Kinematic Turbulent Dynamo]
{The Kinematic Turbulent Dynamo}
\label{KINEMATIC_DYNAMO}

The main assumption of the kinematic turbulent dynamo theory is that the magnetic field
is weak and it does not affect the turbulent motions in plasma. Thus, the magnetic 
field $\bf B$, frozen into plasma, is evolved as a passive vector by the turbulent
velocities $\bf U$~\cite{LL_84}:
\beq
\partial_t B_\alpha=U_{\alpha,\beta} B_\beta - U_\beta B_{\alpha,\beta}.
\label{B_EVOLUTION_KA}
\eeq
Here and below we always assume summation over repeated indices. In order to shorten 
notations, we frequently use $\partial_t\define\partial/\partial t$, and spatial 
derivatives are assumed to be taken with respect to all indices that are listed 
after ``$\,,\,$'' signs~\footnote{ 
For example, $\partial_t B_\alpha\equiv\partial B_\alpha/\partial t$, 
$(U_\alpha B_\beta)_{,\gamma}\equiv B_\beta(\partial U_{\alpha}/\partial x_\gamma)+
U_\alpha(\partial B_{\beta}/\partial x_\gamma)$, and
$U_{\alpha,\beta\gamma}\equiv\partial^2U_{\alpha}/\partial x_\beta\partial x_\gamma$.
}.
We also assume that the turbulence is incompressible (see Section~\ref{ORIGIN_OF_FIELDS}).
In the kinematic dynamo theory the turbulent velocities $\bf U$ are assumed to be 
hydrodynamic and to be independent of the magnetic field. 

A vast number of papers have been written about the kinematic turbulent dynamo in order
to answer three main questions: first, ``How fast is the total magnetic energy built up 
by the kinematic dynamo action'', second, ``What is the magnetic energy spectrum on 
different scales?'', and third, ``What is the resulting geometrical structure of the 
magnetic field?''~\cite{BS_00,C_99,K_68,KN_67,KA_92,M_01,P_70,P_71,PFL_76,SBK_01,
SCMM_01, SK_01,V_70,V_82,VZ_72}. It turns out that in most cases the evolution of 
magnetic field is entirely determined by the specified two-point statistics of the 
turbulent velocities. 

Following the assumptions usually made in the kinematic dynamo theory, we assume that 
the turbulence is incompressible, homogeneous, isotropic and stationary (as it actually 
is in protogalaxies). We also neglect the helical part of the 
turbulence~\footnote{
The helicity is negligible in galaxies on the scales of the smallest turbulent eddy, 
which is the principal driver of the field evolution~\cite{KA_92}.
The helicity in protogalaxies is even smaller than that in galaxies.
}.
We assume that the statistics of the Fourier coefficients of the turbulent velocities,
\beq
{\tilde U}_{{\bf k}\alpha} (t)&=&
\frac{1}{L^3}\int\limits_{-L/2}^{L/2}
U_\alpha(t,{\bf r})\,e^{-i{\bf kr}}\,d^3{\bf r},
\label{FOURIER_K}
\\
{\tilde U}_{{\bf k}\alpha}(\omega)&=&
\frac{1}{\sqrt{2\pi}}\int\limits_{-\infty}^\infty 
{\tilde U}_{{\bf k}\alpha}(t)\,e^{i\omega t}\,dt,
\label{FOURIER_K_OMEGA}
\eeq
is given by the following formulas~\cite{KA_92}:
\beq
\langle {\tilde U}_{{\bf k}\alpha}(\omega)\rangle &=& 0, 
\label{U_STATISTICS}
\\
\langle {\tilde U}_{{\bf k}\alpha}(\omega){\tilde U}_{{\bf k'}\beta}(\omega')\rangle =
\langle {\tilde U}^*_{-{\bf k}\alpha}(-\omega){\tilde U}_{{\bf k'}\beta}(\omega')\rangle 
&=& J_{\omega k}\,\delta^\perp_{\alpha\beta}\,
\delta_{{\bf k'},{\bf -k}}\,\delta(\omega'+\omega). 
\qquad
\label{U_U_STATISTICS}
\eeq
Here and below $\langle ... \rangle$ means ensemble average over all realizations of 
the turbulence, $\delta_{{\bf k'},{\bf -k}}$ is the three-dimensional Kronecker 
symbol (equal to unity only if ${\bf k'}=-{\bf k}$),
$\delta(\omega'+\omega)$ is the Dirac $\delta$-function, 
\beq
\delta^\perp_{\alpha\beta}\define\delta_{\alpha\beta}-\k_\alpha\k_\beta,
\label{DELTA_PERP}
\eeq
$\delta_{\alpha\beta}$ is the one-dimensional Kronecker symbol, and ${\bf \k}={\bf k}/k$ 
is the unit vector along the wave vector $\bf k$. It is important that function 
$J_{\omega k}$, which stands for the normal (non-helical) part of the turbulence, 
depends only on the absolute values of $\omega$ and $\bf k$. Note that 
equations~(\ref{FOURIER_K}) and~(\ref{FOURIER_K_OMEGA}) represent the discrete 
three-dimensional Fourier transformation in space and the continuous one-dimensional 
Fourier transformation in time (see Appendix~\ref{TRANSFORMS}), so that the inverse 
Fourier transformations are
\beq
U_\alpha(t,{\bf r}) &=&
\sum_{\bf k} {\tilde U}_{{\bf k}\alpha}(t)\,e^{i{\bf kr}} = 
{\left(\frac{L}{2\pi}\right)}^{\!3}
\int\limits_{-\infty}^\infty {\tilde U}_\alpha(t,{\bf k})\,e^{i{\bf kr}}\,d^3{\bf k},
\label{INV_FOURIER_K}
\\
{\tilde U}_{{\bf k}\alpha}(t) &=&
\frac{1}{\sqrt{2\pi}}\int\limits_{-\infty}^\infty 
{\tilde U}_{{\bf k}\alpha}(\omega)\,e^{-i\omega t}\,d\omega.
\label{INV_FOURIER_K_OMEGA}
\eeq
Here the summation is done over discrete values of vector $\bf k$: $k_x=(2\pi/L)n_x$,
$k_y=(2\pi/L)n_y$, $k_z=(2\pi/L)n_z$, $n_x\in Z$, $n_y\in Z$, $n_z\in Z$; and
this summation can be replaced by integration of the appropriately defined function 
${\tilde U}_\alpha(t,{\bf k})$ that is continuous over $\bf k$, see
Appendix~\ref{TRANSFORMS}. Below we will use both summation and integration over 
$\bf k$, whichever is more convenient to use.

Let us further assume that the turbulence is Kolmogorov, and that the time correlation 
function of the turbulent velocities has an exponential 
profile~\footnote{
Using a Gaussian time correlation profile, 
$\langle {\bf U}(t) {\bf U}(t')\rangle\propto e^{-(t-t')^2/2\tau^2}$, would be 
more appropriate. In this case equation~(\ref{J_OMEGA_K}) would become 
$J_{\omega k}=J_{0k}e^{-\tau^2\omega^2/2}$. However, we prefer the exponential profile 
because it is easier to deal with. (In particular, for Gaussian integrals it is not 
possible to close integration contours at infinity in the complex plane.)
},
i.~e.~that $\langle {\bf U}(t) {\bf U}(t')\rangle\propto e^{-|t-t'|/\tau}$, 
where $\tau$ is the eddy decorrelation time~\cite{KA_92}, which depends 
on $k$,
\beq
\tau(k)=\tau(0)\left(\frac{k}{k_0}\right)^{\!-2/3}
=\frac{2\pi}{k_0U_0}\left(\frac{k}{k_0}\right)^{\!-2/3}.
\label{TAU}
\eeq
Here, $k_0=2\pi/L$ is the smallest wave number of the turbulence, and $U_0\sim V_T$ is 
the largest eddy velocity. In this case we have
\beq
J_{\omega k} &=& \frac{J_{0k}}{1+\tau^2\omega^2},
\label{J_OMEGA_K}
\\
J_{0k} &\approx& 
\cases{
(U_0/2k_0){(k/k_0)}^{\!-13/3},  &  $k_0\le k\le k_\nu$, \cr
0,  &  $k<k_0$, $k>k_\nu$,
}
\label{J_ZERO_K}
\eeq
and, carrying out the inverse Fourier transformations of equation~(\ref{U_U_STATISTICS}),
in time and in space, we have
\beq
\langle {\tilde U}_{{\bf k}\alpha}(t){\tilde U}_{{\bf k'}\beta}(t')\rangle &=&
\frac{J_{0k}}{2\tau} e^{-|t-t'|/\tau}\:
\delta^\perp_{\alpha\beta}\,\delta_{{\bf k'},{\bf -k}},
\label{U_U_TIME_CORRELATION}
\\
\langle U_\alpha(t,{\bf r})U_\beta(t',{\bf r'})\rangle &=&
\sum_{\bf k} \frac{J_{0k}}{2\tau}\,e^{-|t-t'|/\tau}\,\delta^\perp_{\alpha\beta}
\,e^{i{\bf k}({\bf r}-{\bf r'})}.
\label{U_U_TIME_SPACE_CORRELATION}
\eeq
The Kolmogorov kinetic energy spectrum, shown by the thick 
solid line in Figure~\ref{FIGURE_LENGTH_SCALES}, is~\cite{KA_92}
\beq
I(k) &\approx& 
\cases{
(U_0^2/k_0){(k/k_0)}^{-5/3},  &  $k_0\le k\le k_\nu$, \cr
0,  &  $k<k_0$, $k>k_\nu$,
}
\label{I_K}
\eeq
so that the total kinetic energy of the fluid motions, per unit mass, is
\beq
\frac{1}{2}\langle[{\bf U}(t,{\bf r})]^2\rangle = 
\frac{1}{2}\int\limits_{k_0}^{k_\nu} I(k)\,dk.
\label{TOTAL_KINETIC_ENERGY}
\eeq

Let us now return to the discussion of the magnetic field energy and of the magnetic 
field structure, the topics which are of most importance in the kinematic dynamo 
theory. The evolution of the magnetic field energy spectrum in a turbulent conducting 
plasma was first independently studied by Kazantsev~\cite{K_68}, and by Kraichnan and 
Nagarajan~\cite{KN_67} in 1967. Kraichnan and Nagarajan were interested in the limit 
of small Prandtl numbers, which is not the case in galaxies and protogalaxies.
Kazantsev obtained the correct equation for the evolution of the magnetic energy 
spectrum on small subviscous scales in the limit of large Prandtl numbers. 
He assumed that the turbulent velocities are $\delta$-correlated in time, 
$\langle{\bf U}(t){\bf U}(t')\rangle\propto\delta(t'-t)$. In 1982 Vainshtein 
derived a universal equation for the longitudinal correlation function of magnetic 
field~\cite{V_82}. He also proved, making only very general assumptions about the
turbulence, that the magnetic field is exponentially amplified by the turbulent 
motions (i.~e.~that the fast dynamo inductive action does exist). It is interesting 
that in spite of more general assumptions Vainshtein used, his equation for the 
magnetic energy spectrum on subviscous scales basically coincides with that of 
Kazantsev~\footnote{
The equation~(25) of Vainshtein and the equation~(31) of Kazantsev~\cite{K_68} 
are the same in the absence of resistivity, except there are different constant 
factors in front of their right-hand-side parts. The 
equation~(\ref{SMALL_SCALES_MODE_COUPLING_KA}) in this thesis, which was derived by 
Kulsrud and Anderson~\cite{KA_92}, would also be the same if converted to the 
differential equation for function $k^{-2}M$.
}.

The complete correct theory for the evolution of the magnetic field energy spectrum 
and for the growth of the total magnetic energy in the limit of large Prandtl numbers 
was first developed by Kulsrud and Anderson in 1992~\cite{KA_92}. They used the 
kinematic dynamo model and solved equation~(\ref{B_EVOLUTION_KA}) by the quasilinear 
expansion in time (by iterating it twice in time). Below we list the main results 
obtained by Kulsrud and Anderson, because in this thesis we will use calculational 
methods similar to those that they used, and the results we will obtain reduce to 
their results if the plasma is not magnetized (when the magnetic field is so weak, 
that $\omega_i t_i\ll 1$, and the kinematic dynamo model is valid, see 
Figure~\ref{FIGURE_FIELD_SCALES}).

Kulsrud and Anderson found that the ensemble averaged total magnetic energy 
per unit mass, ${\cal E}=\langle B^2\rangle/8\pi\rho$ ($\rho$ is the plasma density), 
grows exponentially in time:
\beq
\frac{d{\cal E}}{dt} &=& 2\gamma_{\rm o}{\cal E},
\label{ENERGY_GROWTH_KA}
\\
\gamma_{\rm o} &=& \frac{1}{3}\sum_{\bf k}k^2 J_{0k}.
\label{GAMMA_0}
\eeq
They defined the magnetic energy spectrum $M(t, k)$ as
\beq
M(t, k)\define \frac{1}{4\pi\rho}{\left(\frac{L}{2\pi}\right)}^{\!3}
\int k^2\,\langle|{\bf{\tilde B}}(t,{\bf k})|^2\rangle\,d^2{\bf\k}, 
\label{M_K}
\eeq
where the integration is carried out over all directions of ${\bf\k}={\bf k}/k$.
Here, we consider that the function ${\bf{\tilde B}_k}(t)$ is the Fourier coefficient 
of the magnetic field $\bf B$, while ${\bf{\tilde B}}(t,{\bf k})$ is the appropriately 
defined function continuous in $\bf k$, so that 
\beq
B_\alpha(t,{\bf r}) &=&
\sum_{\bf k} {\tilde B}_{{\bf k}\alpha}(t)\,e^{i{\bf kr}} = 
{\left(\frac{L}{2\pi}\right)}^{\!3}
\int\limits_{-\infty}^\infty {\tilde B}_\alpha(t,{\bf k})\,e^{i{\bf kr}}\,d^3{\bf k}
\label{INV_FOURIER_FIELD}
\eeq
(see definition of the Fourier transformations and coefficients in 
Appendix~\ref{TRANSFORMS}). The averaged total magnetic energy per unit mass is 
obviously equal to
\beq
{\cal E}=\frac{1}{2}\int\limits_0^\infty M(t,k)\,dk.
\label{TOTAL_MAGNETIC_ENERGY}
\eeq

Kulsrud and Anderson derived the equation for the evolution of the magnetic energy 
spectrum $M(t, k)$~\cite{A_92,KA_92}. They called it {\it the mode coupling equation}, 
which, in the absence of helicity, is 
\beq
\frac{\partial M}{\partial t}=\int\limits_0^\infty K_{\rm o}(k,k')M(t,k')\,dk' -
2\,\frac{\eta_{\mbox{\tiny$T$}{\rm o}}}{4\pi}\,k^2 M(t,k),
\label{MODE_COUPLING_KA}
\eeq
where~\footnote{
Note, that Kulsrud and Anderson~\cite{KA_92} used a slightly different definition of the
continuous Fourier transformation in time. As a result, they have additional factors
$2\pi$ in their formulas. 
} 
\beq
K_{\rm o}(k,k') &=& 2\pi k^4\,{\left(\frac{L}{2\pi}\right)}^{\!3}
\int\limits_0^\pi\,J_{0k''}\frac{k^2+k'^2-kk'\cos\theta}{k''^2}\,\sin^3\theta\,d\theta,
\label{COUPLING_KERNEL_KA}
\\
k'' &=& (k^2+k'^2-2kk'\cos\theta)^{1/2},
\label{K''}
\eeq
and $\theta$ is the angle between $\bf k$ and $\bf k'$, see 
Figure~\ref{FIGURE_k'_perp_bo_and_k}. The constant $\eta_{\mbox{\tiny$T$}{\rm o}}$ is 
the turbulent magnetic diffusivity:
\beq
\frac{\eta_{\mbox{\tiny$T$}{\rm o}}}{4\pi}=\frac{1}{3}\sum_{\bf k} J_{0k}.
\label{ETA_T_KA}
\eeq
Kulsrud and Anderson found that the magnetic energy cascades down to very small 
subviscous scales via the mode coupling equation~(\ref{MODE_COUPLING_KA}). In 
particular, this mode coupling equation greatly simplifies for scales much less 
than the viscous cutoff scale, $k\gg k_\nu$, and becomes a simple differential 
equation, homogeneous in $k$,
\beq
\frac{\partial M}{\partial t} = 
\frac{\gamma_{\rm o}}{5}\left(k^2\frac{\partial^2 M}{\partial k^2}-
2k\frac{\partial M}{\partial k}+6M\right).
\label{SMALL_SCALES_MODE_COUPLING_KA}
\eeq
If $M(t,k_{\rm ref})$ is known as a function of time at some reference wave number 
$k=k_{\rm ref}$, then the solution of~(\ref{SMALL_SCALES_MODE_COUPLING_KA}) is
\beq
M(t,k)=\int_{-\infty}^t M(t',k_{\rm ref})\,G_{\rm o}(k/k_{\rm ref},t-t')\,dt',
\label{SMALL_SCALES_SOLUTION_KA}
\eeq
where the Green's function $G_{\rm o}(k,t)$ is
\beq
G_{\rm o}(k,t)={\left(\frac{5}{4\pi}\right)}^{\!1/2}\,
\frac{k^{3/2}\ln{k}}{\gamma_{\rm o}^{1/2}t^{3/2}}\,
e^{(3/4)\gamma_{\rm o}t}\,e^{-5\ln^2k\big/4\gamma_{\rm o}t}.
\label{GREENS_FUNCTION_KA}
\eeq
A ``signal'' $M(t,k_{\rm ref})$ at zero time will increase exponentially as 
$e^{(3/4)\gamma_{\rm o} t}$ and will extend down to the scale 
$k_{\rm peak}\approx e^{\gamma_{\rm o} t}k_{\rm ref}$, where $k_{\rm peak}$ is the 
peak of function $kG_{\rm o}(k,t)$. As a result, the magnetic energy tends to quickly 
propagate to very small subviscous scales~\cite{A_92,KA_92}$\,$!

\begin{figure}[!p]
\vspace{11.5truecm}
\includegraphics{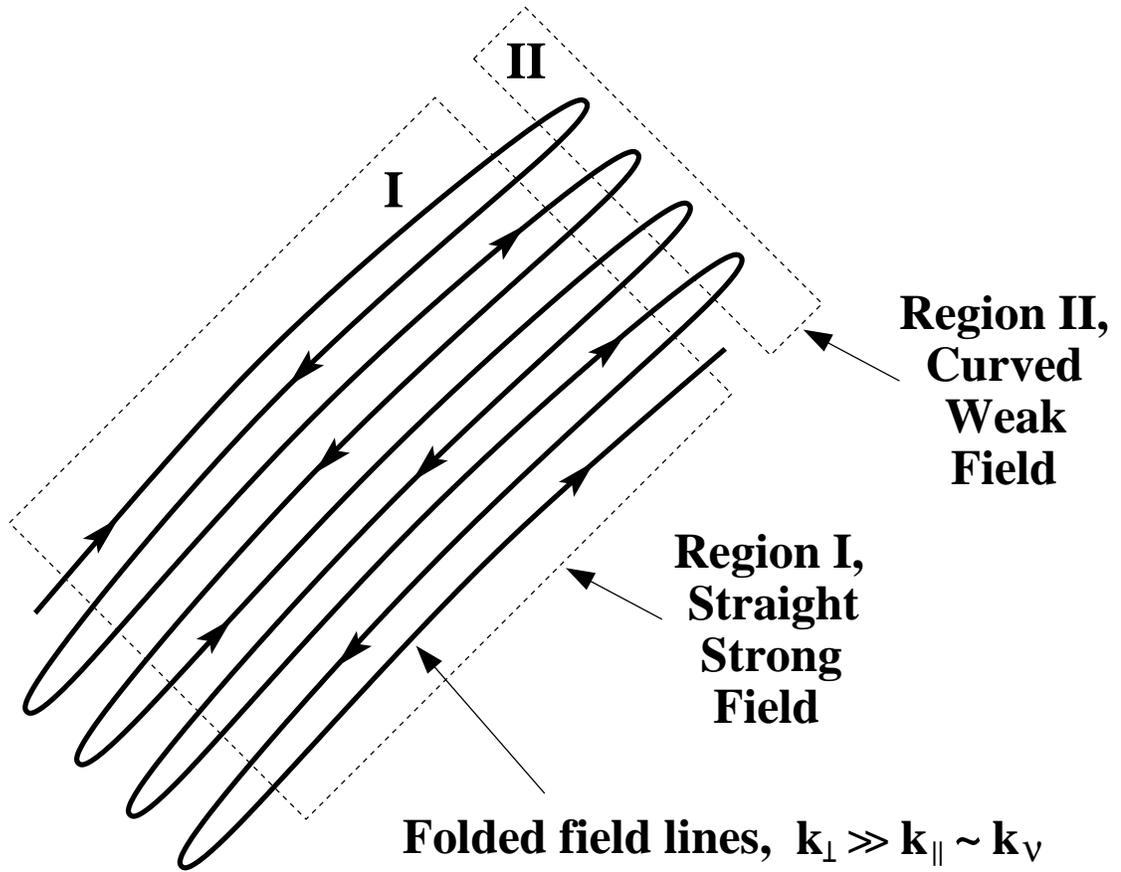}
\caption{
The folding structure of the magnetic fields produced by the kinematic dynamo 
(for simplicity shown in two-dimensions). The field is nearly straight and strong 
in Region I. The field is very curved but weak in Region II. 
}
\label{FIGURE_FOLDING}
\end{figure}

The next important question concerns the magnetic field geometrical structure,
which is produced by the kinematic dynamo action. This structure has recently been 
extensively studied. It was first found that the mean square curvature of the field 
grows exponentially in time, $\propto e^{(16/5)\gamma_{\rm o} t}$, and quickly becomes
much larger than the square of the viscous wave number $k_\nu$~\cite{M_01}. Even if 
the curvature is initially zero, it starts to grow because of the second order spatial 
derivatives of the turbulent velocities. Schekochihin~\etal then found the actual 
probability distribution for the magnetic field curvature and showed that the curvature 
stays of the order of the viscous wave number in most of the volume, but becomes 
extremely large in a small fraction of the volume~\cite{SCMM_01}. Moreover, it 
turns out that there is an anti-correlation between the value of the curvature and 
the strength of the magnetic field! Namely, the magnetic field is weak in those 
(small) regions of the volume, where the field curvature is large, and the field 
is strong in those (large) regions of the volume, where the curvature is weak. Thus,
the curvature stays finite (${}\sim{}$ the smallest eddy size) over the bulk of the 
volume, while the magnetic field energy becomes concentrated on very small 
subviscous scales, as found by Kulsrud and Anderson. Therefore, the field has 
{\it the folding structure} with the wave number perpendicular to the field lines 
much larger than the wave number parallel to the field lines, 
$k_\perp\gg k_\parallel\sim k_\nu$, as demonstrated in Figure~\ref{FIGURE_FOLDING}.
This folding structure was first suggested by Cowley in 1999~\cite{C_99} on  
intuitive grounds and was later supported by MHD numerical simulations. The 
folding nature of magnetic fields is also theoretically supported by the Barnes 
collisionless damping~\cite{B_67}, which states that small perturbations of 
the field (MHD waves) are strongly damped by collisionless energy transfer to 
the electrons~\footnote{
It works as follows. Sinusoidal MHD waves slightly perturb the field strength. 
As a result, the perpendicular kinetic energy of electrons oscillates in time. 
However, because the electrons reflect from magnetic mirrors, a part of the fluid 
kinetic energy is irreversibly transfered into the parallel kinetic energy of 
electrons (it is the second order effect). Thus, the electrons are heated, and 
the waves are damped.
}.
The folding structure of the magnetic field lines is very important for our
calculations of the magnetized dynamo!

\section
[The Magnetized Turbulent Dynamo]
{The Magnetized Turbulent Dynamo}
\label{MAGNETIZED_DYNAMO}

In a protogalaxy the magnetic field is quickly amplified by the kinematic dynamo action. 
When the field strength becomes of the order of ${}\sim 10^{-16}\,{\rm G}$ or higher, 
the ion cyclotron time in the magnetic field becomes less than the ion collision 
time, $\omega_i t_i\gg 1$, (and correspondingly, the ion gyroradius becomes shorter 
than the ion mean free path). As a result, the plasma becomes strongly magnetized, 
and the dynamics of the turbulent velocities on the viscous scales starts to be 
controlled by the magnetic field. These are the scales on which the velocities most 
rapidly amplify the field (see footnote~\ref{VISCOUS_IMPORTANT_SCALES} on 
page~\pageref{VISCOUS_IMPORTANT_SCALES}). As a result, the main assumption of the 
kinematic dynamo theory, the absence of the back reaction influence of the field on the 
turbulence, becomes invalid. The turbulent dynamo becomes what we call the magnetized 
turbulent dynamo, see Figure~\ref{FIGURE_FIELD_SCALES}.

It is important to know that the magnetic Lorentz forces can be neglected as long as 
the magnetic energy stays less than the kinetic energy of the smallest turbulent 
eddies, i.~e.~as long as the field strength stays less than ${}\sim 10^{-7}\,{\rm G}$ 
in a protogalaxy. During the magnetized dynamo stage in the protogalaxy, when 
$\omega_i t_i\gg 1$, but the magnetic energy is small, see 
Figure~\ref{FIGURE_FIELD_SCALES}, the magnetic field strongly affects 
the turbulent motions on the viscous scales by changing the viscous 
forces, while the Lorentz forces are negligible! Indeed, under the
condition that $\omega_i t_i\gg 1$, the transport of both the transverse and the 
longitudinal components of ion momentum in the direction perpendicular to the magnetic 
field lines is inhibited. The transport of the transverse component of ion momentum 
along magnetic field lines is also inhibited, because gyrating ions quickly lose 
their ``memory'' of the transverse ordered velocity in a time equal to the gyrorotation
time~\cite{B_65}. The transport of the longitudinal component of ion momentum along 
the magnetic field lines is the same as in the absence of the field.  As a result, 
the viscous forces acting on turbulent velocities $\bf V$ in an incompressible fully 
ionized plasma are determined by the Braginskii viscosity stress tensor~\cite{B_65}
\beq
\pi_{\alpha\beta}=-\nu(3\b_\alpha\b_\beta-\delta_{\alpha\beta})
\b_\mu\b_\nu V_{\mu,\nu}\;,
\label{PI}
\eeq
where ${\bf\b}={\bf B}/B$ is the unit vector along the magnetic field. Note, that 
this stress tensor depends on the field unit vector $\bf\b$, but is independent of 
the magnetic strength $B=|{\bf B}|$, i.~e.~the viscous forces depend only on 
the magnetic field direction (as long as the plasma is strongly magnetized,
and $\omega_i t_i\gg 1$). 

The MHD equations for the turbulent velocities $\bf V$ in an incompressible 
strongly magnetized plasma are~\cite{B_65,LL_84}
\beq
\partial_t V_\alpha &=&
-P'_{,\alpha}+f_\alpha-\pi_{\alpha\beta,\beta}-(V_\alpha V_\beta)_{,\beta}
\nonumber\\
{}&=&{}-P''_{,\alpha}+f_\alpha+3\nu(\b_\alpha\b_\beta\b_\mu\b_\nu V_{\mu,\nu})_{,\beta}
-(V_\alpha V_\beta)_{,\beta}\;,
\label{EQUATION_FOR_V}
\\
V_{\alpha,\alpha} &=& 0,
\label{DIV_V}
\eeq
where $\bf f$ is the force driving the turbulence. In particular, it can be thought 
as a force, driving a turbulent eddy, that comes from larger eddies. The hydrodynamic 
pressure $P'$ can be determined by the incompressibility condition~(\ref{DIV_V}). To 
obtain the second line in equation~(\ref{EQUATION_FOR_V}), we use formula~(\ref{PI})
for the viscous stress, and we incorporate the isotropic part of the viscous stress 
into the pressure $P''$ (for the same reason the force $\bf f$ can be assumed to be 
solenoidal).

It is difficult to solve equations~(\ref{EQUATION_FOR_V}) and~(\ref{DIV_V}) directly 
because they are non-linear and they include the complicated Braginskii viscous 
forces. We also do not know the exact expression of the driving force $\bf f$, but 
its statistics is the same as for an unmagnetized plasma. Therefore, let us proceed 
as follows. We introduce subsidiary incompressible turbulent velocities $\bf U$ which, 
by definition, satisfy equations
\beq 
\partial_t U_\alpha &=&
-P'''_{,\alpha}+f_\alpha+\frac{1}{5}\nu\triangle U_\alpha-(U_\alpha U_\beta)_{,\beta}\;,
\label{EQUATION_FOR_U}
\\
U_{\alpha,\alpha} &=& 0
\label{DIV_U}
\eeq
($\triangle U_\alpha=U_{\alpha,\beta\beta}\,$, and the pressure $P'''$ can be 
determined by the incompressibility condition). Let us analyze and compare 
equations~(\ref{EQUATION_FOR_V}) and~(\ref{EQUATION_FOR_U}). 
\begin{itemize}
\item
First, let us for the moment formally average the Braginskii viscosity term 
$3\nu(\b_\alpha\b_\beta\b_\mu\b_\nu V_{\mu,\nu})_{,\beta}$ in 
equation~(\ref{EQUATION_FOR_V}) over all directions of an isotropic magnetic field,
i.~e.~let us apply formula $\langle\b_\alpha\b_\beta\b_\mu\b_\nu\rangle_{\bf\b}=
(1/15)(\delta_{\alpha\beta}\delta_{\mu\nu}+\delta_{\alpha\mu}\delta_{\beta\nu}
+\delta_{\alpha\nu}\delta_{\beta\mu})$ to it. Then this term reduces to 
$(1/5)\nu\triangle V_\alpha$, which coincides with the isotropic viscosity 
term in equation~(\ref{EQUATION_FOR_U}). In other words, 
$\nu_{\rm eff}=(1/5)\nu$ could be considered as an effective reduced viscosity 
for an incompressible fully ionized plasma in the presence of a magnetic field 
that is isotropically tangled on subviscous scales.
\item
Second, note that equations~(\ref{EQUATION_FOR_V}) and~(\ref{EQUATION_FOR_U}) 
have the same driving force $\bf f$. By taking the driving force to be 
the same, we assume that this force comes from larger turbulent eddies. These 
larger eddies are on scales larger than the viscous scales, and therefore, 
these eddies ``do not know'' whether the viscous forces are of the Braginskii 
type or of the standard isotropic type.
\item
Third, note that equation~(\ref{EQUATION_FOR_U}) is a familiar hydrodynamic 
equation with a standard isotropic viscosity term. However, is has a reduced 
molecular viscosity, $(1/5)\nu$ instead of $\nu$. Therefore, we can suppose 
that the solution of equation~(\ref{EQUATION_FOR_U}) is the Kolmogorov 
turbulence with the effective reduced viscosity
\beq
\nu_{\rm eff}=\frac{1}{5}\nu.
\label{NU_EFF}
\eeq
As a result, we suppose that the statistical properties and the kinetic energy 
spectrum of the subsidiary velocities $\bf U$ are given by 
formulas~(\ref{U_STATISTICS}),~(\ref{U_U_STATISTICS}),~(\ref{J_OMEGA_K})--
(\ref{TOTAL_KINETIC_ENERGY}), where the Reynolds number $R$ and the viscous 
cutoff wave number $k_\nu$ are now determined by the effective viscosity, 
$\nu_{\rm eff}=(1/5)\nu$, i.~e.~$R\sim V_0L/\nu_{\rm eff}=5V_0L/\nu$, 
$k_\nu\sim R^{3/4}k_0=(5V_0L/\nu)^{3/4}k_0$.
\item
Fourth, we can say that because of the presence of the magnetic field, the 
Braginskii viscous forces ``are doing worse'' at dissipating the turbulent 
motions, as compared with the standard isotropic viscous forces in a field-free 
plasma. This is why the effective viscosity in equation~(\ref{EQUATION_FOR_U}) 
is reduced by a factor $1/5$. Thus, we can draw the important conclusion that 
the spectrum of turbulent velocities in a strongly magnetized plasma extends to 
a smaller scale, $2\pi k_\nu^{-1}\sim (5V_0L/\nu)^{-3/4}L$, as compared to the 
cutoff scale in an unmagnetized hydrodynamic plasma, 
$2\pi k_\nu^{-1}\sim (V_0L/\nu)^{-3/4}L$!~\footnote{
This conclusion is important because it means that in a magnetized plasma 
the smallest turbulent eddies, which are located on the viscous scales, 
have turnover times shorter than those of the smallest eddies in an 
unmagnetized plasma. As a result, the magnetic field amplification rate 
in the magnetized plasma is expected to be higher, compared to that in the 
unmagnetized plasma. See also footnote~\ref{VISCOUS_IMPORTANT_SCALES} on 
page~\pageref{VISCOUS_IMPORTANT_SCALES}.
}
\end{itemize}

Now let us subtract equations~(\ref{EQUATION_FOR_U}) and~(\ref{DIV_U}) from 
equations~(\ref{EQUATION_FOR_V}) and~(\ref{DIV_V}) to eliminate the unknown 
driving force $\bf f$, and let us introduce {\em the back-reaction velocity}~ 
$\vbf\define{\bf V}-{\bf U}$. We have
\beq
\partial_t \v_\alpha &=& {}-P_{,\alpha}
+3\nu(b_{\alpha\beta\mu\nu}\v_{\mu,\nu})_{,\beta}
+3\nu(b_{\alpha\beta\mu\nu}U_{\mu,\nu})_{,\beta}
-\frac{1}{5}\nu\triangle U_\alpha
\nonumber\\
&&{}-(\v_\alpha U_\beta+U_\alpha\v_\beta+\v_\alpha\v_\beta)_{,\beta}\;,
\label{v_EVOLUTION}
\\
\v_{\alpha,\alpha} &=& 0,
\label{v_DIV}
\eeq
where the pressure $P=P''-P'''$. Here and below we use the following symmetric 
tensors (which vary in time and in space):
\beq
b_{\alpha\beta\gamma\delta} &\!\!\define\!\!& 
\b_\alpha\b_\beta\b_\gamma\b_\delta\;,
\label{bbbb}
\\
b_{\alpha\beta\gamma} &\!\!\define\!\!& \b_\alpha\b_\beta\b_\gamma\;,
\label{bbb}
\\
b_{\alpha\beta} &\!\!\define\!\!& \b_\alpha\b_\beta\;.
\label{bb}
\eeq
Velocity $\vbf$, which satisfies equations~(\ref{v_EVOLUTION}) 
and~(\ref{v_DIV}), can be considered as the correction to the Kolmogorov 
velocity $\bf U$. This correction results from the strong influence of the 
magnetic field direction $\bf\b$ on the turbulent motions through the Braginskii 
viscosity tensor~(\ref{PI}). The evolution of the magnetic field $\bf B$
is given by the following standard MHD equation~\cite{LL_84}
\beq
\partial_t B_\alpha &=& 
V_{\alpha,\beta} B_\beta - V_\beta B_{\alpha,\beta}\;,
\label{B_EVOLUTION}
\eeq
where the plasma velocities ${\bf V}={\bf U}+\vbf$ are incompressible [compare 
with equation~(\ref{B_EVOLUTION_KA})], and we neglect resistivity.
Consequently, the equations for the magnetic field squared, $B^2$, and for the 
magnetic unit vector $\bf\b$ are
\beq
\partial_t B^2 &=& 
2V_{\alpha,\beta} B_\alpha B_\beta - V_\beta (B^2)_{,\beta}\;,
\label{B_STRENGTH_EVOLUTION}
\\
\partial_t\,\b_\alpha &=& V_{\alpha,\beta}\b_\beta
-V_{\beta,\gamma}b_{\alpha\beta\gamma}-V_\beta\b_{\alpha,\beta}\;.
\label{b_EVOLUTION}
\eeq

In the end of this section, it should be noted that the magnetized turbulent 
dynamo does exist in protogalaxies, but does not in galaxies. The reason for 
this is that, in contrast to temperatures in protogalaxies, temperatures in galaxies 
are low enough ($T\simlt 10^{5}\,{\rm K}$) for a considerable number of neutral 
particles to be present. These neutral particles have mean free path much longer 
than the ions have. Therefore, the viscosity in galaxies is dominated by the neutral 
particles, and the viscous forces are given by the standard familiar formula, 
$\nu_{\rm n}\triangle{\bf V}$ (here $\nu_{\rm n}$ is the neutral viscosity). As a 
result, in galaxies the turbulence is Kolmogorov and the kinematic dynamo theory can 
be used~\footnote{
Temperatures can be very high in the coronal regions of galaxies, and in these 
regions the magnetized turbulent dynamo can be present.
}.

%% file: chapter_3.tex
\chapter
[Statistics of Turbulent Velocities in Strongly Magnetized Plasmas]
{Statistics of Turbulent Velocities\\ in Strongly Magnetized Plasmas}
\label{VELOCITIES}

The statistics of the turbulent velocities in an unmagnetized plasma is given by
formulas~(\ref{U_STATISTICS}) and~(\ref{U_U_STATISTICS}). In this chapter we 
derive similar equations for the statistics of the turbulent velocities in a 
strongly magnetized plasma by making use of the quasi-linear (up to the second order) 
expansion in time, to solve the MHD equations. The final results of the derivations 
are given by equations~(\ref{ViT_AVERAGED_NEW})--(\ref{Vii_AVERAGE_NEW}). 
We then use these results in Chapter~\ref{MAGNETIC_SPECTRUM} to derive formulas 
for the evolution of the magnetic field total energy and energy spectrum in the 
magnetized turbulent dynamo theory.

\section
[The Quasi-linear Expansion of the MHD Equations]
{The Quasi-linear Expansion of\\ the MHD Equations}
\label{EXPANSION}

Let us assume that we know the magnetic field at zero time, 
${{\bf B}|}_{t=0}=\Bobf({\bf r})$ and ${{\bf\b}|}_{t=0}=\bobf({\bf r})$, 
and that the back-reaction velocity $\vbf$ is initially zero, 
${\vbf|}_{t=0}=\vobf({\bf r})\equiv 0$.~\footnote{
A nonzero initial back-reaction velocity would lead to transients which would be 
dissipated anyway.
}
We advance the magnetic field and the back-reaction velocity to some future time $t>0$
by the nonlinear terms, i.~e.~by integrating 
equations~(\ref{v_EVOLUTION}),~(\ref{B_EVOLUTION}) and~(\ref{b_EVOLUTION})
twice in time. This quasi-linear expansion procedure is similar to the calculations of 
Kulsrud and Anderson~\cite{KA_92}. Considering $t$ as the expansion 
parameter~\footnote{
If we would like to be more formal, then we need to introduce a dimensionless 
variable $\xi=t/\Delta t$, and to consider $U_{\alpha,\beta}\Delta t$ and 
$V_{\alpha,\beta}\Delta t$, which are dimensionless, as the expansion parameters.
}, 
up to the second order, we have
\beq
{\bf B}(t,{\bf r}) &=& \Bobf({\bf r})+\Bibf(t,{\bf r})+\Biibf(t,{\bf r}),
\label{B_EXPANSION}
\\
{\bf\b}(t,{\bf r}) &=& \bobf({\bf r})+\bibf(t,{\bf r})+\biibf(t,{\bf r}),
\label{b_EXPANSION}
\\
\vbf(t,{\bf r}) &=& \vibf(t,{\bf r})+\viibf(t,{\bf r}),
\label{v_EXPANSION}
\\
{\bf V}(t,{\bf r}) &=& \Vibf(t,{\bf r})+\Viibf(t,{\bf r}) 
= \Big[{\bf U}(t,{\bf r})+\vibf(t,{\bf r})\Big]+\viibf(t,{\bf r}),
\label{V_EXPANSION}
\eeq
where ${\bf V}={\bf U}+\vbf$ is the total fluid velocity, and the incompressible 
Kolmogorov turbulent velocities ${\bf U}$ are considered to be given and to be 
of the first order~\cite{KA_92,V_70}. All first and second order quantities, except
${\bf U}$, are initially zero, e.~g.~$\Bibf(0,{\bf r})=0$ and~$\vibf(0,{\bf r})=0$.

Now, we substitute the above expansion formulas into 
equations~(\ref{B_EVOLUTION}),~(\ref{b_EVOLUTION}) 
and into equations~(\ref{v_EVOLUTION})--(\ref{bbb}). We find that 
the zero order equations are
\beq
\partial_t\,\Bo_\alpha &=& 0, 
\qquad\qquad\qquad\qquad\qquad\qquad\qquad\qquad\qquad\qquad\qquad\qquad\:
\label{B_EVOLUTION_o}
\\
\partial_t\,\bo_\alpha &=& 0,
\label{b_EVOLUTION_o}
\\
\vo_\alpha &=& 0,
\label{v_EVOLUTION_o}
\\
\Vo_\alpha &=& 0,
\label{V_EVOLUTION_o}
\\
\bbo_{\alpha\beta\gamma} &=& \bo_\alpha\bo_\beta\bo_\gamma\;,
\label{bbbo}
\\
\bbo_{\alpha\beta\mu\nu} &=& \bo_\alpha\bo_\beta\bo_\mu\bo_\nu\;;
\label{bbbbo}
\eeq
the first order equations are
\beq
\partial_t\,\Bi_\alpha &=& \Vi_{\alpha,\beta}\Bo_\beta 
-\Vi_\beta\Bo_{\alpha,\beta}\;,
\label{B_EVOLUTION_i}
\\
\partial_t\,\bi_\alpha &=& \Vi_{\alpha,\beta}\bo_\beta
-\Vi_{\beta,\gamma}\bbo_{\alpha\beta\gamma}-\Vi_\beta\bo_{\alpha,\beta}\;,
\label{b_EVOLUTION_i}
\\
\partial_t\,\vi_\alpha &=& -\:\Pi_{,\alpha}
+3\nu(\bbo_{\alpha\beta\mu\nu}\vi_{\mu,\nu})_{,\beta}
+3\nu(\bbo_{\alpha\beta\mu\nu}U_{\mu,\nu})_{,\beta}
-\frac{1}{5}\nu\triangle U_\alpha\;,
\;\,
\label{v_EVOLUTION_i}
\\
\vi_{\alpha,\alpha} &=& 0,
\label{v_DIV_i}
\\
\Vi_\alpha &=& U_\alpha+\vi_\alpha\;,
\label{Vi}
\\
\bbi_{\alpha\beta\gamma} &=& \bi_\alpha\bo_\beta\bo_\gamma
+\bo_\alpha\bi_\beta\bo_\gamma+\bo_\alpha\bo_\beta\bi_\gamma\;,
\label{bbbi}
\\
\bbi_{\alpha\beta\mu\nu} &=& \bi_\alpha\bo_\beta\bo_\mu\bo_\nu
+\bo_\alpha\bi_\beta\bo_\mu\bo_\nu+\bo_\alpha\bo_\beta\bi_\mu\bo_\nu
+\bo_\alpha\bo_\beta\bo_\mu\bi_\nu\;;
\label{bbbbi}
\eeq
and the second order equations are
\beq
\partial_t\,\Bii_\alpha &=& \Vi_{\alpha,\beta}\Bi_\beta 
+\Vii_{\alpha,\beta}\Bo_\beta -\Vi_\beta\Bi_{\alpha,\beta}
-\Vii_\beta\Bo_{\alpha,\beta}\;,
\label{B_EVOLUTION_ii}
\\
\phantom{}&&
\nonumber\\
\partial_t\,\bii_\alpha &=& \Vi_{\alpha,\beta}\bi_\beta 
+\Vii_{\alpha,\beta}\bo_\beta -\Vi_{\beta,\gamma}\bbi_{\alpha\beta\gamma}
-\Vii_{\beta,\gamma}\bbo_{\alpha\beta\gamma} 
\nonumber\\
{}&-& \Vi_\beta\bi_{\alpha,\beta}
-\Vii_\beta\bo_{\alpha,\beta}\;,
\label{b_EVOLUTION_ii}
\\
\partial_t\,\vii_\alpha &=& -\:\Pii_{,\alpha}
+\Big(3\nu\bbi_{\alpha\beta\mu\nu}\vi_{\mu,\nu}
+3\nu\bbo_{\alpha\beta\mu\nu}\vii_{\mu,\nu}
+3\nu\bbi_{\alpha\beta\mu\nu}U_{\mu,\nu}
\nonumber\\
{}&-&\vi_\alpha U_\beta-U_\alpha\vi_\beta-\vi_\alpha\vi_\beta\Big)_{\!,\beta}\;,
\label{v_EVOLUTION_ii}
\\
\vii_{\alpha,\alpha} &=& 0,
\label{v_DIV_ii}
\\
\Vii_\alpha &=& \vii_\alpha\;.
\label{Vii}
\eeq
Here, of course, the pressure is also expanded, $P=\Pi+\Pii$, and it
is completely determined by the incompressibility 
conditions~(\ref{v_DIV_i}) and~(\ref{v_DIV_ii}).

\section
[The Time Laplace Transform Solution for the Velocities]
{The Time Laplace Transform Solution\\ 
for the Velocities}
\label{LAPLACE_TRANSFORM}

Let us first solve the first order equations~(\ref{v_EVOLUTION_i})--(\ref{Vi})
to find the first order back-reaction velocity $\vibf$ and the first order
total velocity $\Vibf$. It is convenient to Fourier transform these equations 
in space, ${\bf r}\rightarrow{\bf k}$, by making use of the discrete Fourier 
transformation, and to Laplace transform them in time, $t\rightarrow s$, by making 
use of the continuous Laplace transformation (see Appendix~\ref{TRANSFORMS}). We have
\beq
s  \viT_{{\bf k}\alpha}(s) &=& -ik_\alpha  \PiT_{\bf k}(s)
+3\nu ik_\beta\sum_{{\bf k'}\atop{{\bf k}''=\,{\bf k}-{\bf k}'}}
ik'_\nu  \viT_{{\bf k}'\mu}(s)  \bboT_{{\bf k}''\alpha\beta\mu\nu}
\nonumber\\
{}&+&3\nu ik_\beta\sum_{{\bf k'}\atop{{\bf k}''=\,{\bf k}-{\bf k}'}}
ik'_\nu  {\tilde U}_{{\bf k}'\mu}(s)  \bboT_{{\bf k}''\alpha\beta\mu\nu}
+\frac{1}{5}\nu k^2  {\tilde U}_{{\bf k}\alpha}(s),
\label{viT_EVOLUTION_PRESSURE}
\\
k_\alpha\viT_{{\bf k}\alpha}(s) &=& 0,
\label{viT_DIV}
\\
\ViT_{{\bf k}\alpha}(s) &=& {\tilde U}_{{\bf k}\alpha}(s) + \viT_{{\bf k}\alpha}(s),
\label{ViT}
\eeq
where $\viT_{{\bf k}\alpha}$, $\PiT_{\bf k}$, $\viT_{{\bf k}'\mu}$, 
${\tilde U}_{{\bf k}'\mu}$, ${\tilde U}_{{\bf k}\alpha}$ and $\ViT_{{\bf k}\alpha}$ 
are the Fourier-Laplace coefficients, and $\bboT_{{\bf k}''\alpha\beta\mu\nu}$ is the 
Fourier coefficient ($\bboT_{{\bf k}''\alpha\beta\mu\nu}$ is constant in time because 
it is of the zero order). In the derivation of equation~(\ref{viT_EVOLUTION_PRESSURE}) 
we also make use of the zero initial condition, ${\vibf|}_{t=0}=0$ [see 
equation~(\ref{LAPLACE_OF_DERIVATIVE}) in Appendix~\ref{TRANSFORMS}]. Now, we 
multiply equation~(\ref{viT_EVOLUTION_PRESSURE}) on the left by tensor
$\delta^\perp_{\gamma\alpha}=\delta_{\gamma\alpha}-\k_\gamma\k_\alpha$, 
which is perpendicular to the unit vector ${\bf\k}={\bf k}/k$, see 
equation~(\ref{DELTA_PERP}). The pressure term goes away. Using the 
incompressibility condition~(\ref{viT_DIV}), interchanging indices, 
and using the symmetry property of tensor $\bboT_{{\bf k}''\alpha\beta\mu\nu}$
over indices $\alpha$, $\beta$, $\mu$, and $\nu$, we obtain 
\beq
\MoT_{{\bf k}\alpha,{\bf k'}\beta}\viT_{{\bf k'}\beta} &=& \FiT_{{\bf k}\alpha}\;,
\label{viT_EVOLUTION}
\\
\MoT_{{\bf k}\alpha,{\bf k'}\beta} &=& s\delta_{{\bf k}\alpha,{\bf k'}\beta}
+3\nu\delta^\perp_{\alpha\gamma}k_\mu k'_\nu \bboT_{{\bf k''}\gamma\mu\nu\beta}\;,
\quad{\bf k''}={\bf k}-{\bf k'},\qquad
\label{MoT_DEFINITION}
\\
\FiT_{{\bf k}\alpha} &=& \left[-\MoT_{{\bf k}\alpha,{\bf k'}\beta}
+\left(s+\frac{1}{5}\nu k^2\right)\delta_{{\bf k}\alpha,{\bf k'}\beta}\right]
{\tilde U}_{{\bf k'}\beta}\;.
\label{FiT_DEFINITION}
\eeq

Here we use convenient matrix notation, so that summation is implicitly assumed 
not only over repeated spatial indices but also over repeated wave numbers 
[e.~g.~a summation over $\beta=x,y,z$ and an infinite summation over all discrete 
values of $\bf k'$ are assumed on the left-hand-side of 
equation~(\ref{viT_EVOLUTION})]. Function 
$\delta_{{\bf k}\alpha,{\bf k'}\beta}=\delta_{{\bf k},{\bf k'}}\delta_{\alpha\beta}$ 
is the Kronecker tensor, it is equal to unity if $\alpha=\beta$, ${\bf k}={\bf k'}$,
and is zero otherwise. The matrix operator $\MoT_{{\bf k}\alpha,{\bf k'}\beta}(s)$ 
is of the zero order, while ``the driving force'' $\FiT_{{\bf k}\alpha}(s)$ is of 
the first order.

If there exist matrix $\MoT_{{\bf k}\alpha,{\bf k'}\beta}^{-1}$ that is 
inverse to the matrix $\MoT_{{\bf k}\alpha,{\bf k'}\beta}$, then the solution
of equation~(\ref{viT_EVOLUTION}) is
\beq
\viT_{{\bf k}\alpha} &=& \MoT_{{\bf k}\alpha,{\bf k'}\beta}^{-1}\FiT_{{\bf k'}\beta}\;.
\label{viT_SOLUTION}
\eeq
Using this formula and equations~(\ref{ViT}),~(\ref{FiT_DEFINITION}), we obtain 
the Fourier-Laplace coefficient of the first order total velocity $\Vibf$,
\beq 
\ViT_{{\bf k}\alpha}(s) &=& {\tilde U}_{{\bf k}\alpha} + \viT_{{\bf k}\alpha} 
= \left(s+\frac{1}{5}\nu k^2\right)\MoT_{{\bf k}\alpha,{\bf k'}\beta}^{-1}
{\tilde U}_{{\bf k'}\beta}(s)\;.
\label{ViT_SOLUTION}
\eeq

Now, let us consider an important case for which we can invert matrix 
$\MoT_{{\bf k}\alpha,{\bf k'}\beta}$, given by formula~(\ref{MoT_DEFINITION}).
Let $(\Bobf\cdot\nabla)\Bobf=0$, i.~e.~let the magnetic field at zero time, 
$\Bobf$, vary only in the direction perpendicular to itself. This is equivalent to 
the case when the magnetic field lines are initially straight, the initial curvature
of the field is zero, and $\bbo_{\alpha\beta}=\bo_\alpha\bo_\beta\equiv{\rm const}$. 
This model of straight magnetic field lines is not as artificial as it seems at 
first glance, because of the following arguments. Remember, that in the kinematic 
dynamo case, the magnetic field lines have the folding pattern, shown in 
Figure~\ref{FIGURE_FOLDING}. This folding pattern implies that in the bulk of 
the volume the field is strong and has small curvature, 
$k_\perp\gg k_\parallel\sim k_\nu$, while in a small fraction of the volume 
the field is weak and curved, $k_\perp\sim k_\parallel\gg k_\nu$. The regions
of weak and curved field, Region~II in Figure~\ref{FIGURE_FOLDING}, can be 
disregarded as long as we consider the volume averaged magnetic field energy 
spectrum and are not interested in the field curvature. As for the regions of 
strong magnetic field with small curvature, Region~I in Figure~\ref{FIGURE_FOLDING}, 
the field in these regions can be well approximated by our straight magnetic 
field lines model on scales $k$ which satisfy $k\simgt k_\parallel\sim k_\nu$. 
However, it should be made clear that we make an assumption that in the magnetized 
turbulent dynamo case the magnetic field has a folding structure similar to that in 
the kinematic turbulent dynamo case.
This assumption is our {\it first working hypothesis}. We do not prove it, but 
there exist a simple argument by contradiction, supporting it, which is as follows. 
If the magnetic field were not folded, i.~e.~if 
$k_\perp\sim k_\parallel\simgt k_\nu$ everywhere in space, then the field would 
be isotropically tangled on viscous and subviscous scales. In this case 
the Braginskii viscosity could be averaged over the field directions, and it 
would probably rapidly reduce to the isotropically averaged viscosity 
$\nu_{\rm eff}=\nu/5$, see equation~(\ref{NU_EFF}). As a result, in this case the 
magnetized dynamo would probably possess the properties of the kinematic dynamo 
with this effective viscosity, and would develop the folding structure of the 
magnetic field.

In the case when $\bbo_{\alpha\beta}\equiv{\rm const}$, we have
$\bboT_{{\bf k''}\gamma\mu\nu\beta}=
\delta_{{\bf k''},0}\bo_\gamma\bo_\mu\bo_\nu\bo_\beta$, and
matrix~(\ref{MoT_DEFINITION}) reduces to
\beq
\MoT_{{\bf k}\alpha,{\bf k'}\beta} &=& \delta_{{\bf k},{\bf k'}}
\left[s\delta_{\alpha\beta}+3\nu k^2\mu^2
\delta^\perp_{\alpha\gamma}\bo_\gamma\bo_\beta\right],
\label{MoT}
\\
\mu &\define& {\bf\bobf}\cdot{\bf\k}.
\label{MU}
\eeq
Here $\mu$ is the cosine of the angle between vectors $\bf\bobf$ and $\bf\k$. 
The matrix $\MoT_{{\bf k}\alpha,{\bf k'}\beta}$
is now diagonal in $\bf k$, so we can easily invert it:
\beq
\MoT_{{\bf k}\alpha,{\bf k'}\beta}^{-1} &=& \frac{\delta_{{\bf k},{\bf k'}}}{s} 
\left[\delta_{\alpha\beta}-
\frac{3\nu k^2\mu^2(1-\mu^2)}{s+3\nu k^2\mu^2(1-\mu^2)}
\,\frac{\delta^\perp_{\alpha\gamma}\bo_\gamma\bo_\beta}{1-\mu^2}
\right].
\label{INVERSE_MATRIX}
\eeq
Next, we substitute this inverse matrix into equation~(\ref{ViT_SOLUTION}) and obtain
the Fourier-Laplace coefficient of the first order total velocity
\beq
\ViT_{{\bf k}\alpha}(s) &=& \ViT'_{{\bf k}\alpha}(s)+\ViT''_{{\bf k}\alpha}(s)\;,
\label{ViT_L}
\\
\ViT'_{{\bf k}\alpha}(s) &=& \frac{s+{\bar\Omega}}{s} 
\left[\delta_{\alpha\beta}-
\frac{\delta^\perp_{\alpha\gamma}\bo_\gamma\bo_\beta}{1-\mu^2}\right]
{\tilde U}_{{\bf k}\beta}(s)\;,
\label{ViT_L_FIRST}
\\
\ViT''_{{\bf k}\alpha}(s) &=& \frac{s+{\bar\Omega}}{s+2\Omega}\;
\frac{\delta^\perp_{\alpha\gamma}\bo_\gamma\bo_\beta}{1-\mu^2}\:
{\tilde U}_{{\bf k}\beta}(s)\;,
\label{ViT_L_SECOND}
\eeq
where, we introduce the following notations for the viscous damping frequencies
\beq
{\bar\Omega} &\define& \frac{1}{5}\nu k^2 = \nu_{\rm eff} k^2,
\label{OMEGA_AVERAGED}
\\
\Omega &\define& \frac{3}{2}\nu k^2 \mu^2(1-\mu^2) 
= \frac{15}{2}{\bar\Omega}\mu^2(1-\mu^2).
\label{OMEGA}
\eeq
The frequency ${\bar\Omega}$ represents the effective averaged rate of the Braginskii 
viscous dissipation, see equation~(\ref{NU_EFF}). The frequency $\Omega$ depends
on $\mu^2=({\bf \bobf\k})^2$, and this dependence reflects the anisotropy of the 
Braginskii viscous stress tensor. Note, that ${\bar\Omega}$ is equal to $\Omega$ 
averaged over $\mu$.

Now, we calculate $\langle\ViT_{{\bf k}\alpha}(t)\rangle$ and 
$\langle\ViT_{{\bf k}\alpha}(t)\ViT_{{\bf k'}\beta}(t')\rangle$, which are the ensemble
averages of the $\bf V$'s over all possible realizations of the turbulent motions. 
For this, first, we need the ensemble averages of the $\bf U$'s. We 
use the Laplace and Fourier transformation formulas (see Appendix~\ref{TRANSFORMS}), 
and equations~(\ref{U_STATISTICS}),~(\ref{U_U_STATISTICS}) to obtain
\beq
\langle{\tilde U}_{{\bf k}\alpha}(s)\rangle 
\!\!&=&\!\! \int\limits_0^\infty \langle{\tilde U}_{{\bf k}\alpha}(t)\rangle\,
e^{-st}\,dt
\nonumber\\
\!\!&=&\!\! \int\limits_0^\infty e^{-st}dt\;\,
\frac{1}{\sqrt{2\pi}}\!\int\limits_{-\infty}^\infty
\langle{\tilde U}_{{\bf k}\alpha}(\omega)\rangle\,e^{-i\omega t}\,d\omega 
\:=\: 0,
\label{U_LAPLACE_STATISTICS}
\\
\langle{\tilde U}_{{\bf k}\alpha}(s){\tilde U}_{{\bf k'}\beta}(s')\rangle
\!\!&=&\!\! \int\limits_0^\infty\!\int\limits_0^\infty 
\langle{\tilde U}_{{\bf k}\alpha}(t){\tilde U}_{{\bf k'}\beta}(t')\rangle\,
e^{-st-s't'}\,dt'\,dt
\nonumber\\
{}\!\!&=&\!\! \int\limits_0^\infty\!\int\limits_0^\infty e^{-st-s't'}dt'dt\;\,
\frac{1}{2\pi}\!\int\limits_{-\infty}^\infty\!\int\limits_{-\infty}^\infty
\langle{\tilde U}_{{\bf k}\alpha}(\omega){\tilde U}_{{\bf k'}\beta}(\omega')\rangle\,
e^{-i\omega t-i\omega' t'}d\omega'd\omega
\nonumber\\
{}\!\!&=&\!\! \delta^\perp_{\alpha\beta}\delta_{{\bf k'},-{\bf k}}\;\frac{1}{2\pi}\!
\int\limits_{-\infty}^\infty J_{\omega k}\,d\omega
\int\limits_0^\infty\!\int\limits_0^\infty e^{-st-i\omega t}\,e^{-s't'+i\omega t'}dt'dt
\nonumber\\
{}\!\!&=&\!\! \delta^\perp_{\alpha\beta}\delta_{{\bf k'},-{\bf k}}\;\frac{1}{2\pi}\!
\int\limits_{-\infty}^\infty J_{\omega k}\:\frac{d\omega}{(s+i\omega)(s'-i\omega)}\;.
\label{U_U_LAPLACE_STATISTICS}
\eeq
Second, we use these formulas to calculate the ensemble averages of the two modes 
of the velocity Fourier-Laplace coefficient in equation~(\ref{ViT_L}), 
$\ViT'_{{\bf k}\alpha}(s)$ and $\ViT''_{{\bf k}\alpha}(s)$. Using 
equations~(\ref{ViT_L_FIRST}) and~(\ref{ViT_L_SECOND}), we obtain
\beq
\langle\ViT'_{{\bf k}\alpha}(s)\rangle \!\!&=&\!\! 0,
\label{ViT'}
\\
\langle\ViT''_{{\bf k}\alpha}(s)\rangle \!\!&=&\!\! 0, 
\\
\langle\ViT'_{{\bf k}\alpha}(s)\ViT''_{{\bf k'}\beta}(s')\rangle \!\!&=&\!\! 0,
\label{ViT'_ViT''}
\\
\langle\ViT'_{{\bf k}\alpha}(s)\ViT'_{{\bf k'}\beta}(s')\rangle \!\!&=&\!\!
\delta_{{\bf k'},-{\bf k}}\;\frac{(s+{\bar\Omega})(s'+{\bar\Omega})}{ss'}
\left[\delta^\perp_{\alpha\beta}-
\frac{\delta^\perp_{\alpha\gamma}\delta^\perp_{\beta\tau}\bo_\gamma\bo_\tau}
{1-\mu^2}\right]
\nonumber\\
\!\!&&\!\!{}\times\frac{1}{2\pi}
\int\limits_{-\infty}^\infty \frac{J_{\omega k}\,d\omega}{(s+i\omega)(s'-i\omega)}\;,
\label{ViT'_ViT'}
\\
\langle\ViT''_{{\bf k}\alpha}(s)\ViT''_{{\bf k'}\beta}(s')\rangle \!\!&=&\!\!
\delta_{{\bf k'},-{\bf k}}\;
\frac{(s+{\bar\Omega})(s'+{\bar\Omega})}{(s+2\Omega)(s'+2\Omega)}\;
\frac{\delta^\perp_{\alpha\gamma}\delta^\perp_{\beta\tau}\bo_\gamma\bo_\tau}
{1-\mu^2} 
\nonumber\\
\!\!&&\!\!{}\times\frac{1}{2\pi}
\int\limits_{-\infty}^\infty\frac{J_{\omega k}\,d\omega}{(s+i\omega)(s'-i\omega)}
\label{ViT''_ViT''}\;.
\eeq
Third, using these last equations and equation~(\ref{ViT_L}), we easily obtain
\beq
\langle\ViT_{{\bf k}\alpha}(s)\rangle \!\!&=&\!\! 0,
\label{ViT_L_AVERAGED}
\\
\langle\ViT_{{\bf k}\alpha}(s)\ViT_{{\bf k'}\beta}(s')\rangle \!\!&=&\!\!
\langle\ViT'_{{\bf k}\alpha}(s)\ViT'_{{\bf k'}\beta}(s')\rangle
+\langle\ViT''_{{\bf k}\alpha}(s)\ViT''_{{\bf k'}\beta}(s')\rangle
\nonumber\\
{} &=& \delta_{{\bf k'},-{\bf k}}\;\frac{1}{2\pi}\!
\int\limits_{-\infty}^\infty J_{\omega k}\,d\omega\,
\nonumber\\
\!\!&&\!\!{}\times
\left[{\tilde H}_L(s;\omega,0){\tilde H}_L(s';-\omega,0)
\left(\delta^\perp_{\alpha\beta}-
\frac{\delta^\perp_{\alpha\gamma}\delta^\perp_{\beta\tau}\bo_\gamma\bo_\tau}
{1-\mu^2}\right)\right.
\nonumber\\
\!\!&&\!\!\left.\qquad{}+{\tilde H}_L(s;\omega,\Omega){\tilde H}_L(s';-\omega,\Omega)\:
\frac{\delta^\perp_{\alpha\gamma}\delta^\perp_{\beta\tau}\bo_\gamma\bo_\tau}
{1-\mu^2}
\right],
\label{ViT_ViT_L_AVERAGED}
\eeq
where 
\beq
{\tilde H}_L(p;q_1,q_2) &=& \frac{p+{\bar\Omega}}{(p+iq_1)(p+2q_2)}\;.
\label{H_L_T}
\eeq
The inverse Laplace transformation $p\rightarrow t$ of function ${\tilde H}_L(p;q_1,q_2)$ 
is
\beq
H_L(t,q_1,q_2) \!\!&=&\!\! \frac{1}{2\pi i}\int\limits_{c-i\infty}^{c+i\infty}
{\tilde H}_L(p;q_1,q_2)\,e^{tp}\,dp
\nonumber\\
\!\!{}&=&\!\! \frac{1}{2\pi i}\int\limits_{c-i\infty}^{c+i\infty}
\frac{p+{\bar\Omega}}{(p+iq_1)(p+2q_2)}\,e^{tp}\,dp
\nonumber\\
\!\!{}&=&{}\!\!\frac{({\bar\Omega}-iq_1)e^{-iq_1t}-({\bar\Omega}-2q_2)e^{-2q_2t}}
{2q_2-iq_1},
\label{H_L_T_qq}
\eeq
where we carry out the integral by closing the integration contour in the complex 
plane and by calculating the residues at $p=-iq_1,\:-2q_2$.
Next, we apply the inverse Laplace transformations $s\rightarrow t$ and 
$s'\rightarrow t'$ to equations~(\ref{ViT_L_AVERAGED}) and~(\ref{ViT_ViT_L_AVERAGED}). 
We finally obtain the desired results
\beq
\langle\ViT_{{\bf k}\alpha}(t)\rangle \!\!&=&\!\! 0,
\label{ViT_AVERAGED}
\\
\langle\ViT_{{\bf k}\alpha}(t)\ViT_{{\bf k'}\beta}(t')\rangle \!\!&=&\!\!
\langle\ViT'_{{\bf k}\alpha}(t)\ViT'_{{\bf k'}\beta}(t')\rangle
+\langle\ViT''_{{\bf k}\alpha}(t)\ViT''_{{\bf k'}\beta}(t')\rangle
\nonumber\\
\!\!&=&\!\! \delta_{{\bf k'},-{\bf k}}\;\frac{1}{2\pi}\!
\int\limits_{-\infty}^\infty J_{\omega k}\,d\omega\,
\nonumber\\
\!\!&&\!\!{}\times 
\left[H_L(t;\omega,0)H_L(t';-\omega,0)\left(\delta^\perp_{\alpha\beta}-
\frac{\delta^\perp_{\alpha\gamma}\delta^\perp_{\beta\tau}\bo_\gamma\bo_\tau}
{1-\mu^2}\right)\right.
\nonumber\\
\!\!&&\!\!\left.\qquad{}+H_L(t;\omega,\Omega)H_L(t';-\omega,\Omega)\:
\frac{\delta^\perp_{\alpha\gamma}\delta^\perp_{\beta\tau}\bo_\gamma\bo_\tau}
{1-\mu^2}
\right],
\label{ViT_ViT_AVERAGED}
\eeq
where function $H_L(t,q_1,q_2)$ is given by equation~(\ref{H_L_T_qq}). As one might
expect, the ensemble averaged first order turbulent velocity, 
$\langle\ViT_{{\bf k}\alpha}(t)\rangle$, is 
zero~\footnote{
This is because there is no preferred direction. Of course, there is a preferred 
axis in space, which is along the magnetic field unit vector. However, there is 
no preferred direction because the Braginskii viscous stress tensor~(\ref{PI}) 
is invariant with respect to a substitution ${\bf\b}\rightarrow{}-{\bf\b}$ 
(gyrating ions ``do not care'' about the exact direction of ${\bf\b}$).
},
while the averaged velocity correlation tensor, 
$\langle\ViT_{{\bf k}\alpha}(t)\ViT_{{\bf k'}\beta}(t')\rangle$ 
is not zero because it is of the second order.

So far we have considered only the first order velocity, $\Vibf$, and have found its 
statistics, given by equations~(\ref{ViT_AVERAGED}) and~(\ref{ViT_ViT_AVERAGED}). In 
order to find the second order velocity, $\Viibf$, we need to solve the complicated 
second order equations~(\ref{v_EVOLUTION_ii})--(\ref{Vii}). Fortunately, we will 
need only the ensemble average of the second order velocity, 
$\langle\ViiT_{{\bf k}\alpha}(t)\rangle$. It turns out that in our case of a straight 
initial field, $\bbo_{\alpha\beta}={\rm const}$, which we consider here, this average 
is zero,
\beq
\langle\ViiT_{{\bf k}\alpha}(t)\rangle &=& 0,
\label{Vii_AVERAGE}
\eeq
see Appendix~\ref{Vii_ENSEMBLE_AVERAGE}. The reason for this simple result is that 
$\bbo_{\alpha\beta}$ is constant in space, and the hydrodynamic turbulence $\bf U$ is 
homogeneous. Therefore, the ensemble average of the term in the parentheses in 
equation~(\ref{v_EVOLUTION_ii}) is constant in space, and its spatial derivative is 
zero. As a result, the ensemble averaged velocity 
$\langle\ViiT_{{\bf k}\alpha}(t)\rangle=\langle\viiT_{{\bf k}\alpha}(t)\rangle$ is zero 
(see Appendix~\ref{Vii_ENSEMBLE_AVERAGE} for the proof).

Now, let us return to the first order velocities, $\Vibf$. According to 
equation~(\ref{ViT_AVERAGED}) the ensemble average of $\Vibf$ is zero, 
$\langle\Vibf\rangle=0$. Let us calculate the ensemble average of $\Vibf$ squared. 
We have
\beq
\langle\Vibf^2\rangle &=&
\langle\Vi_\alpha(t,{\bf r})\Vi_\alpha(t,{\bf r})\rangle
= \sum\limits_{{\bf k},{\bf k'}}
\langle\ViT_{{\bf k}\alpha}(t)\ViT_{{\bf k'}\alpha}(t)\rangle\,
e^{i({\bf k}+{\bf k'}){\bf r}}
\nonumber\\
{}&=& \sum\limits_{{\bf k}}
\langle\ViT_{{\bf k}\alpha}(t)\ViT_{-{\bf k}\alpha}(t)\rangle
=\sum\limits_{{\bf k}}\langle|\ViTbf_{\bf k}(t)|^2\rangle\;.
\label{Vi_SQUARED}
\eeq
Here we use the fact that $\langle\ViT_{{\bf k}\alpha}\ViT_{{\bf k'}\beta}\rangle
=\langle\ViT_{{\bf k}\alpha}\ViT^*_{-{\bf k'}\beta}\rangle
\propto\delta_{{\bf k'},-{\bf k}}$, see equation~(\ref{ViT_ViT_AVERAGED}).
Using equations~(\ref{H_L_T_qq}) and~(\ref{ViT_ViT_AVERAGED}), 
we obtain
\beq
\langle|\ViTbf_{\bf k}(t)|^2\rangle &=&
\frac{1}{2\pi} \int\limits_{-\infty}^\infty J_{\omega k}\,d\omega\,
\left[H_L(t;\omega,0)H_L(t;-\omega,0)+H_L(t;\omega,\Omega)H_L(t;-\omega,\Omega)\right]
\nonumber\\
{}&=& \frac{1}{2\pi} \int\limits_{-\infty}^\infty
J_{\omega k}\,d\omega\,
\Bigg\{ 2{\bar\Omega}^2\,\frac{1-\cos{\omega t}}{\omega^2}
+2{\bar\Omega}\,\frac{\sin{\omega t}}{\omega}+1
\nonumber\\
&&\qquad\qquad {}+ \frac{{[{\bar\Omega}-({\bar\Omega}-2\Omega)e^{-2\Omega t}]}^2}
{\omega^2+4\Omega^2}
+2{\bar\Omega}({\bar\Omega}-2\Omega)e^{-2\Omega t}\,\frac{1-\cos{\omega t}}
{\omega^2+4\Omega^2}
\nonumber\\
&&\qquad\qquad {}+ 2({\bar\Omega}-2\Omega)e^{-2\Omega t}\,\frac{\omega\sin{\omega t}}
{\omega^2+4\Omega^2}
+\frac{\omega^2}{\omega^2+4\Omega^2}\Bigg\}\;.
\label{ViT_SQUARED}
\eeq
Let us analyze this last equation. The width of function $J_{\omega k}$ in $\omega$
space is about $1\Big/\tau(k)$, where $\tau(k)$ is the decorrelation time of 
hydrodynamic eddies on scale $2\pi k^{-1}$, e.~g.~see equation~(\ref{J_OMEGA_K}). We 
have an obvious estimate $1\Big/\tau(k)\simlt 1\Big/\tau(k_\nu)\sim\nu_{\rm eff}k_\nu^2$,
where $k_\nu$ is the viscous cutoff wave number, and $\nu_{\rm eff}$ is the effective 
viscosity~(\ref{NU_EFF}). In order to make an estimate of the right-hand-side of 
equation~(\ref{ViT_SQUARED}), let us for the moment consider the case when the 
angle between $\bf\k$ and $\bf\bo$ is neither very close to zero or $\pi$, nor very 
close to $\pm \pi/2$, so that $2\Omega=15{\bar\Omega}\mu^2(1-\mu^2)$ is not small 
($\mu={\bf\k}\cdot{\bf\bo}$). Then, in the limit $t\gg\tau$ we have $\Omega t\gg 1$, 
${\bar\Omega}t\gg 1$ and $\omega^{-1}\sin{\omega t}\sim \pi\delta(\omega)$. Therefore, 
we can drop the exponential terms in equation~(\ref{ViT_SQUARED}), and the first term 
in the brackets $\{...\}$ in this equation gives the main contribution 
to the integral over $\omega$. As a result, equation~(\ref{ViT_SQUARED}) reduces to
\beq
\langle|\ViTbf_{\bf k}(t)|^2\rangle \!&\approx&\!
J_{0k}{\bar\Omega}^2t \:\propto\: t, 
\qquad t\gg\tau. \qquad
\label{ViT_GROWTH}
\eeq
The velocity modes, which have the angle between $\bf\k$ and $\bf\bo$ not close to 
$0$, $\pi$, or $\pm\pi/2$, grow in 
time~\footnote{
If the angle between $\bf\k$ and $\bf\bo$ is $0$, $\pi$, or $\pm\pi/2$, then 
$\Omega=0$. In this case equation~(\ref{ViT_SQUARED}) obviously reduces to 
$\langle|\ViTbf_{\bf k}(t)|^2\rangle=2J_{0k}{\bar\Omega}^2t\propto t$, and the 
growth is even faster.
}. 
Because all terms under the sum in equation~(\ref{Vi_SQUARED}) are nonnegative, the 
ensemble average of the first order velocity squared, $\langle\Vibf^2\rangle$, also 
grows in time. It is clear that in reality this growth can not be as unrestricted, as
it appears to be, according to equations~(\ref{Vi_SQUARED}) and~(\ref{ViT_GROWTH}). 
In the next section we will discuss this growth in more details. We will see that this
growth is a feature of the anisotropy of the Braginskii viscous forces and of our 
quasi-linear expansion procedure, and is indeed restricted.

\section
[The Effective Rotational Damping of Velocities]
{The Effective Rotational Damping\\
of Velocities}
\label{DUMPING}

Let us try to understand the reasons of the growth of the ensemble averaged first order
velocity squared, $\langle\Vibf^2\rangle$, which we found in the previous section. 
First, refer to equations~(\ref{ViT'_ViT''})--(\ref{ViT''_ViT''}) and 
equations~(\ref{ViT_ViT_L_AVERAGED}),~(\ref{ViT_ViT_AVERAGED}). From calculations of 
the first order velocity growth, which follow equation~(\ref{ViT_ViT_AVERAGED}), it is 
clear that in the case when the angle between unit vectors $\bf\k$ and $\bobf$ 
is not equal to $0$, $\pi$ or $\pm\pi/2$, the growth of $\Vibf(t)=\Vibf'(t)+\Vibf''(t)$ 
happens due to the growth of the $\ViT'_{{\bf k}\alpha}(t)$ modes, while the 
$\ViT''_{{\bf k}\alpha}(t)$ modes do not grow 
unrestrictively~\footnote{
This can also be seen because of the following simple arguments. The poles 
of function $\langle\ViT'_{{\bf k}\alpha}(s)\ViT'_{{\bf k'}\beta}(s')\rangle$ 
are $s=0$, $s'=0$, $s=-i\omega$ and $s'=i\omega$, see equation~(\ref{ViT'_ViT'}). 
All these poles have nonnegative real parts, and there is no viscous damping of 
$\langle|\ViTbf'_{\bf k}(t)|^2\rangle$. On the other hand, the poles of function 
$\langle\ViT''_{{\bf k}\alpha}(s)\ViT''_{{\bf k'}\beta}(s')\rangle$ are $s=-2\Omega$, 
$s'=-2\Omega$, $s=-i\omega$ and $s'=i\omega$, see equation~(\ref{ViT''_ViT''}). Two 
poles are negative, and $\langle|\ViTbf''_{\bf k}(t)|^2\rangle$ is viscously damped
with a rate ${}\sim\Omega$.
}.
Next, let us refer to equations~(\ref{ViT_L_FIRST}) and~(\ref{ViT_L_SECOND}) for  
$\ViT'_{{\bf k}\alpha}$ and $\ViT''_{{\bf k}\alpha}$. On one hand, we have
\beq
\k_\alpha\ViT'_{{\bf k}\alpha}(s) &=& 0,
\\
\k_\alpha\ViT''_{{\bf k}\alpha}(s) &=& 0,
\eeq
as it should be because plasma velocities are incompressible. On the other hand, we have
\beq
\bo_\alpha\ViT'_{{\bf k}\alpha}(s) &=& 0,
\label{ViT'_bo}
\\
\bo_\alpha\ViT''_{{\bf k}\alpha}(s) &=& \frac{s+{\bar\Omega}}{s+2\Omega}\;
\bo_\alpha{\tilde U}_{{\bf k}\alpha}(s)\;.
\eeq
Thus, the growing velocity modes $\ViTbf'_{\bf k}$ are perpendicular to both 
vectors $\bf\k$ and $\bf\bo$ (of course, all modes must be perpendicular to 
$\bf\k$ because of the fluid incompressibility condition). At the same time, 
the other, non-growing modes, $\ViTbf''_{\bf k}$, have nonzero components along 
the initial magnetic field unit vector $\bf\bo$. In addition, 
equation~(\ref{ViT'_ViT''}) means that $\ViTbf'_{\bf k}$ and $\ViTbf''_{\bf k}$ 
are perpendicular on average. 

These results can be understood by an independent physical argument that appeals
to the Braginskii viscous dissipation process. The viscous dissipation into heat 
is~\cite{B_65}
\beq
Q_{\rm vis} = 
\pi_{\alpha\beta}V_{\alpha,\beta}
= {}-3\nu\left(\Vi_{\alpha,\beta}\bo_\alpha\bo_\beta\right)^2
= {}-3\nu\left(\Vi''_{\alpha,\beta}\bo_\alpha\bo_\beta\right)^2,
\eeq
where we keep only the first order terms in the parentheses, and make use of 
equation~(\ref{ViT'_bo}). Using this formula, the inverse Laplace transform 
of equation~(\ref{ViT''_ViT''}) [which is the second term in the square 
brackets in equation~(\ref{ViT_ViT_AVERAGED})] and definition~(\ref{OMEGA}), 
we obtain the ensemble averaged dissipation
\beq
\langle Q_{\rm vis}\rangle &=&
{}-3\nu\langle\Vi_{\alpha,\beta}\Vi_{\gamma,\tau}\rangle\bbo_{\alpha\beta\gamma\tau}
= 3\nu\sum\limits_{{\bf k},{\bf k'}} k_\beta k'_\tau
\left\langle\ViT''_{{\bf k}\alpha}\ViT''_{{\bf k'}\gamma}\right\rangle\,
\bbo_{\alpha\beta\gamma\tau}\, e^{i({\bf k}+{\bf k'}){\bf r}}
\nonumber\\
{} &=& {}-3\nu\sum\limits_{\bf k} k^2\mu^2\,
\bo_\alpha\langle\ViT''_{{\bf k}\alpha}\ViT''_{-{\bf k}\gamma}\rangle\bo_\gamma
\nonumber\\
{} &=& {}-\sum\limits_{\bf k}2\Omega\;
\frac{1}{2\pi}\!\int\limits_{-\infty}^\infty J_{\omega k}\,
H_L(t;\omega,\Omega)H_L(t';-\omega,\Omega)\,d\omega\;.
\eeq
We see that first, only $\ViTbf''_{\bf k}$ velocity modes are dissipated by the 
Braginskii viscous forces, and second, the viscous dissipation is proportional to 
$2\Omega=3\nu k^2\mu^2(1-\mu^2)$. As a result, we can conclude that the growth of 
the first order velocity happens because the $\ViTbf'_{\bf k}$ velocity modes, which 
are perpendicular to $\bf\bo$, are not damped by the Braginskii viscous dissipation
process. Of course, in the degenerate cases, when ${\bf\k}\perp\bobf$ 
(i.~e.~$\mu^2=0$) or ${\bf\k}\parallel\bobf$ (i.~e.~$1-\mu^2=0$), both velocity 
modes $\ViTbf'_{\bf k}$ and $\ViTbf''_{\bf k}$ are undamped.

It is clear that the growing velocity modes can not grow unrestrictedly. What are 
the mechanisms which stop this growth? To answer this question, let us again note 
that the only growing velocity modes are those which are perpendicular to $\bf\b$. 
All other modes are damped by the Braginskii viscosity. For example, 
in our quasi-linear expansion the zero order magnetic field unit vector $\bo$ is 
assumed to be stationary, see equation~(\ref{b_EVOLUTION_o}), and therefore, the 
first order velocity mode $\ViTbf'_{\bf k}(t)$, which is perpendicular to 
$\bf\bo$, can unrestrictedly grow in time. However, in reality, any velocity 
vector continuously rotates relative to the magnetic field unit vector. As a result, 
the growing modes do not stay perpendicular to $\bf\b$, and therefore, they do 
not grow in time forever. We can understand this rotation from different standpoints.
First, in the laboratory reference frame the velocity vector and the magnetic 
field unit vector both rotate, but they must rotate differently because 
different ``forces'' are acting on them [in other words, because they satisfy
different differential equations]. Second, we can go to a reference frame that
has the origin at a given point of space and rotates together with the field unit 
vector at this point. In this rotating reference frame there exist Coriolis 
forces which act on velocity vectors and force them to rotate relative to the 
non-rotating field vector. These Coriolis forces are caused by the non-linear
coupling of velocity modes via the inertial term $({\bf V\nabla}){\bf V}$ of the 
MHD equation~(\ref{EQUATION_FOR_V}). Consequently, the rotation is produced by 
the non-linear coupling of velocity modes.

Thus, a growing velocity mode, which is perpendicular to the magnetic field unit 
vector, rotates out of its initial direction. After this, the velocity mode is 
viscously damped, and stops growing. We call this process the ``effective 
rotational damping'' of velocity modes. Of course, the effective 
rotational damping operates only on the viscous scale, on which the Braginskii
viscous dissipation is significant. On larger scales the viscous dissipation is 
small and the rotation of velocities does not make any difference.
Now, let us estimate the effective damping rate of the growing velocity modes, 
which is associated with the rotational damping process.
First, the square of the angular velocity of the rotation can be estimated as
\beq
\omega_{\rm rot}^2\sim \frac{1}{3}\,\frac{1}{\tau_\nu^2},
\label{OMEGA_ROT}
\eeq
where $\tau_\nu$ is the velocity decorrelation time on the viscous scale. The 
factor $1/3$ in this equation comes from the fact that only one of the three 
angular velocity components contributes to the deviation of the velocity 
$\bf V$ from its original direction perpendicular to the magnetic field unit 
vector $\bf\b$. Namely, only the component along the vector product 
${\bf\b}\times{\bf V}$ contributes. 
Second, the typical Braginskii viscous damping rate is about 
$1/\tau_\nu\sim\nu_{\rm eff}k_\nu^2$ because of the definition of the 
viscous wave number $k_\nu$, as the wave number at which the viscous dissipation 
becomes important. Thus, the typical angular velocity of the rotation of the velocity 
modes is less than the typical viscous damping rate. Therefore, we assume that after 
a time interval $\Delta t$ the growing velocity mode rotates by an angle $\Delta\phi$ 
relative to its original direction perpendicular to $\bf\b$, only the 
projection of the velocity mode on an undamped direction (which is in the 
plane perpendicular to $\bf\b$) survives, and all other components of the 
velocity mode are immediately viscously dissipated. 
As a result, the effective rotational damping rate $\Omega_{\rm rd}$ can 
be estimated as
\beq
\frac{dV}{dt} &\sim& \frac{\Delta V}{\Delta t}
\sim \frac{V(\cos{\Delta\phi}-1)}{\Delta t} 
\sim {}-\frac{V}{2\Delta t}\:{\Delta\phi}^2
\sim {}-\frac{V}{2\Delta t}\;(\omega_{\rm rot}\tau_\nu)^2\frac{\Delta t}{\tau_\nu}
\nonumber\\
&\sim& {}-\frac{1}{6\tau_\nu}V
\sim \frac{1}{6}\nu_{\rm eff} k_\nu^2 V^2
\sim \frac{1}{30}\nu k_\nu^2 V^2
= {}-\Omega_{\rm rd}V^2,
\label{ROTATIONAL_DAMPING}
\\
\Omega_{\rm rd} &=& \frac{1}{6}\nu_{\rm eff} k_\nu^2
=\frac{1}{30}\nu k_\nu^2.
\label{OMEGA_DAMP}
\eeq
To obtain the last result in the first line of equation~(\ref{ROTATIONAL_DAMPING}), 
we use the random-walk approximation for the estimate of 
${\Delta\phi}^2$.~\footnote{
Basically, here we assume that $\omega_{\rm rot}^2\tau_\nu^2\sim 1/3\ll 1$.
In this case we can choose such time interval $\Delta t\gg\tau_\nu$ that 
$\Delta\phi\sim\omega_{\rm rot}\Delta t\ll 1$. As a result, on one hand, we 
can expand the cosine in equation~(\ref{ROTATIONAL_DAMPING}). On the other 
hand, the turbulent eddies on the viscous scale are correlated on time intervals 
${}\sim\tau_\nu\ll\Delta t$, and we can use the random-walk approximation to
estimate ${\Delta\phi}^2$.
}
We also use formula~(\ref{OMEGA_ROT}) for $\omega_{\rm rot}$.

As we said, the effective rotational damping is associated with non-linear coupling
of velocity modes. On one hand, this non-linearity is hard to deal with directly. 
On the other hand, the rotational damping is physical and is very important, it 
has to be included into the theory in order to restrict the growth of the velocity 
modes undamped by the Braginskii viscous forces. Therefore, we are going to 
incorporate the rotational damping into our equations in a simple way as
follows. First, we go back to the original equations for velocities, 
equations~(\ref{EQUATION_FOR_V}) and~(\ref{v_EVOLUTION}).
We recall that the total velocity ${\bf V}$ is the sum of the
Kolmogorov hydrodynamic turbulent velocity $\bf U$, which is assumed to be given,
and of the back-reaction velocity $\vbf$, which we obtain by solving 
equation~(\ref{v_EVOLUTION}). The Kolmogorov turbulence is assumed to be 
stationary (and homogeneous). Thus, the growth of the undamped modes of the 
total velocity ${\bf V}={\bf U}+\vbf$ happens because the corresponding modes 
of the back-reaction velocity $\vbf$ are undamped and
grow~\footnote{
One can obtain the Fourier-Laplace coefficient of the first order back-reaction
velocity, $\viTbf_{\bf k}$, from equation~(\ref{ViT_SOLUTION}), and then derive
formulas for the growth of $\langle\vibf^2(t)\rangle$, which are very similar 
to equations~(\ref{Vi_SQUARED})--(\ref{ViT_GROWTH}).
}.
The rotational damping restricts the growth of these back-reaction velocity
modes on the viscous scale. As for the scales larger than the viscous scales, 
the turbulence is Kolmogorov on these large scales, ${\bf V}={\bf U}$, and the back 
reaction velocities are zero [see the discussion in Section~\ref{MAGNETIZED_DYNAMO}, 
also note that the driving force~(\ref{FiT_DEFINITION}), which is ${}\propto k^2$, 
becomes small on large scales]. As a result, on all scales we add the rotational 
non-linear damping into our equations  for the velocities $\vbf$. We do this by 
replacing the time derivative $\partial/\partial t$ by 
$\partial/\partial t+\Omega_{\rm rd}$ in equation~(\ref{v_EVOLUTION}):
\beq
(\partial_t+\Omega_{\rm rd}) \v_\alpha &=& {}-P_{,\alpha}
+3\nu(b_{\alpha\beta\mu\nu}\v_{\mu,\nu})_{,\beta}
+3\nu(b_{\alpha\beta\mu\nu}U_{\mu,\nu})_{,\beta}
-\frac{1}{5}\nu\triangle U_\alpha
\nonumber\\
&&{}-(\v_\alpha U_\beta+U_\alpha\v_\beta+\v_\alpha\v_\beta)_{,\beta}\;,
\label{v_EVOLUTION_NEW}
\\
\v_{\alpha,\alpha} &=& 0.
\label{v_DIV_NEW}
\eeq
Once again, the purpose of this replacement is to incorporate the rotational non-linear 
damping, which stops the unrestricted growth of the velocity modes undamped by the 
Braginskii viscosity, into our equations in a simple way, by including the linear 
damping term $\Omega_{\rm rd}\v_\alpha$. This is our {\it second working hypothesis}.
A possible check of it can be intensive MHD numerical simulations, including 
the Braginskii viscosity. However, such simulations are rather 
complicated~\footnote{
In particular because the Braginskii forces are anisotropic and there are 
undamped velocity modes, it is not possible to drop the inertial terms in the 
MHD equations (i.~e.~not possible to assume the viscosity dominated regime). For 
example, if we drop the inertial term $s\delta_{\alpha\beta}$ in equation~(\ref{MoT}), 
then the matrix $\MoT_{{\bf k}\alpha,{\bf k'}\beta}$ will not have an inverse matrix. 
},
and they are beyond the scope of this thesis.
Note, that in equation~(\ref{v_EVOLUTION_NEW}), the effective rotational damping term
$\Omega_{\rm rd}\v_\alpha$ is isotropic, and therefore, it damps not only the growing 
modes of $\vbf$ (which are undamped by the Braginskii viscosity) but all $\vbf$ modes. 
This is not a serious problem though, because the rotational damping is smaller than 
the Braginskii damping by a factor ${}\sim 1/6$ [see equation~(\ref{OMEGA_DAMP})], and 
our results should be valid within a factor of order unity.

\section
[The Time Fourier Transform Solution for the Velocities]
{The Time Fourier Transform Solution\\
for the Velocities}
\label{FOURIER_TRANSFORM}

In this section we carry out calculations similar to those of 
Section~\ref{LAPLACE_TRANSFORM}, again assuming our first working hypothesis that the
initial field can be taken to be straight, $\bbo_{\alpha\beta}={\rm const}$. We also 
again use the quasi-linear expansion~(\ref{B_EXPANSION})--(\ref{V_EXPANSION}). The 
main difference is that now, instead of using equation~(\ref{v_EVOLUTION}) for the 
back-reaction velocity, we use equation~(\ref{v_EVOLUTION_NEW}) with the effective 
rotational damping~(\ref{OMEGA_DAMP}) in it. In the end of this section we finally 
obtain convenient formulas for the statistics of turbulent velocities in a strongly 
magnetized plasma.

The first and the second order equations~(\ref{v_EVOLUTION_i}) and~(\ref{v_EVOLUTION_ii}) 
for the back-reaction velocity become
\beq
(\partial_t+\Omega_{\rm rd})\,\vi_\alpha &=& -\:\Pi_{,\alpha}
+3\nu(\bbo_{\alpha\beta\mu\nu}\vi_{\mu,\nu})_{,\beta}
+3\nu(\bbo_{\alpha\beta\mu\nu}U_{\mu,\nu})_{,\beta}
\nonumber\\
{}&-&\frac{1}{5}\nu\triangle U_\alpha\;,
\;\,
\label{v_EVOLUTION_i_NEW}
\\
(\partial_t+\Omega_{\rm rd})\,\vii_\alpha &=& -\:\Pii_{,\alpha}
+\Big(3\nu\bbi_{\alpha\beta\mu\nu}\vi_{\mu,\nu}
+3\nu\bbo_{\alpha\beta\mu\nu}\vii_{\mu,\nu}
+3\nu\bbi_{\alpha\beta\mu\nu}U_{\mu,\nu}
\nonumber\\
{}&-&\vi_\alpha U_\beta-U_\alpha\vi_\beta-\vi_\alpha\vi_\beta\Big)_{\!,\beta}\;,
\label{v_EVOLUTION_ii_NEW}
\eeq
and they now include the rotational damping terms in them. All other first and second 
order equations~(\ref{B_EVOLUTION_i})--(\ref{b_EVOLUTION_i}),~(\ref{v_DIV_i})--
(\ref{b_EVOLUTION_ii}) and~(\ref{v_DIV_ii})--(\ref{Vii}) stay unchanged.
In Section~\ref{LAPLACE_TRANSFORM}, to solve the first and the second order 
equations, we use the Fourier and the Laplace transformations in space and in time 
respectively. In this section it is convenient to use the Fourier transformations
both in space and in time~\footnote{
Using the Fourier transformation in time is actually better than using the Laplace
transformation. The reason is that in the case of the Laplace transformation we 
have to assume the initial condition on the back-reaction velocity (in 
Section~\ref{LAPLACE_TRANSFORM} we assume ${\vibf|}_{t=0}$), while the Fourier 
transformation ``does not care'' about the initial condition. As a result, 
we do not have to worry about any velocity transients, which are eventually 
damped away by the Braginskii viscosity and by the rotational damping.
}.
We use the discrete Fourier transformation ${\bf r}\rightarrow{\bf k}$ in space and 
the continuous Fourier transformation $t\rightarrow \omega$ in time (see 
Appendix~\ref{TRANSFORMS}). Comparing equations~(\ref{v_EVOLUTION_i}) 
and~(\ref{v_EVOLUTION_i_NEW}), and equations~(\ref{TIME_FOURIER_A}) 
and~(\ref{TIME_LAPLACE_A}), we easily see that the double Fourier transformation 
(in space and in time) of equation~(\ref{v_EVOLUTION_i_NEW}), which is the first 
order equation for the back-reaction velocity Fourier coefficient 
$\viT_{{\bf k},\alpha}(\omega)$, coincides with equation~(\ref{viT_EVOLUTION_PRESSURE}) 
where the Laplace variable $s$ is replaced by $-i\omega+\Omega_{\rm rd}$. As a result, 
there is no need to rederive all equations of Section~\ref{LAPLACE_TRANSFORM}. In most 
cases we can obtain the new formulas just by making the replacement 
$s\rightarrow -i\omega+\Omega_{\rm rd}$ in the corresponding formulas of 
Section~\ref{LAPLACE_TRANSFORM}. However, let us be more general, and in the 
calculations below assume that the effective rotational damping rate $\Omega_{\rm rd}$ 
is a function of $k=|{\bf k}|$ and of 
$({\bf\b}\cdot{\bf\k})^2$,~\footnote{
We choose $\Omega_{\rm rd}$ to depend on the square of ${\bf\b}\cdot{\bf\k}$
because the Braginskii viscosity is invariant under reflection 
${\bf\b}\rightarrow{}-{\bf\b}$.
\label{BRAGINSKII_IS_EVEN_IN_b}
}
e.~g.~$\Omega_{\rm rd}=\Omega_{\rm rd}(k,\mu^2)$ up to the zero order, where 
$\mu$ is given by equation~(\ref{MU}). We can always substitute our simple 
estimate~(\ref{OMEGA_DAMP}) for $\Omega_{\rm rd}$ into our final formulas.

Equations~(\ref{ViT_L})--(\ref{ViT_L_SECOND}) for the Fourier 
coefficient of the first order velocity now become
\beq
\ViT_{{\bf k}\alpha}(\omega) &=& \ViT'_{{\bf k}\alpha}(\omega)
+\ViT''_{{\bf k}\alpha}(\omega)\;,
\label{ViT_F}
\\
\ViT'_{{\bf k}\alpha}(\omega) &=& 
\frac{-i\omega+{\bar\Omega}+\Omega_{\rm rd}}{-i\omega+\Omega_{\rm rd}} 
\left[\delta_{\alpha\beta}-
\frac{\delta^\perp_{\alpha\gamma}\bo_\gamma\bo_\beta}{1-\mu^2}\right]
{\tilde U}_{{\bf k}\beta}(\omega)\;,
\label{ViT_F_FIRST}
\\
\ViT''_{{\bf k}\alpha}(\omega) &=& 
\frac{-i\omega+{\bar\Omega}+\Omega_{\rm rd}}{-i\omega+2\Omega+\Omega_{\rm rd}}\;
\frac{\delta^\perp_{\alpha\gamma}\bo_\gamma\bo_\beta}{1-\mu^2}\:
{\tilde U}_{{\bf k}\beta}(\omega)\;.
\label{ViT_F_SECOND}
\eeq
In these formulas the effective rotational damping rate, $\Omega_{\rm rd}$, 
is added to the viscous damping frequencies~(\ref{OMEGA_AVERAGED}) and~(\ref{OMEGA}).
Thus, the overall damping is produced by both the rotational damping rate and
the viscous damping (of course, there is only one physical dissipation process 
present in the fluid --- the viscous dissipation).
Now, we use formulas~(\ref{U_STATISTICS}),~(\ref{U_U_STATISTICS}) to obtain averages 
similar to those given by equations~(\ref{ViT'})--(\ref{ViT''_ViT''}). We have
\beq
\langle\ViT'_{{\bf k}\alpha}(\omega)\rangle \!\!&=&\!\! 0,
\label{ViT'_NEW}
\\
\langle\ViT''_{{\bf k}\alpha}(\omega)\rangle \!\!&=&\!\! 0, 
\\
\langle\ViT'_{{\bf k}\alpha}(\omega)\ViT''_{{\bf k'}\beta}(\omega')\rangle \!\!&=&\!\! 0,
\label{ViT'_ViT''_NEW}
\\
\langle\ViT'_{{\bf k}\alpha}(\omega)\ViT'_{{\bf k'}\beta}(\omega')\rangle \!\!&=&\!\!
J_{\omega k}\,\delta_{{\bf k'},-{\bf k}}\,\delta(\omega'+\omega)\;
\frac{\omega^2+({\bar\Omega}+\Omega_{\rm rd})^2}{\omega^2+\Omega_{\rm rd}^2}
\nonumber\\
{}&\times&
\left[\delta^\perp_{\alpha\beta}-
\frac{\delta^\perp_{\alpha\gamma}\delta^\perp_{\beta\tau}\bo_\gamma\bo_\tau}
{1-\mu^2}\right],
\label{ViT'_ViT'_NEW}
\\
\langle\ViT''_{{\bf k}\alpha}(\omega)\ViT''_{{\bf k'}\beta}(\omega')\rangle \!\!&=&\!\!
J_{\omega k}\,\delta_{{\bf k'},-{\bf k}}\,\delta(\omega'+\omega)\;
\frac{\omega^2+({\bar\Omega}+\Omega_{\rm rd})^2}{\omega^2+(2\Omega+\Omega_{\rm rd})^2}
\nonumber\\
{}&\times&\frac{\delta^\perp_{\alpha\gamma}\delta^\perp_{\beta\tau}\bo_\gamma\bo_\tau}
{1-\mu^2} 
\label{ViT''_ViT''_NEW}\;.
\eeq
Using these equations and equation~(\ref{ViT_F}), we obtain equations, which are similar 
to equations~(\ref{ViT_L_AVERAGED}) and~(\ref{ViT_ViT_L_AVERAGED}) of 
Section~\ref{LAPLACE_TRANSFORM},
\beq
\langle\ViT_{{\bf k}\alpha}(\omega)\rangle \!\!&=&\!\! 0,
\label{ViT_F_AVERAGED}
\\
\langle\ViT_{{\bf k}\alpha}(\omega)\ViT_{{\bf k'}\beta}(\omega')\rangle \!\!&=&\!\!
\langle\ViT'_{{\bf k}\alpha}(\omega)\ViT'_{{\bf k'}\beta}(\omega')\rangle
+\langle\ViT''_{{\bf k}\alpha}(\omega)\ViT''_{{\bf k'}\beta}(\omega')\rangle
\nonumber\\
{} \!\!&=&\!\! J_{\omega k}\,\delta_{{\bf k'},-{\bf k}}\,\delta(\omega'+\omega)
\nonumber\\
{} \!\!&\times&\!\! 
\left[{\tilde H}_F(\omega;{\bar\Omega}+\Omega_{\rm rd},\Omega_{\rm rd})
\left(\delta^\perp_{\alpha\beta}-
\frac{\delta^\perp_{\alpha\gamma}\delta^\perp_{\beta\tau}\bo_\gamma\bo_\tau}
{1-\mu^2}\right)\right.
\nonumber\\
\!\!&&\!\!
{}+\left.{\tilde H}_F(\omega;{\bar\Omega}+\Omega_{\rm rd},2\Omega+\Omega_{\rm rd})\:
\frac{\delta^\perp_{\alpha\gamma}\delta^\perp_{\beta\tau}\bo_\gamma\bo_\tau}
{1-\mu^2}
\right],
\label{ViT_ViT_F_AVERAGED}
\eeq
where function ${\tilde H}_F(\omega;q_1,q_2)$ is
\beq
{\tilde H}_F(\omega;q_1,q_2) &=& 
\frac{\omega^2+q_1^2}{\omega^2+q_2^2}
=1+\frac{q_1^2-q_2^2}{\omega^2+q_2^2}\;.
\label{H_F_T}
\eeq
Next, we apply the inverse Fourier transformations $\omega\rightarrow t$ and 
$\omega'\rightarrow t'$ to equations~(\ref{ViT_F_AVERAGED}) 
and~(\ref{ViT_ViT_F_AVERAGED}). 
We have
\beq
\langle\ViT_{{\bf k}\alpha}(t)\rangle \!\!&=&\!\! 0,
\label{ViT_AVERAGED_NEW}
\\
\langle\ViT_{{\bf k}\alpha}(t)\ViT_{{\bf k'}\beta}(t')\rangle \!\!&=&\!\!
\langle\ViT_{{\bf k}\alpha}(t)\ViT^*_{-{\bf k'}\beta}(t')\rangle
\nonumber\\
\!\!&=&\!\! \langle\ViT'_{{\bf k}\alpha}(t)\ViT'_{{\bf k'}\beta}(t')\rangle
+\langle\ViT''_{{\bf k}\alpha}(t)\ViT''_{{\bf k'}\beta}(t')\rangle
\nonumber\\
\!\!&=&\!\! \delta_{{\bf k'},-{\bf k}}\;
\left[H_F(t-t';{\bar\Omega}+\Omega_{\rm rd},\Omega_{\rm rd})
\left(\delta^\perp_{\alpha\beta}-
\frac{\delta^\perp_{\alpha\gamma}\delta^\perp_{\beta\tau}\bo_\gamma\bo_\tau}
{1-\mu^2}\right)\right.
\nonumber\\
\!\!&&\!\!\left.\quad{}
+H_F(t-t';{\bar\Omega}+\Omega_{\rm rd},2\Omega+\Omega_{\rm rd})\:
\frac{\delta^\perp_{\alpha\gamma}\delta^\perp_{\beta\tau}\bo_\gamma\bo_\tau}
{1-\mu^2}
\right],
\label{ViT_ViT_AVERAGED_NEW}
\eeq
where function
\beq
H_F(t-t';q_1,q_2) &=& \frac{1}{2\pi}\!
\int\limits_{-\infty}^\infty J_{\omega k}\,{\tilde H}_F(\omega;q_1,q_2)\:
e^{-i\omega(t-t')}\,d\omega\:
\nonumber\\
{}&=& \frac{1}{2\pi}\!
\int\limits_{-\infty}^\infty J_{\omega k}\,{\tilde H}_F(\omega;q_1,q_2)\,
\cos{[\omega(t-t')]}\,d\omega\:
\label{H_F}
\eeq
depends only on the absolute value of the time difference $t-t'$ because $J_{\omega k}$ 
and ${\tilde H}_F(\omega;q_1,q_2)$ both are even functions of $\omega$. 

Equations~(\ref{ViT_AVERAGED_NEW})--(\ref{ViT_ViT_AVERAGED_NEW}) are similar to
equations~(\ref{ViT_AVERAGED})--(\ref{ViT_ViT_AVERAGED}) of 
Section~\ref{LAPLACE_TRANSFORM},~\footnote{
It is interesting, that we can obtain equations~(\ref{ViT_AVERAGED_NEW})--(\ref{H_F}) 
from equations~(\ref{H_L_T_qq})--(\ref{ViT_ViT_AVERAGED}) by first, making replacements 
${\bar\Omega}\rightarrow{\bar\Omega}+\Omega_{\rm rd}$, 
$2q_2\rightarrow 2q_2+\Omega_{\rm rd}$ in equation~(\ref{H_L_T_qq}), and second, by 
taking a limit $t\rightarrow\infty$ in it. This is not surprising, because as time goes 
on, the velocity transients, which ``remember'' the initial conditions, are damped away 
by the Braginskii viscosity and by the rotational damping.
}
and they give the statistics of the first order turbulent velocities, which we 
will use in the next chapter. As, for the second order turbulent velocity,
equation~(\ref{Vii_AVERAGE}) stay the same,
\beq
\langle\ViiT_{{\bf k}\alpha}(t)\rangle &=& 0,
\label{Vii_AVERAGE_NEW}
\eeq
because the proof of this formula, given in Appendix~\ref{Vii_ENSEMBLE_AVERAGE},  
does not involve any transformation in time.

In the end of this section let us substitute equation~(\ref{J_OMEGA_K}) for 
$J_{\omega k}$ and equation~(\ref{H_F_T}) for ${\tilde H}_F(\omega;q_1,q_2)$ into 
equation~(\ref{H_F}), and obtain the following formulas
\beq
H_F(0;q_1,q_2) \!\!&=&\!\!
\frac{1}{2\pi}\!\int\limits_{-\infty}^\infty \frac{J_{0k}}{1+\tau^2\omega^2}
\left[1+\frac{q_1^2-q_2^2}{\omega^2+q_2^2}\right]
d\omega
\nonumber\\
{}\!\!&=&\!\! \frac{J_{0k}}{2\tau}
+\frac{J_{0k}}{2}\frac{q_1^2-q_2^2}{q_2(1+\tau q_2)},
\label{H_F_ZERO}
\\
\int\limits_0^t\! H_F(t-t')dt' \!\!&=&\!\!
\frac{1}{2\pi}\!\int\limits_{-\infty}^\infty \frac{J_{0k}}{1+\tau^2\omega^2}
\left[1+\frac{q_1^2-q_2^2}{\omega^2+q_2^2}\right]
\frac{\sin{\omega t}}{\omega}
\,d\omega
\nonumber\\
{}\!\!&=&\!\! \frac{J_{0k}}{2}(1-e^{-t/\tau})
+\frac{J_{0k}}{2}\frac{q_1^2-q_2^2}{1-\tau^2q_2^2}
\left[\frac{1-e^{-q_2t}}{q_2^2}-\tau^2(1-e^{-t/\tau})\right],
\nonumber\\
{}\!\!&\rightarrow&\!\! \frac{J_{0k}}{2}+\frac{J_{0k}}{2}\frac{q_1^2-q_2^2}{q_2^2}
= \frac{J_{0k}}{2}\,\frac{q_1^2}{q_2^2}\:,
\label{H_F_INTEGRAL_T}
\\
\int\limits_0^t dt'\!\int\limits_0^{t'}\! H_F(t'-t'')dt'' \!\!&=&\!\!
\frac{1}{\pi}\!\int\limits_{-\infty}^\infty \frac{J_{0k}}{1+\tau^2\omega^2}
\left[1+\frac{q_1^2-q_2^2}{\omega^2+q_2^2}\right]
\frac{\sin^2{(\omega t/2)}}{\omega^2}
\,d\omega
\nonumber\\
{}\!\!&=&\!\! \frac{J_{0k}}{2}\left[t-\tau(1-e^{-t/\tau})\right]
\nonumber\\
\!\!&&\!\!{}+ \frac{J_{0k}}{2}\frac{q_1^2-q_2^2}{1-\tau^2q_2^2}
\left[\frac{q_2t-1+e^{-q_2t}}{q_2^3}-\tau^2t+\tau^3(1-e^{-t/\tau})\right]
\nonumber\\
{}\!\!&\rightarrow&\!\! \frac{J_{0k}}{2}\,t
+\frac{J_{0k}}{2}\,\frac{q_1^2-q_2^2}{q_2^2}\,t
= \frac{J_{0k}}{2}\frac{q_1^2}{q_2^2}\,t\:,
\label{H_F_INTEGRAL_T'T}
\\
\int\limits_0^t dt'\!\int\limits_0^{t}\! H_F(t'-t'')dt'' \!\!&=&\!\!
2\int\limits_0^t dt'\!\int\limits_0^{t'}\! H_F(t'-t'')dt''
\rightarrow J_{0k}\,\frac{q_1^2}{q_2^2}\,t\:,
\label{H_F_INTEGRAL_TT}
\eeq
which we will use below.
The integrals over $\omega$ can be done by closing the integration contours
in the complex plane and by evaluating the residues. The final answers in
formulas~(\ref{H_F_INTEGRAL_T})--(\ref{H_F_INTEGRAL_TT}), which are given after 
the right arrows $\rightarrow$, give the results in the limit $t\gg \tau$, 
$t\gg q_2^{-1}$. 

Using equations~(\ref{ViT_ViT_AVERAGED_NEW}) and~(\ref{H_F_ZERO}), we easily 
obtain
\beq
\langle|\ViTbf_{\bf k}(t)|^2\rangle &=&
H_F(0;{\bar\Omega}+\Omega_{\rm rd},\Omega_{\rm rd})
+H_F(0;{\bar\Omega}+\Omega_{\rm rd},2\Omega+\Omega_{\rm rd})
\nonumber\\
{}&=& 
\frac{J_{0k}}{\tau}
+\frac{J_{0k}}{2}\frac{({\bar\Omega}+\Omega_{\rm rd})^2-\Omega_{\rm rd}^2}
{\Omega_{\rm rd}(1+\tau \Omega_{\rm rd})}
\nonumber\\
&&{}+\frac{J_{0k}}{2}\frac{({\bar\Omega}+\Omega_{\rm rd})^2-(2\Omega+\Omega_{\rm rd})^2}
{(2\Omega+\Omega_{\rm rd})[1+\tau(2\Omega+\Omega_{\rm rd})]}.
\eeq
Substituting this equation into equation~(\ref{Vi_SQUARED}), we see that the ensemble 
average of $\Vibf$ squared is finite, but it would be infinite if the rotational 
damping were absent (i.~e.~if $\Omega_{\rm rd}=0$).


%% file: chapter_4.tex
\chapter
[Energy Spectrum of Random Magnetic Fields]
{Energy Spectrum of Random\\
Magnetic Fields}
\label{MAGNETIC_SPECTRUM}

In this chapter we use equations~(\ref{ViT_AVERAGED_NEW})--(\ref{Vii_AVERAGE_NEW})
for the statistics of turbulent velocities to find the evolution of the magnetic
field energy and of the magnetic energy spectrum in the magnetized turbulent dynamo
theory. The principal results of this chapter (and of the thesis) are given by
the following equations. The growth of the total magnetic energy is given by 
equations~(\ref{ENERGY_GROWTH})--(\ref{GAMMA_RESULT}). The evolution of the magnetic 
energy spectrum is given by 
equations~(\ref{MODE_COUPLING})--(\ref{OMEGA_DAMP_RATIO_SECOND}). The evolution of 
the magnetic energy spectrum on small (subviscous) scales is given by 
equations~(\ref{SMALL_SCALES_MODE_COUPLING})--(\ref{GREENS_FUNCTION}).

\section
[The Growth of the Total Magnetic Energy]
{The Growth of the Total Magnetic Energy}
\label{MAGNETIC_ENERGY}

We start with calculation of the magnetic field energy growth because it is of the
greatest interest. The volume averaged and ensemble 
averaged~\footnote{
Ensemble averaged over the ensemble of turbulent forces.
} 
magnetic energy per unit mass is
\beq
{\cal E} &\define& \frac{1}{L^3}\int\limits_{-L/2}^{L/2}
\frac{\langle B^2\rangle}{8\pi\rho}\,d^3{\bf r}
=\frac{1}{8\pi\rho}\Big\langle\BST_{{\bf k}=0}\Big\rangle\:,
\label{AVERAGED_MAGNETIC_ENERGY}
\eeq
where $\rho$ is the plasma density, and $\BST_{{\bf k}=0}$ is the 
${\bf k}=0$ Fourier coefficient of the magnetic field strength squared, $B^2$,
see equation~(\ref{FOURIER_COEFFICIENTS_A}) in Appendix~\ref{TRANSFORMS}. To find
${\cal E}(t)$, it is convenient to introduce the following symmetric tensor
\beq
B_{\alpha\beta}&\define&B_\alpha B_\beta=B^2\,b_{\alpha\beta}\:,
\label{BB}
\eeq
where $b_{\alpha\beta}$ is given by equation~(\ref{bb}).
The differential equation for $B_{\alpha\beta}$ can easily be derived from 
equation~(\ref{B_EVOLUTION}),
\beq
\partial_t B_{\alpha\beta} = B_\alpha\partial_t B_\beta + B_\beta\partial_t B_\alpha
=V_{\alpha,\gamma} B_{\beta\gamma}+V_{\beta,\gamma} B_{\alpha\gamma}
-V_\gamma B_{\alpha\beta,\gamma}\;.
\label{BB_EVOLUTION}
\eeq
Now, we solve this equation and equation~(\ref{B_STRENGTH_EVOLUTION}) by making use of 
the quasi-linear expansion procedure, described in Section~\ref{EXPANSION}.
First, we write the expansion for $B^2$ up to the second order, and for 
$B_{\alpha\beta}$ up to the first 
order~\footnote{
Note, that below the magnetic field strength squared $B^2$ is expanded as a whole. 
Thus, for example, the first order quantity $\BSi$ in equation~(\ref{BS_EXPANSION}) is 
not equal to $\Bibf\cdot\Bibf$, where $\Bibf$ is the first order expansion term in 
equation~(\ref{B_EXPANSION}), the later is of the second order. Of course, $\BSi$ 
is equal to $2\,(\Bobf\cdot\Bibf)$.
},
\beq
B^2(t) &=& \BSo+\BSi(t)+\BSii(t),
\label{BS_EXPANSION}
\\
B_{\alpha\beta}(t) &=& \Bo_{\alpha\beta}+\Bi_{\alpha\beta}(t).
\eeq
Here, at zero time $B^2(0)=\BSo$, $\BSi(0)=0$, $\BSii(0)=0$, 
$B_{\alpha\beta}(0)=\Bo_{\alpha\beta}$ and $\Bi_{\alpha\beta}(0)=0$.
Second, we substitute these expansion formulas into 
equations~(\ref{B_STRENGTH_EVOLUTION}) and~(\ref{BB_EVOLUTION}). We find that the
first order equations are
\beq
\partial_t\BSi &=& 
2\,\Vi_{\alpha,\beta} \Bo_{\alpha\beta} - \Vi_\beta \BSo_{,\beta}\;,
\label{BS_FIRST_ORDER_EVOLUTION}
\\
\partial_t \Bi_{\alpha\beta} &=&
\Vi_{\alpha,\gamma}\Bo_{\beta\gamma}+\Vi_{\beta,\gamma}\Bo_{\alpha\gamma}
-(\Vi_\gamma\Bo_{\alpha\beta})_{,\gamma}\;,
\label{BB_FIRST_ORDER_EVOLUTION}
\eeq
and the second order equation for $\BSii(t)$ is
\beq
\partial_t\BSii &=& 
2\,\Vi_{\alpha,\beta} \Bi_{\alpha\beta}
+2\,\Vii_{\alpha,\beta} \Bo_{\alpha\beta} 
-(\Vi_\beta \BSi)_{,\beta}
-(\Vii_\beta \BSo)_{,\beta}\;.
\label{BS_SECOND_ORDER_EVOLUTION}
\eeq
Here, in the last two equations we transform the last (convective) terms by making 
use of the plasma incompressibility condition, $V_{\alpha,\alpha}=0$.

Next, we first integrate first order equation~(\ref{BS_FIRST_ORDER_EVOLUTION}) 
in time (with the zero initial conditions) and ensemble average the result. Using 
equation~(\ref{ViT_AVERAGED_NEW}), we obviously obtain
\beq
\Big\langle\BSi(t)\Big\rangle &=& 0.
\label{BSi_AVERAGED}
\eeq
Second, we integrate equation~(\ref{BB_FIRST_ORDER_EVOLUTION}) in time (with the 
zero initial conditions), and then Fourier transform the result in space, 
${\bf r}\rightarrow{\bf k}$. We have
\beq
\BiT_{{\bf k}\alpha\beta}(t) \!\!&=&\!\!
i\int\limits_0^t\, \Bigg[\,
k'_\gamma\sum_{{\bf k'}\atop{{\bf k}''=\,{\bf k}-{\bf k}'}}
\ViT_{{\bf k'}\alpha}(t')\,\BoT_{{\bf k''}\beta\gamma}
+k'_\gamma\sum_{{\bf k'}\atop{{\bf k}''=\,{\bf k}-{\bf k}'}}
\ViT_{{\bf k'}\beta}(t')\,\BoT_{{\bf k''}\alpha\gamma}
\nonumber\\
\!\!&&\!\!{}-k_\gamma\sum_{{\bf k'}\atop{{\bf k}''=\,{\bf k}-{\bf k}'}}
\ViT_{{\bf k'}\gamma}(t')\,\BoT_{{\bf k''}\alpha\beta}
\,\Bigg]\,dt'
\nonumber\\
{}\!\!&=&\!\! i\Big[k'_\gamma(\delta_{\alpha\tau}\bbo_{\beta\gamma}
+\delta_{\beta\tau}\bbo_{\alpha\gamma})
-k_\tau\bbo_{\alpha\beta}\Big]
\int\limits_0^t \!\!\sum_{{\bf k'}\atop{{\bf k}''=\,{\bf k}-{\bf k}'}}\!\!\!
\ViT_{{\bf k'}\tau}(t')\,\BSoT_{\bf k''}\,dt'.
\label{BBiT}
\eeq
Here, to obtain the last line of this equation, we use equation~(\ref{BB}) and we 
assume that the magnetic field lines can be considered as initially straight, 
$\bbo_{\alpha\beta}={\rm const}$, (this is our first working hypothesis, see 
Section~\ref{LAPLACE_TRANSFORM}).
Third, we integrate the second order equation~(\ref{BS_SECOND_ORDER_EVOLUTION}) 
in time (with the zero initial conditions) and ensemble average the result. Using 
equation~(\ref{Vii_AVERAGE_NEW}), we obtain
\beq
\Big\langle\BSii(t)\Big\rangle =
\int\limits_0^t \bigg[
2\,\Big\langle\Vi_{\alpha,\beta}(t')\Bi_{\alpha\beta}(t')\Big\rangle
-\Big\langle\Vi_\beta(t')\BSi(t')\Big\rangle_{\!,\beta} \bigg]\,dt'.
\label{BSii_INTEGRATED_AVERAGED}
\eeq
Fourth, we Fourier transform this result in space, ${\bf r}\rightarrow{\bf k}$, 
and set ${\bf k}$ to zero. The second (convective) term in the brackets $[...]$
in equation~(\ref{BSii_INTEGRATED_AVERAGED}) gives zero contribution to the
final result~\footnote{
This is not surprising because the convective term can only redistribute magnetic
energy in space, but can not change the volume averaged energy.
}, 
and we have
\beq
\Big\langle\BSiiT_{{\bf k}=0}(t)\Big\rangle =
2i\int\limits_0^t \sum_{\bf k}
k_\beta\,\Big\langle\ViT_{{\bf k}\alpha}(t')\,\BiT_{-{\bf k}\alpha\beta}(t')\Big\rangle
\,dt'.
\eeq
Fifth, we substitute equation~(\ref{BBiT}) into this last equation, and use 
equation~(\ref{ViT_ViT_AVERAGED_NEW}). We obtain
\beq
\Big\langle\BSiiT_{{\bf k}=0}(t)\Big\rangle \!\!&=&\!\!
2\,\BSoT_{{\bf k}=0} \int\limits_0^t dt'\,\int\limits_0^{t'} \sum_{\bf k}
\mu^2k^2\, \langle\ViT_{{\bf k}\alpha}(t')\ViT_{-{\bf k}\alpha}(t'')\rangle
\,dt''
\nonumber\\
{}\!\!&=&\!\! 2\,\BSoT_{{\bf k}=0}\,\sum_{\bf k}\mu^2k^2\, 
\int\limits_0^t dt'\,\int\limits_0^{t'}
\Big[H_F(t'-t'';{\bar\Omega}+\Omega_{\rm rd},\Omega_{\rm rd})
\nonumber\\
&&\qquad\qquad{}+H_F(t'-t'';{\bar\Omega}+\Omega_{\rm rd},2\Omega+\Omega_{\rm rd})\Big]
\,dt''
\nonumber\\
{}&=& 2\gamma t\,\BSoT_{{\bf k}=0}\:,
\label{BSiiT_AVERAGED}
\eeq
where 
\beq
\gamma = \frac{1}{2}\sum_{\bf k} k^2 J_{0k} 
\left(1+\frac{{\bar\Omega}}{\Omega_{\rm rd}}\right)^2\,
\mu^2\left[\,1+\left(1+\frac{2\Omega}{\Omega_{\rm rd}}\right)^{-2}\,\right].
\label{GAMMA_SUM}
\eeq
Here, we also use equation~(\ref{H_F_INTEGRAL_T'T}) in the limit $t\gg\tau$.

Now, we can finally obtain the differential equation for the averaged magnetic energy
$\cal E$. Following Kulsrud and Anderson~\cite{KA_92}, we choose $t$ small enough 
for the quasi-linear expansion to be valid, but large enough for the limit $t\gg\tau$ 
to be satisfied. As a result, using equations~(\ref{BS_EXPANSION}),~(\ref{BSi_AVERAGED}) 
and~(\ref{BSiiT_AVERAGED}), we obtain
\beq
\frac{\partial}{\partial t}\Big\langle\BST_{{\bf k}=0}(t)\Big\rangle =
\frac{1}{t}\left[\Big\langle\BST_{{\bf k}=0}(t)\Big\rangle-\BST_{{\bf k}=0}(0)\right]
=\frac{1}{t}\Big\langle\BSiiT_{{\bf k}=0}(t)\Big\rangle
=2\gamma\BSoT_{{\bf k}=0}\:,
\eeq
and using equation~(\ref{AVERAGED_MAGNETIC_ENERGY}), we finally obtain
\beq
\frac{\partial{\cal E}}{\partial t} &=& 2\gamma{\cal E},
\label{ENERGY_GROWTH}
\\
\gamma &=& \pi{\left(\frac{L}{2\pi}\right)}^{\!3}
\int\limits_0^\infty k^4 J_{0k}\,dk
\int\limits_{-1}^1 \mu^2\left(1+\frac{{\bar\Omega}}{\Omega_{\rm rd}}\right)^2\,
\left[\,1+\left(1+\frac{2\Omega}{\Omega_{\rm rd}}\right)^{-2}\,\right]d\mu\,,
\qquad
\label{GAMMA}
\\
\frac{{\bar\Omega}}{\Omega_{\rm rd}} \!\!&=&\!\!
6\,\frac{k^2}{k_\nu^2}\,,
\label{OMEGA_BAR_DAMP_RATIO}
\\
\frac{2\Omega}{\Omega_{\rm rd}} \!\!&=&\!\!
90\,\frac{k^2}{k_\nu^2}\,\mu^2(1-\mu^2)\,.
\label{OMEGA_DAMP_RATIO}
\eeq
Here, we replace the summation over ${\bf k}$ in equation~(\ref{GAMMA_SUM}) by 
integration over ${\bf k}$, and then use $d^3{\bf k}=2\pi k^2\,dk\,d\mu$. We also use 
equations~(\ref{OMEGA_AVERAGED}),~(\ref{OMEGA}) and~(\ref{OMEGA_DAMP}).
Equation~(\ref{ENERGY_GROWTH}) coincides with formula~(\ref{ENERGY_GROWTH_KA}) obtained 
by Kulsrud and Anderson in the case of the kinematic turbulent dynamo. However, the 
growth rate $\gamma$, given by equation~(\ref{GAMMA}) in the case of the magnetized 
turbulent dynamo, is different from the growth rate $\gamma_{\rm o}$, given by 
equation~(\ref{GAMMA_0}) in the case of the kinematic dynamo. Let us now compare 
these two magnetic energy growth rates.
First, if we consider the kinematic turbulent dynamo, then in our formulas we need 
to take the limit ${\bar\Omega}\rightarrow 0$, $\Omega\rightarrow 0$, or alternatively, 
the limit $\Omega_{\rm rd}\rightarrow \infty$~\footnote{
In this limit the ``driving'' terms $3\nu(\bbo_{\alpha\beta\mu\nu}U_{\mu,\nu})_{,\beta}$
and $(1/5)\nu\triangle U_\alpha$ in equation~(\ref{v_EVOLUTION_i_NEW}), which in the
Fourier $\bf k$-space result in the two corresponding driving terms proportional to 
$\Omega$ and ${\bar\Omega}$ respectively [e.~g.~see equation~(\ref{v_i_EVOLUTION_B})],
go away. Taking the limit $\Omega_{\rm rd}\rightarrow \infty$ is equivalent to an 
infinitely large damping of the back-reaction velocities. As a result, in both these 
limits the back-reaction velocities are zero, and the turbulence is Kolmogorov.
\label{KA_LIMIT_FOOTNOTE}
}. 
In this case, after averaging over $\mu$, the growth rate~(\ref{GAMMA}) 
[see also~(\ref{GAMMA_SUM})] reduces to the growth rate~(\ref{GAMMA_0}), as one can 
expect. Second, let us use formula~(\ref{OMEGA_DAMP}) to make an estimates of the 
magnetized dynamo growth rate $\gamma$ and of the ratio $\gamma/\gamma_{\rm o}$. The 
calculations are given in Appendix~\ref{CALCULATION_OF_GAMMA}, the result is
\beq
\gamma &\approx& 80\,\left(\frac{U_0L}{\nu}\right)^{1/2}\frac{U_0}{L},
\label{GAMMA_RESULT}
\\
\gamma^{-1} &\approx& 
10^5\:{\rm yrs}\, \left(\frac{\xi}{10}\right)^{\!1/2}
\left(\frac{L}{0.2\,{\rm Mpc}}\right)^{\!3/2},
\label{GAMMA_INVERSE_RESULT}
\\
\gamma\,t_{\rm collapse} &\approx& 
10^4\, \left(\frac{\xi}{10}\right)^{\!-1/2}
\left(\frac{M}{10^{12}\,{\rm M}_\odot}\right)^{\!-1/2},
\label{GAMMA_COLLAPSE_TIME}
\\
\gamma/\gamma_{\rm o} &\approx& 10,
\label{GAMMA_RATIO}
\eeq
where to obtain formulas~(\ref{GAMMA_INVERSE_RESULT}) and~(\ref{GAMMA_COLLAPSE_TIME}), 
we use typical parameters in a protogalaxy given in Table~\ref{TABLE_PARAMETERS} 
(here $\xi$ is the ratio of the total mass $M$ to the baryon mass, $L$ is the 
protogalaxy size). Thus, our prediction is that the magnetic energy growth rate 
in the magnetized dynamo theory is up to ten time larger than that in the 
kinematic dynamo theory. Note, that two different effects contribute to this difference. 
First, the effective viscosity in the magnetized dynamo theory is smaller than the 
molecular viscosity, $\nu_{\rm eff}=(1/5)\nu<\nu$, this effect makes the magnetic 
energy growth rate larger by a factor of square root of five (this factor was included
in Kulsrud \etal~\cite{KCOR_97}). The rest of the contribution to the difference between
the growth rates comes from the local anisotropy of the turbulent velocities in the 
strongly magnetized plasma. 

Please note, that the number of e-foldings of the magnetic energy during the collapse 
time, $\gamma\,t_{\rm collapse}$, given by equation~(\ref{GAMMA_COLLAPSE_TIME}), does 
not depend on $L$, and therefore, does not depend on the redshift.

\section
[The Mode Coupling Equation for the Magnetic Energy Spectrum]
{The Mode Coupling Equation for\\ the Magnetic Energy Spectrum}
\label{MODE_COUPLING_EQUATION}

In this section we derive the mode coupling equation for the evolution
of the magnetic energy spectrum $M(t, k)$, given by equation~(\ref{M_K}). Using the
quasi-linear expansion formula~(\ref{B_EXPANSION}) for the magnetic field, we
obtain the ensemble averaged square of the magnetic field Fourier coefficient, up 
to the second order,
\beq
\langle|{\bf{\tilde B}}_{\bf k}(t)|^2\rangle &=& 
|\BoTbf_{\bf k}(t)|^2 
+\Big[\langle\BiT_{{\bf k}\alpha}(t)\rangle\BoT^*_{{\bf k}\alpha}(t)
+{\rm c.c.}\Big]
\nonumber\\
&&{}+\langle|\BiTbf_{\bf k}(t)|^2\rangle
+\Big[\langle\BiiT_{{\bf k}\alpha}(t)\rangle\BoT^*_{{\bf k}\alpha}(t)
+{\rm c.c.}\Big].
\label{BkS_EXPANSION}
\eeq
We calculate all terms in equation~(\ref{BkS_EXPANSION}) separately, our calculations
are similar to those of Kulsrud and Anderson~\cite{KA_92}.

\subsection
[The $\langle\BiT_{{\bf k}\alpha}(t)\rangle\BoT^*_{{\bf k}\alpha}+{\rm c.c.}$ term]
{The {\boldmath$\langle\BiT_{{\bf k}\alpha}(t)\rangle\BoT^*_{{\bf k}\alpha}+{\rm c.c.}$} term}
\label{Bi_Bo_COUPLING_TERM}

We start with the calculation of the
$\langle\BiT_{{\bf k}\alpha}(t)\rangle\BoT^*_{{\bf k}\alpha}+{\rm c.c.}$ term. 
We integrate the first order equation~(\ref{B_EVOLUTION_i}) in time 
(with the zero initial conditions) and ensemble average the result. Using 
equation~(\ref{ViT_AVERAGED_NEW}), we obviously obtain 
$\langle\Bi_\alpha(t)\rangle=0$. Thus,
\beq
\langle\BiT_{{\bf k}\alpha}(t)\rangle &=& 0,
\label{BiT_AVERAGED}
\eeq
and the $\langle\BiT_{{\bf k}\alpha}\rangle\BoT^*_{{\bf k}\alpha}+{\rm c.c.}$ term 
on the right-hand-side of equation~(\ref{BkS_EXPANSION}) is zero.

\subsection
[The $\langle|\BiTbf_{\bf k}(t)|^2\rangle$ term]
{The {\boldmath$\langle|\BiTbf_{\bf k}(t)|^2\rangle$} term}
\label{Bi_Bi_COUPLING_TERM}

Next, we calculate the $\langle|\BiTbf_{\bf k}(t)|^2\rangle$ term. 
First, we integrate equation~(\ref{B_EVOLUTION_i}) in time (with the zero initial 
conditions), and then Fourier transform the result in space, 
${\bf r}\rightarrow{\bf k}$. We have
\beq
\BiT_{{\bf k}\chi}(t) \!\!&=&\!\! 
ik_\gamma(\delta_{\chi\alpha}\delta_{\gamma\delta}
-\delta_{\gamma\alpha}\delta_{\chi\delta})
\int\limits_0^t \!\!\sum_{{\bf k'}\atop{{\bf k''}=\,{\bf k}-{\bf k'}}}\!\!\!
\ViT_{{\bf k''}\alpha}(t')\,\BoT_{{\bf k'}\delta}\,dt'.
\label{BiT}
\eeq
Here we also use the fact that the fluid velocity and the magnetic field are 
divergence free, and therefore, $k'_\alpha\BoT_{{\bf k'}\alpha}=0$ and
$k''_\alpha\ViT_{{\bf k''}\alpha}=0$. The complex conjugate of formula~(\ref{BiT}) is
\beq
\BiT^*_{{\bf k}\chi}(t) \!\!&=&\!\! 
-ik_\tau(\delta_{\chi\beta}\delta_{\tau\eta}
-\delta_{\tau\beta}\delta_{\chi\eta})
\int\limits_0^t \!\!\sum_{{\bf k'''}\atop{{\bf k^{iv}}=\,{\bf k}-{\bf k}'''}}\!\!\!
\ViT^*_{{\bf k^{iv}}\beta}(t'')\,\BoT^*_{{\bf k'''}\eta}\,dt''.
\label{BiT*}
\eeq
Using these two equations, we obtain
\beq
\langle|\BiTbf_{\bf k}(t)|^2\rangle &=&
k_\gamma k_\tau (\delta_{\alpha\beta}\delta_{\gamma\delta}\delta_{\tau\eta}
-\delta_{\alpha\eta}\delta_{\beta\tau}\delta_{\gamma\delta}
-\delta_{\alpha\gamma}\delta_{\beta\delta}\delta_{\tau\eta}
+\delta_{\alpha\gamma}\delta_{\beta\tau}\delta_{\delta\eta})
\nonumber\\
&&{}\times \int\limits_0^t\!\int\limits_0^t
\!\!\sum_{{\bf k'}\atop{{\bf k''}=\,{\bf k}-{\bf k'}}}
\sum_{{\bf k'''}\atop{{\bf k^{iv}}=\,{\bf k}-{\bf k}'''}}\!\!\!\!
\BoT_{{\bf k'}\delta}\BoT^*_{{\bf k'''}\eta}
\langle\ViT_{{\bf k''}\alpha}(t')\ViT^*_{{\bf k^{iv}}\beta}(t'')\rangle
\,dt'dt''
\nonumber\\
&=&
k^2\,\k_\gamma\k_\tau (\delta_{\alpha\beta}\delta_{\gamma\delta}\delta_{\tau\eta}
-\delta_{\alpha\eta}\delta_{\beta\tau}\delta_{\gamma\delta}
-\delta_{\alpha\gamma}\delta_{\beta\delta}\delta_{\tau\eta}
+\delta_{\alpha\gamma}\delta_{\beta\tau}\delta_{\delta\eta})
\nonumber\\
&&{}\times \!\sum_{{\bf k'}\atop{{\bf k''}=\,{\bf k}-{\bf k'}}}
\int\limits_0^t\!\int\limits_0^t
\bbo_{\delta\eta}\,|\BoT_{\bf k'}|^2\,
\langle\ViT_{{\bf k''}\alpha}(t')\ViT^*_{{\bf k''}\beta}(t'')\rangle
\,dt'dt''.
\label{BiTS}
\eeq
Here, we use $\langle\ViT_{{\bf k''}\alpha}\ViT^*_{{\bf k^{iv}}\beta}\rangle\propto
\delta_{{\bf k''},{\bf k^{iv}}}$, see equation~(\ref{ViT_ViT_AVERAGED_NEW}), and 
therefore, ${\bf k^{iv}}={\bf k''}$ and ${\bf k'''}={\bf k'}$. We also assume that
$\bbo_{\alpha\beta}={\rm const}$ (our first working hypothesis), and therefore,
$\BoT_{{\bf k'}\delta}\BoT^*_{{\bf k'}\eta}=\bbo_{\delta\eta}|\BoT_{\bf k'}|^2$.
Now, according to definition~(\ref{M_K}) for the magnetic energy spectrum 
$M(t, k)$ and equation~(\ref{BkS_EXPANSION}), in order to obtain the ensemble 
averaged energy spectrum, $\langle M(t, k)\rangle$, we need to integrate 
$\langle|\BiTbf_{\bf k}(t)|^2\rangle$ over all directions of unit vector $\bf\k$. 
Equation~(\ref{BiTS}) has four terms on the right-hand-side. Therefore, we have 
four terms for $\langle|\BiTbf_{\bf k}(t)|^2\rangle$ integrated over $\bf\k$,
\beq
\int k^2\,\langle|\BiTbf_{\bf k}(t)|^2\rangle\,d^2{\bf\k} &=&
{\cal T}-{\cal T}'-{\cal T}''+{\cal T}''',
\label{BiTS_INTEGRATED_FOUR_TERMS}
\eeq
where
\beq
{\cal T} \!\!\!\!&=&\!\!\! k^4 {\left(\frac{L}{2\pi}\right)}^{\!3}\!
\int\limits_0^\infty\!dk'\!\!\int\!k'^2\,|\BoT_{\bf k'}|^2 d^2{\bf\k'}\!
\int\!\mu^2 d^2{\bf\k}\!
\int\limits_0^t\!\!\int\limits_0^t 
\langle\ViT_{{\bf k''}\alpha}(t')\ViT^*_{{\bf k''}\alpha}(t'')\rangle 
\,dt'dt'',
\label{T}
\\
{\cal T}' \!\!\!\!&=&\!\!\! k^4 {\left(\frac{L}{2\pi}\right)}^{\!3}\!
\int\limits_0^\infty\!dk'\!\!\int\!k'^2\,|\BoT_{\bf k'}|^2 d^2{\bf\k'}\!
\int\!\mu\,d^2{\bf\k}\!
\int\limits_0^t\!\!\int\limits_0^t 
\bo_\alpha\langle\ViT_{{\bf k''}\alpha}(t')\ViT^*_{{\bf k''}\beta}(t'')\rangle\k_\beta
\,dt'dt'',
\qquad\;\;
\label{T'}
\\
{\cal T}'' \!\!\!\!&=&\!\!\! {{\cal T}'}^* \!= {\cal T}',
\label{T''}
\\
{\cal T}''' \!\!\!\!&=&\!\!\! k^4 {\left(\frac{L}{2\pi}\right)}^{\!3}\!
\int\limits_0^\infty\!dk'\!\!\int\!k'^2\,|\BoT_{\bf k'}|^2 d^2{\bf\k'}\!
\int\!d^2{\bf\k}\!
\int\limits_0^t\!\!\int\limits_0^t 
\k_\alpha\langle\ViT_{{\bf k''}\alpha}(t')\ViT^*_{{\bf k''}\beta}(t'')\rangle\k_\beta
\,dt'dt'',
\label{T'''}
\\
{\bf k''} \!\!\!\!&=&\!\!\! {\bf k}-{\bf k'}. 
\label{k''}
\eeq
Here, $\mu=({\bf\k}\cdot\bobf)$, see equation~(\ref{MU}), and we replace the summation 
over $\bf k'$ by the integration over $\bf k'$ (see Appendix~\ref{TRANSFORMS}), which, 
in turn, is replaced by the double integration over $k'=|{\bf k'}|$ and over 
${\bf\k'}={\bf k'}/k'$.

[In equations~(\ref{BiTS_INTEGRATED_FOUR_TERMS})--(\ref{T'''}) and below, in order 
to shorten our notations, we use the same notation for a Fourier coefficient, no 
matter whether it is a discrete or a continuous function of $\bf k$. For example, 
the function $\langle|\BiTbf_{\bf k}(t)|^2\rangle$ in equation~(\ref{BiTS}) is a 
discrete function of $\bf k$, while the function $\langle|\BiTbf_{\bf k}(t)|^2\rangle$ 
in equation~(\ref{BiTS_INTEGRATED_FOUR_TERMS}), strictly speaking, is the appropriately 
defined continuous function of $\bf k$, $\langle|\BiTbf(t,{\bf k})|^2\rangle$, see 
Appendix~\ref{TRANSFORMS}. However, it is convenient to use the same notation for 
the both functions.]

\begin{figure}[!p]
\vspace{12.0truecm}
\includegraphics{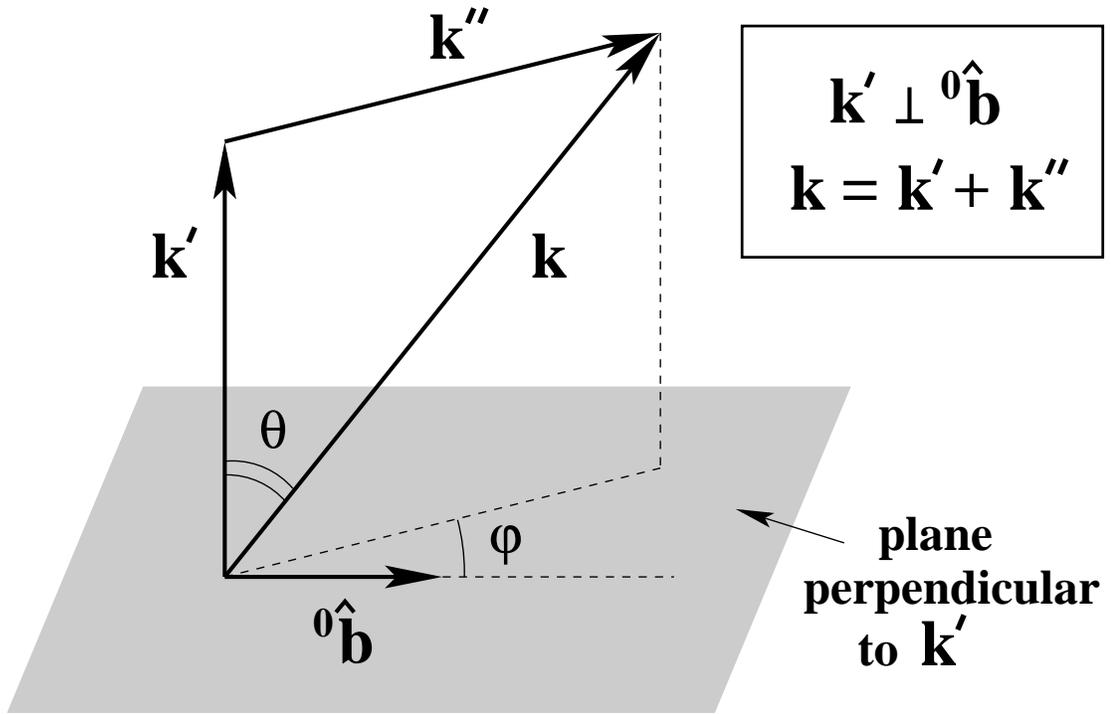}
\caption{
This plot shows relative position of vectors $\bobf$, $\bf k$, $\bf k'$ 
and ${\bf k''}={\bf k}-{\bf k'}$ in space for the mode-coupling 
kernel~(\ref{COUPLING_KERNEL}). The $k''$ modes of the turbulence
interact with the $k'$ modes of the magnetic field to change the 
energy in the $k$ modes of the magnetic field.
(In our case of an initially straight magnetic field vector $\bf k'$ is 
perpendicular to $\bobf$ because the field is divergence free.)
}
\label{FIGURE_k'_perp_bo_and_k}
\end{figure}

Now, refer to Figure~\ref{FIGURE_k'_perp_bo_and_k}. Note that vector $\bf k'$ is 
perpendicular to the zero order magnetic field unit vector $\bobf$ because the 
field is divergence free, $k'_\alpha\BoT_{{\bf k'}\alpha}=0$. We also have
${\bf k''}={\bf k}-{\bf k'}$, see equation~(\ref{k''}). Therefore, the following 
useful equations are valid, see also Figure~\ref{FIGURE_k'_perp_bo_and_k},
\beq
{\bf\k}\cdot{\bf\k'} &=& \cos\theta\,,
\label{kk'}
\\
k'' &=& |{\bf\k''}| = (k^2+k'^2-2kk'\cos\theta)^{1/2},
\label{k''_VALUE}
\\
{\bf\k}\cdot{\bf\k''} &=& \frac{{\bf\k}\cdot{\bf k''}}{k''}
=\frac{{\bf\k}\cdot({\bf k}-{\bf k'})}{k''}
=\frac{k-k'\cos\theta}{k''},
\\
\mu &=& {\bf\k}\cdot\bobf = \sin\theta\,\cos\varphi\,,
\label{MU_RESULT}
\\
\mu'' &=& \bobf\cdot{\bf\k''} =\frac{\bobf\cdot{\bf k''}}{k''}
=\frac{\bobf\cdot{\bf k}}{k''}=\frac{k}{k''}\,\sin\theta\cos\varphi\,,
\label{MU''}
\\
1-\mu''^2 &=& \frac{k''^2-k^2\sin^2\theta\cos^2\varphi}{k''^2}
=\frac{k^2+k'^2-2kk'\cos\theta-k^2\sin^2\theta\cos^2\varphi}{k''^2}
\nonumber\\
{}&=&\frac{k'^2-2kk'\cos\theta+k^2-k^2\sin^2\theta+k^2\sin^2\theta\sin^2\varphi}{k''^2}
\nonumber\\
{}&=& \frac{(k'-k\cos\theta)^2+k^2\sin^2\theta\sin^2\varphi}{k''^2},
\label{ONE_MINUS_MU''S}
\\
d^2{\bf\k} &=& \sin\theta\;d\theta\:d\varphi\,.
\label{dk}
\eeq
Using these formulas, equation~(\ref{ViT_ViT_AVERAGED_NEW}), and 
equation~(\ref{H_F_INTEGRAL_TT}) in the limit $t\gg\tau$, it is straightforward 
to calculate the double time integral terms in equations~(\ref{T})--(\ref{T'''}),
\beq
\int\limits_0^t\!\!\int\limits_0^t 
\langle\ViT_{{\bf k''}\alpha}(t')\ViT^*_{{\bf k''}\alpha}(t'')\rangle
\,dt'dt'' \!\!&=&\!\!
J_{0k''}t\left(1+\frac{{\bar\Omega}''}{\Omega''_{\rm rd}}\right)^{\!2}\!
\left[1+\left(1+\frac{2\Omega''}{\Omega''_{\rm rd}}\right)^{\!-2}\right],
\qquad\quad\!
\label{FIRST_DOUBLE_TIME_INTEGRAL}
\\
\int\limits_0^t\!\!\int\limits_0^t 
\bo_\alpha\langle\ViT_{{\bf k''}\alpha}(t')\ViT^*_{{\bf k''}\beta}(t'')\rangle\k_\beta
\,dt'dt'' \!\!&=&\!\!
J_{0k''}t\left(1+\frac{{\bar\Omega}''}{\Omega_{\rm rd}''}\right)^{\!2}\!
\left(1+\frac{2\Omega''}{\Omega''_{\rm rd}}\right)^{\!-2}
\nonumber\\
{}\!\!&\times&\!\! \frac{k'(k'-k\cos\theta)}{k''^2}\,\sin\theta\cos\varphi
\qquad
\\
\int\limits_0^t\!\!\int\limits_0^t 
\k_\alpha\langle\ViT_{{\bf k''}\alpha}(t')\ViT^*_{{\bf k''}\beta}(t'')\rangle\k_\beta
\,dt'dt'' \!\!&=&\!\!
J_{0k''}t\left(1+\frac{{\bar\Omega}''}{\Omega_{\rm rd}''}\right)^{\!2}\!
\nonumber\\
{}\!\!&\times&\!\! \left[\frac{k'^2\sin^2\theta\sin^2\varphi}
{(k'-k\cos\theta)^2+k^2\sin^2\theta\sin^2\varphi}\right.
\nonumber\\
\!\!&&\!\!{}+\left(1+\frac{2\Omega''}{\Omega''_{\rm rd}}\right)^{\!-2}\frac{k'^2}{k''^2}
\nonumber\\
\!\!&&\!\!{}\times
\left.\frac{(k'-k\cos\theta)^2\sin^2\theta\cos^2\varphi}
{(k'-k\cos\theta)^2+k^2\sin^2\theta\sin^2\varphi}\right].
\label{THIRD_DOUBLE_TIME_INTEGRAL}
\eeq
Here, ${\bar\Omega}''$, $\Omega''$ and $\Omega_{\rm rd}''$ depend on 
$k''$ and on $\mu''^2$, see equations~(\ref{OMEGA_AVERAGED}),~(\ref{OMEGA}) 
and~(\ref{OMEGA_DAMP}). In turn, $k''$ and on $\mu''$ are functions of $k$, $k'$,
$\theta$ and $\varphi$, given by equations~(\ref{k''_VALUE}) and~(\ref{MU''}).

Now, we substitute formulas~(\ref{MU_RESULT}),~(\ref{dk}) 
and~(\ref{FIRST_DOUBLE_TIME_INTEGRAL})--(\ref{THIRD_DOUBLE_TIME_INTEGRAL}) into 
equations~(\ref{T}), (\ref{T'}) and~(\ref{T'''}). The factors that we obtain 
in these equations after integration over $d^2{\bf\k}=\sin\theta\;d\theta\:d\varphi$ 
depend only on $k$ and $k'$, so they can be exchanged with the integrations over 
$d^2{\bf\k'}$. Combining the results together in 
equation~(\ref{BiTS_INTEGRATED_FOUR_TERMS}), we obtain
\beq
\int k^2\,\langle|\BiTbf_{\bf k}(t)|^2\rangle\,d^2{\bf\k} &=&
\int\limits_0^\infty\!dk'\:{\cal K}_1(k,k')\,\int\!k'^2\,|\BoT_{\bf k'}|^2\,d^2{\bf\k'},
\label{BiTS_TERM}
\eeq
where
\beq
{\cal K}_1(k,k') \!\!&=&\!\! t\, k^4 {\left(\frac{L}{2\pi}\right)}^{\!3}\!
\int\limits_0^\pi d\theta\,\sin^3\theta\,J_{0k''} \int\limits_0^{2\pi} d\varphi\, 
\left(1+\frac{{\bar\Omega}''}{\Omega''_{\rm rd}}\right)^{\!2}\!
\nonumber\\
\!\!&&\!\! {}\times \left\{\frac{k'^2+2k(k-k'\cos\theta)\cos^2\varphi}{k''^2}
\:-\: \Bigg[1-\left(1+\frac{2\Omega''}{\Omega''_{\rm rd}}\right)^{\!-2}\Bigg]
\,\frac{k^2}{k''^2} \right.
\nonumber\\
\!\!&&\!\! \qquad{}\times \left.
\frac{(k'-k\cos\theta)^2+(k^2-k'^2)\sin^2\theta\sin^2\varphi}
{(k'-k\cos\theta)^2+k^2\sin^2\theta\sin^2\varphi}\cos^2\varphi
\vphantom{\Bigg[1-\left(1+\frac{2\Omega''}{\Omega''_{\rm rd}}\right)^{\!-2}\Bigg]}
\right\}\!,
\qquad
\eeq
and $k''$ is given by equation~(\ref{k''_VALUE}).

\subsection
[The $\langle\BiiT_{{\bf k}\alpha}(t)\rangle\BoT^*_{{\bf k}\alpha}+{\rm c.c.}$ term]
{The {\boldmath$\langle\BiiT_{{\bf k}\alpha}(t)\rangle\BoT^*_{{\bf k}\alpha}+{\rm c.c.}$} term}
\label{Bii_Bo_COUPLING_TERM}

Next, we calculate the 
$\langle\BiiT_{{\bf k}\alpha}(t)\rangle\BoT^*_{{\bf k}\alpha}+{\rm c.c.}$ term
of expansion equation~(\ref{BkS_EXPANSION}). First, we integrate 
equation~(\ref{B_EVOLUTION_ii}) in time (with the zero initial conditions), and then 
Fourier transform the result in space, ${\bf r}\rightarrow{\bf k}$. We have
\beq
\BiiT_{{\bf k}\eta}(t) = ik_\tau(\delta_{\eta\beta}\delta_{\tau\chi}
-\delta_{\tau\beta}\delta_{\eta\chi})
\int\limits_0^t \!\!\!\sum_{{\bf k'}\atop{{\bf k''}=\,{\bf k}-{\bf k'}}}\!\!\!
\!\left[\ViT_{{\bf k''}\beta}(t')\,\BiT_{{\bf k'}\chi}(t')
+\ViiT_{{\bf k''}\beta}(t')\,\BoT_{{\bf k'}\chi}\right] dt',
\quad
\label{BiiT}
\eeq
where we use the divergence free conditions $k_\alpha\BoT_{{\bf k}\alpha}=0$, 
$k_\alpha\BiT_{{\bf k}\alpha}=0$, $k_\alpha\ViT_{{\bf k}\alpha}=0$ and 
$k_\alpha\ViiT_{{\bf k}\alpha}=0$. 
Second, we ensemble average his equation. The second term in the brackets $[...]$ 
averages out because of equation~(\ref{Vii_AVERAGE_NEW}). Then, we multiply the 
resulting equation by $\BoT^*_{{\bf k}\eta}$, add the complex conjugate, and
use formula~(\ref{BiT}) for $\BiT_{{\bf k'}\chi}$. We have
\beq
\langle\BiiT_{{\bf k}\eta}\rangle\BoT^*_{{\bf k}\eta}+{\rm c.c.} \!\!&=&\!\!
ik_\tau(\delta_{\eta\beta}\delta_{\tau\chi}
-\delta_{\tau\beta}\delta_{\eta\chi})
\int\limits_0^t \!\!\!\sum_{{\bf k'}\atop{{\bf k''}=\,{\bf k}-{\bf k'}}}\!\!\!
\langle\ViT_{{\bf k''}\beta}(t')\,\BiT_{{\bf k'}\chi}(t')\rangle\,
\BoT^*_{{\bf k}\eta}\:dt' + {\rm c.c.}
\nonumber\\
{}\!\!&=&\!\! ik_\tau(\delta_{\eta\beta}\delta_{\tau\chi}
-\delta_{\tau\beta}\delta_{\eta\chi})\,
i(\delta_{\chi\alpha}\delta_{\gamma\delta}
-\delta_{\gamma\alpha}\delta_{\chi\delta})
\nonumber\\
{}\!\!&\times&\!\! \!\!\!\!\!\sum_{{\bf k'}\atop{{\bf k''}=\,{\bf k}-{\bf k'}}}
\sum_{{\bf k'''}\atop{{\bf k^{iv}}=\,{\bf k'}-{\bf k'''}}} \!\!\!\!\! k'_\gamma
\int\limits_0^t\!dt'\!\int\limits_0^{t'}\!dt''\,
\langle\ViT_{{\bf k^{iv}}\alpha}(t'')\,\ViT_{{\bf k''}\beta}(t')\rangle\,
\BoT^*_{{\bf k}\eta}\BoT_{{\bf k'''}\delta} + {\rm c.c.}
\nonumber\\
{}\!\!&=&\!\!{}-k_\tau(\delta_{\alpha\tau}\delta_{\eta\beta}\delta_{\gamma\delta}
-\delta_{\alpha\eta}\delta_{\tau\beta}\delta_{\gamma\delta}
+\delta_{\delta\eta}\delta_{\gamma\alpha}\delta_{\tau\beta})\,
\BoT^*_{{\bf k}\eta}\BoT_{{\bf k}\delta}
\nonumber\\
{}\!\!&\times&\!\!
\!\!\sum_{{\bf k'}\atop{{\bf k''}=\,{\bf k}-{\bf k'}}} \!\! k'_\gamma
\int\limits_0^t\!dt'\!\int\limits_0^{t'}\!dt''\,
\langle\ViT_{-{\bf k''}\alpha}(t'')\,\ViT_{{\bf k''}\beta}(t')\rangle + {\rm c.c.}
\label{BiiT_AVERAGED_FIRST}
\eeq
Here, we use $\langle\ViT_{{\bf k^{iv}}\alpha}\ViT_{{\bf k''}\beta}\rangle\propto
\delta_{{\bf k^{iv}},-{\bf k''}}$, see equation~(\ref{ViT_ViT_AVERAGED_NEW}), and 
therefore, ${\bf k^{iv}}=-{\bf k''}$ and ${\bf k'''}={\bf k}$. We also use
the field divergence free condition, $k_\alpha\BoT_{{\bf k}\alpha}=0$,
this is why we have only three terms at the end. 
Third, we change the summation over ${\bf k'}$ to summation over ${\bf k''}$
in equation~(\ref{BiiT_AVERAGED_FIRST}). We have
\beq
\langle\BiiT_{{\bf k}\eta}\rangle\BoT^*_{{\bf k}\eta}+{\rm c.c.} \!\!&=&\!\!
{}-k_\tau(\delta_{\alpha\tau}\delta_{\eta\beta}\delta_{\gamma\delta}
-\delta_{\alpha\eta}\delta_{\tau\beta}\delta_{\gamma\delta}
+\delta_{\delta\eta}\delta_{\gamma\alpha}\delta_{\tau\beta})\,
\BoT^*_{{\bf k}\eta}\BoT_{{\bf k}\delta}
\nonumber\\
\!\!&&\!\!{}\times \sum_{\bf k''} \,(k_\gamma-k''_\gamma)
\int\limits_0^t\!dt'\!\int\limits_0^{t'}\!dt''\,
\langle\ViT_{-{\bf k''}\alpha}(t'')\,\ViT_{{\bf k''}\beta}(t')\rangle + {\rm c.c.}
\nonumber\\
{}\!\!&=&\!\! {}-2k_\alpha k_\beta\,|\BoTbf_{\bf k}|^2
\sum_{\bf k''} \int\limits_0^t\!dt'\!\int\limits_0^{t'}\!dt''\,
\langle\ViT_{-{\bf k''}\alpha}(t'')\,\ViT_{{\bf k''}\beta}(t')\rangle
\nonumber\\
\!\!&&\!\!{}+\, 2k_\tau(\delta_{\alpha\tau}\delta_{\eta\beta}
-\delta_{\alpha\eta}\delta_{\tau\beta})\,\bbo_{\eta\gamma}\,|\BoT_{\bf k}|^2
\nonumber\\
\!\!&&\!\!\quad{}\times \sum_{\bf k''} \,k''_\gamma 
\int\limits_0^t\!dt'\!\int\limits_0^{t'}\!dt''\,
\langle\ViT_{-{\bf k''}\alpha}(t'')\,\ViT_{{\bf k''}\beta}(t')\rangle
\nonumber\\
{}\!\!&=&\!\!
{}-2k^2\,|\BoTbf_{\bf k}|^2
\sum_{\bf k''} \int\limits_0^t\!dt'\!\int\limits_0^{t'}\!dt''\,
\k_\alpha\langle\ViT_{-{\bf k''}\alpha}(t'')\,\ViT_{{\bf k''}\beta}(t')\rangle\k_\beta
\nonumber\\
\!\!&&\!\!{}+\, 2k\,|\BoT_{\bf k}|^2 \sum_{\bf k''} \bo_\gamma k''_\gamma
\int\limits_0^t\!dt'\!\int\limits_0^{t'}\!dt''\,
\k_\alpha\langle\ViT_{-{\bf k''}\alpha}(t'')\,\ViT_{{\bf k''}\beta}(t')\rangle\bo_\beta
\nonumber\\
\nonumber\\
\!\!&&\!\!{}-\, 2k\,|\BoT_{\bf k}|^2 \sum_{\bf k''} \bo_\gamma k''_\gamma
\int\limits_0^t\!dt'\!\int\limits_0^{t'}\!dt''\,
\bo_\alpha\langle\ViT_{-{\bf k''}\alpha}(t'')\,\ViT_{{\bf k''}\beta}(t')\rangle\k_\beta
\nonumber\\
{}\!\!&=&\!\! 
{}-2k^2|\BoTbf_{\bf k}|^2\,
{\left(\frac{L}{2\pi}\right)}^{\!3}\int\limits_{-\infty}^\infty d^3{\bf k''}
\nonumber\\
\!\!&&\!\!\quad{}\times
\int\limits_0^t\!dt'\int\limits_0^{t'}\!dt''\;
\k_\alpha\langle\ViT_{-{\bf k''}\alpha}(t'')\,\ViT_{{\bf k''}\beta}(t')\rangle\k_\beta\:.
\qquad
\label{BiiT_AVERAGED_SECOND}
\eeq
Here, we again used the divergence free conditions on the field and on the fluid
velocities, and formula
$\BoT_{{\bf k}\eta}\BoT^*_{{\bf k}\delta}=\bbo_{\eta\delta}|\BoT_{\bf k}|^2$ 
($\bbo_{\eta\delta}={\rm const}$, according to our first working hypothesis).
Two terms on the seventh and eighth lines of  equation~(\ref{BiiT_AVERAGED_SECOND})
cancel each other because of the symmetry of tensor 
$\langle\ViT_{-{\bf k''}\alpha}\,\ViT_{{\bf k''}\beta}\rangle$ with respect to 
exchange $\alpha\leftrightarrow\beta$ and ${\bf k''}\leftrightarrow-{\bf k''}$, 
see equation~(\ref{ViT_ViT_AVERAGED_NEW}). We change the summation over ${\bf k''}$
to the integration over ${\bf k''}$ in the last line of 
equation~(\ref{BiiT_AVERAGED_SECOND}).

\begin{figure}[!p]
\vspace{12.0truecm}
\includegraphics{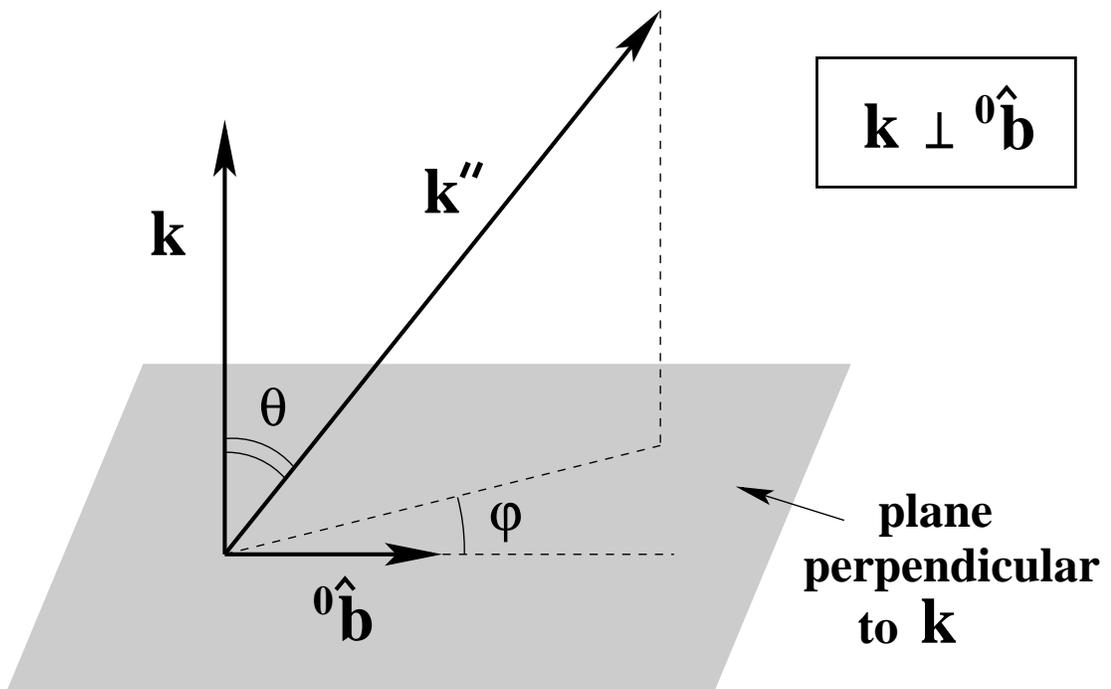}
\caption{
This plot shows relative position of vectors $\bobf$, $\bf k$ 
and ${\bf k''}$ in space for equations~(\ref{BiiT_INTEGRATED})--(\ref{BiiT_TERM}).
}
\label{FIGURE_k_perp_bo_and_k''}
\end{figure}

Now, we integrate 
$\langle\BiiT_{{\bf k}\eta}\rangle\BoT^*_{{\bf k}\eta}+{\rm c.c.}$, given by 
equation~(\ref{BiiT_AVERAGED_SECOND}), over all directions of unit vector $\bf\k$. 
We have
\beq
\int\!k^2\left[\langle\BiiT_{{\bf k}\eta}\rangle\BoT^*_{{\bf k}\eta}+{\rm c.c.}\right]
d^2{\bf\k} \!\!&=&\!\! 
{}- 2k^2 \int k^2\,|\BoTbf_{\bf k}|^2\,d^2{\bf\k}\;\,
{\left(\frac{L}{2\pi}\right)}^{\!3}\!\int\limits_{-\infty}^\infty d^3{\bf k''}
\nonumber\\
\!\!&&\!\!{}\times \int\limits_0^t\!dt'\int\limits_0^{t'}\!dt''\;
\k_\alpha\langle\ViT_{-{\bf k''}\alpha}(t'')\,\ViT_{{\bf k''}\beta}(t')\rangle\k_\beta
\:.\qquad
\label{BiiT_INTEGRATED}
\eeq
Let us refer to Figure~\ref{FIGURE_k_perp_bo_and_k''}. Note that vector $\bf k$ is 
perpendicular to the zero order magnetic field unit vector $\bobf$ because the 
field is divergence free, $k_\alpha\BoT_{{\bf k}\alpha}=0$. We also have
\beq
{\bf\k''}\cdot{\bf\k} &=& \cos\theta,
\label{k''_k}
\\
\mu'' &=& {\bf\k''}\cdot\bobf = \sin\theta\cos\varphi,
\label{MU''_SECOND}
\\
d^3{\bf k''} &=& k''^2\,dk''\,\sin\theta\,d\theta\,d\varphi.
\label{dk''}
\eeq
Using the first two of these equations, equation~(\ref{ViT_ViT_AVERAGED_NEW}), 
and equation~(\ref{H_F_INTEGRAL_T'T}) in the limit $t\gg\tau$, we calculate the 
double time integral in equation~(\ref{BiiT_INTEGRATED}),
\beq
\int\limits_0^t\!dt'\int\limits_0^{t'}\!dt''\;
\k_\alpha\langle\ViT_{-{\bf k''}\alpha}(t'')\,\ViT_{{\bf k''}\beta}(t')\rangle\k_\beta
\!\!&=&\!\!
\frac{1}{2}J_{0k''}t\left(1+\frac{{\bar\Omega}''}{\Omega''_{\rm rd}}\right)^{\!2}
\sin^2\theta
\nonumber\\
\!\!&&\!\! {}\times
\left\{ 
\vphantom{\Bigg[1-\left(1+\frac{2\Omega''}{\Omega''_{\rm rd}}\right)^{\!-2}\Bigg]}
1-\Bigg[1-\left(1+\frac{2\Omega''}{\Omega''_{\rm rd}}\right)^{\!-2}\Bigg] \right.
\nonumber\\
\!\!&&\!\!\qquad{}\times \left.
\vphantom{\Bigg[1-\left(1+\frac{2\Omega''}{\Omega''_{\rm rd}}\right)^{\!-2}\Bigg]}
\frac{\cos^2\theta\cos^2\varphi}{\cos^2\theta\cos^2\varphi+\sin^2\varphi}\right\}.
\qquad
\label{TT'_INTEGRAL_FOR_BiiT}
\eeq
Here, ${\bar\Omega}''$, $\Omega''$ and $\Omega''_{\rm rd}$ depend on $k''$ and on 
$\mu''^2$, see equations~(\ref{MU''_SECOND}),~(\ref{OMEGA_AVERAGED}),~(\ref{OMEGA}) 
and~(\ref{OMEGA_DAMP}). Finally, substituting equation~(\ref{TT'_INTEGRAL_FOR_BiiT}) into 
equation~(\ref{BiiT_INTEGRATED}), and using equation~(\ref{dk''}), we obtain
\beq
\int\!k^2\left[\langle\BiiT_{{\bf k}\eta}\rangle\BoT^*_{{\bf k}\eta}+{\rm c.c.}\right]
d^2{\bf\k} \!\!&=&\!\! 
{}- t\,k^2 \int k^2\,|\BoTbf_{\bf k}|^2\,d^2{\bf\k}
\nonumber\\
\!\!&&\!\!{}\times {\left(\frac{L}{2\pi}\right)}^{\!3}\!
\int\limits_0^\infty k''^2J_{0k''}\,dk'' 
\int\limits_0^\pi \sin^3\theta\:d\theta \int\limits_0^{2\pi} d\varphi\, 
\nonumber\\
\!\!&&\!\!{}\times 
\left(1+\frac{{\bar\Omega}''}{\Omega''_{\rm rd}}\right)^{\!2}
\left\{
\vphantom{\Bigg[1-\left(1+\frac{2\Omega''}{\Omega''_{\rm rd}}\right)^{\!-2}\Bigg]}
1-\Bigg[1-\left(1+\frac{2\Omega''}{\Omega''_{\rm rd}}\right)^{\!-2}\Bigg] \right.
\nonumber\\
\!\!&&\!\!\qquad{}\times \left.
\vphantom{\Bigg[1-\left(1+\frac{2\Omega''}{\Omega''_{\rm rd}}\right)^{\!-2}\Bigg]}
\frac{\cos^2\theta\cos^2\varphi}{\cos^2\theta\cos^2\varphi+\sin^2\varphi}\right\}.
\label{BiiT_TERM}
\eeq

\subsection
[Collecting the terms together]
{Collecting the terms together}
\label{COLLECTING_TERMS}

Next, we substitute equations~(\ref{BkS_EXPANSION}),~(\ref{BiTS_TERM}) 
and~(\ref{BiiT_TERM}) into equation~(\ref{M_K}) for the magnetic
energy spectrum $M(t, k)$. We also make use of equation~(\ref{BiT_AVERAGED}). 
We choose $t$ small enough for the quasi-linear expansion to be valid, so that 
$\partial_t\langle M(t, k)\rangle=[M(t, k)-M(0, k)]/t$. As a result, we finally 
obtain {\it the mode coupling equation} for the magnetic energy spectrum,
\beq
\frac{\partial M}{\partial t} = \int\limits_0^\infty K(k,k')M(t,k')\,dk' -
2\frac{\eta_{\mbox{\tiny$T$}}}{4\pi}\,k^2 M(t,k).
\label{MODE_COUPLING}
\eeq
Here the mode coupling kernel $K(k,k')$ is
\beq
K(k,k') &=& 
k^4 {\left(\frac{L}{2\pi}\right)}^{\!3}
\int\limits_0^\pi d\theta\,\sin^3\theta\,J_{0k''} \int\limits_0^{2\pi} d\varphi\, 
\left(1+\frac{{\bar\Omega}''}{\Omega''_{\rm rd}}\right)^{\!2}
\nonumber\\
&& {}\times \left\{\frac{k'^2+2k(k-k'\cos\theta)\cos^2\varphi}{k''^2}
\:-\: \Bigg[1-\left(1+\frac{2\Omega''}{\Omega''_{\rm rd}}\right)^{\!-2}\Bigg]
\,\frac{k^2}{k''^2} \right.
\nonumber\\
&& \qquad{}\times \left.
\frac{(k'-k\cos\theta)^2+(k^2-k'^2)\sin^2\theta\sin^2\varphi}
{(k'-k\cos\theta)^2+k^2\sin^2\theta\sin^2\varphi}\cos^2\varphi
\vphantom{\Bigg[1-\left(1+\frac{2\Omega''}{\Omega''_{\rm rd}}\right)^{\!-2}\Bigg]}
\right\},
\qquad
\label{COUPLING_KERNEL}
\\
k'' &=& (k^2+k'^2-2kk'\cos\theta)^{1/2},
\\
\frac{{\bar\Omega}''}{\Omega''_{\rm rd}} &=& 6\,\frac{k''^2}{k_\nu^2},
\label{OMEGA_BAR_DAMP_RATIO_FIRST}
\\
\frac{2\Omega''}{\Omega''_{\rm rd}} &=& 
90\,\frac{k^2}{k_\nu^2}\,\frac{(k'-k\cos\theta)^2+k^2\sin^2\theta\sin^2\varphi}{k''^2}
\sin^2\theta\cos^2\varphi,
\label{OMEGA_DAMP_RATIO_FIRST}
\eeq
and the turbulent diffusion constant
\beq
\frac{\eta_{\mbox{\tiny$T$}}}{4\pi} &=&
\frac{1}{2}{\left(\frac{L}{2\pi}\right)}^{\!3}
\int\limits_0^\infty k'''^2J_{0k'''}\,dk''' 
\int\limits_0^\pi \sin^3\theta\:d\theta \int\limits_0^{2\pi} d\varphi\,
\left(1+\frac{{\bar\Omega}'''}{\Omega'''_{\rm rd}}\right)^{\!2}
\nonumber\\
&&\quad{}\times \left\{
1-\Bigg[1-\left(1+\frac{2\Omega'''}{\Omega'''_{\rm rd}}\right)^{\!-2}\Bigg]
\frac{\cos^2\theta\cos^2\varphi}{\cos^2\theta\cos^2\varphi+\sin^2\varphi}
\vphantom{\Bigg[1-\left(1+\frac{2\Omega'''}{\Omega'''_{\rm rd}}\right)^{\!-2}\Bigg]}
\right\},
\label{ETA_T}
\\
\frac{{\bar\Omega}'''}{\Omega'''_{\rm rd}} &=& 6\,\frac{k'''^2}{k_\nu^2},
\label{OMEGA_BAR_DAMP_RATIO_SECOND}
\\
\frac{2\Omega'''}{\Omega'''_{\rm rd}} &=& 
90\,\frac{k'''^2}{k_\nu^2}\,\sin^2\theta\cos^2\varphi\,(1-\sin^2\theta\cos^2\varphi).
\label{OMEGA_DAMP_RATIO_SECOND}
\eeq
The function $J_{0k}$ is given by equation~(\ref{J_ZERO_K}).
To derive the above formulas, we use equations~(\ref{OMEGA_AVERAGED}), (\ref{OMEGA}) 
and~(\ref{OMEGA_DAMP}). To obtain equation~(\ref{OMEGA_DAMP_RATIO_FIRST}) we use 
equations~(\ref{MU''}) and~(\ref{ONE_MINUS_MU''S}). To obtain 
equation~(\ref{OMEGA_DAMP_RATIO_SECOND}), we use equation~(\ref{MU''_SECOND}). In 
the last three equations for the turbulent diffusion constant we replace $k''$ by 
$k'''$ in order to distinguish it from $k''$ in the equations for the coupling 
kernel.

If we consider the kinematic turbulent dynamo, then we need to take the limit 
${\bar\Omega}\rightarrow 0$, $\Omega\rightarrow 0$, or alternatively, 
the limit $\Omega_{\rm rd}\rightarrow \infty$ (see the footnote~\ref{KA_LIMIT_FOOTNOTE}
on page~\pageref{KA_LIMIT_FOOTNOTE}). In this case, as one might expect, after 
integrating over $\varphi$, equation~(\ref{COUPLING_KERNEL}) reduces to 
equation~(\ref{COUPLING_KERNEL_KA}), and after integrating over both $\theta$ and 
$\varphi$, equation~(\ref{ETA_T}) reduces to equation~(\ref{ETA_T_KA}).

\section
[The Magnetic Energy Spectrum on Subviscous Scales]
{The Magnetic Energy Spectrum\\ 
on Subviscous Scales}
\label{SMALL_SCALES}

Equations~(\ref{MODE_COUPLING})--(\ref{OMEGA_DAMP_RATIO_SECOND}), which give the 
evolution of the magnetic energy spectrum in the magnetized turbulent dynamo theory, 
are the principal result of this thesis. However, these equations are rather 
complicated for easy interpretation. In this section we limit ourselves to 
the evolution of the magnetic energy spectrum on small subviscous scales. 
In this limit, $k\gg k_\nu$, and the integro-differential equation for the 
magnetic spectrum evolution simplifies to an ordinary differential equation. At the 
same time, the evolution of the magnetic energy spectrum on subviscous scales 
is of great interest for the origin of cosmic magnetic fields (see 
Chapter~\ref{DISCUSSION}).

\begin{figure}[!p]
\vspace{12.0truecm}
\includegraphics{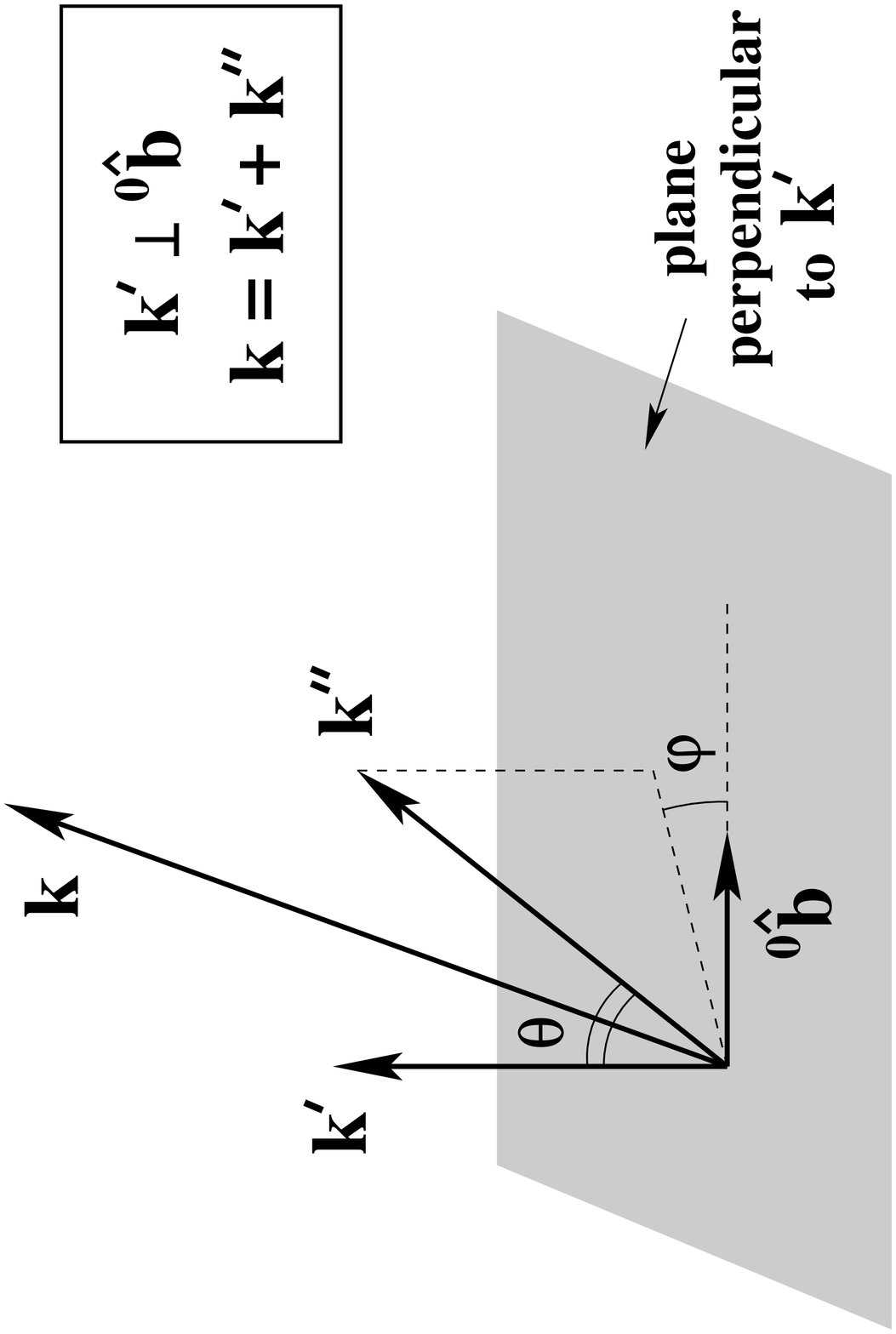}
\caption{
This plot shows relative position of vectors $\bobf$, $\bf k'$, $\bf k''$ 
and ${\bf k}={\bf k'}+{\bf k''}$ in space for 
equations~(\ref{T_d})--(\ref{k_k'_k''}).
}
\label{FIGURE_k'_perp_bo_and_k''}
\end{figure}

Let us refer to equation~(\ref{COUPLING_KERNEL}) for the mode coupling kernel 
$K(k,k')$. The function $J_{0k''}$ cuts off at the viscous wave number $k_\nu$
[see equation~(\ref{J_ZERO_K})].
Therefore, in the large-$k$ limit, $k\gg k_\nu$, we have $k''\sim|k-k'|\ll k,k'$, 
and can expand the kernel $K(k,k')$. However, the simplest way of calculations is 
to introduce an arbitrary function of $k$, $F(k)$, which varies slowly in the 
region $k\gg k_\nu$, and vanishes outside of this region~\cite{KA_92}.
To derive the mode coupling equation on small (subviscous) scales, we calculate 
the following integral
\beq
\int\limits_0^\infty\!F(k){\left.\frac{\partial M}{\partial t}\right|}_{t=0} dk 
\!\!&=&\!\!
\frac{1}{4\pi\rho}{\left(\frac{L}{2\pi}\right)}^{\!3}\!\!\int\limits_{-\infty}^\infty\!
F(k) 
{\left.\frac{\partial\langle|{\bf{\tilde B}}_{\bf k}|^2\rangle}{\partial t}\right|}_{t=0}
\,d^3{\bf k}
\nonumber\\
{}\!\!&=&\!\! 
\frac{1}{4\pi\rho}{\left(\frac{L}{2\pi}\right)}^{\!3}\!\!\int\limits_{-\infty}^\infty\!
F(k) \frac{\langle|{\bf{\tilde B}}_{\bf k}(t)|^2\rangle
-|{\bf{\tilde B}}_{\bf k}(0)|^2}{t}\,d^3{\bf k}
\nonumber\\
{}\!\!&=&\!\! 
\frac{1}{4\pi\rho}{\left(\frac{L}{2\pi}\right)}^{\!3}\:\frac{1}{t} \left\{\;
\int\limits_{-\infty}^\infty\! 
F(k)\,\langle|\BiTbf_{\bf k}(t)|^2\rangle\,d^3{\bf k} \right.
\nonumber\\
\!\!&&\!\! \qquad\left.{}+\int\limits_{-\infty}^\infty\!
F(k)\,\Big[\langle\BiiT_{{\bf k}\alpha}(t)\rangle\BoT^*_{{\bf k}\alpha}+{\rm c.c.}\Big]
\,d^3{\bf k} \,\right\}.
\label{F_M_INTEGRAL}
\eeq
To obtain the first line of this equation, we use equation~(\ref{M_K}). To obtain the 
second line, we replace the time derivative by the time finite difference, assuming 
that $t$ is small, and our quasi-linear expansion is valid. To obtain the final result
in equation~(\ref{F_M_INTEGRAL}) [the third and the fourth lines], we use 
equations~(\ref{BkS_EXPANSION}) and~(\ref{BiT_AVERAGED}). 
Next, we use equations~(\ref{BiTS_INTEGRATED_FOUR_TERMS})--(\ref{k''})
and equation~(\ref{BiiT_INTEGRATED}) to obtain the term in the brackets $\{...\}$
in equation~(\ref{F_M_INTEGRAL}),
\beq
&&\int\limits_{-\infty}^\infty\! F(k)\,\langle|\BiTbf_{\bf k}(t)|^2\rangle\,d^3{\bf k}
+\int\limits_{-\infty}^\infty\! F(k)\,
\Big[\langle\BiiT_{{\bf k}\alpha}(t)\rangle\BoT^*_{{\bf k}\alpha}+{\rm c.c.}\Big]
\,d^3{\bf k}
\nonumber\\
&&\quad{} = \int\limits_0^\infty\!F(k)\, dk 
\int\! k^2\,\langle|\BiTbf_{\bf k}(t)|^2\rangle\,d^2{\bf\k}
+\int\limits_0^\infty\!F(k)\, dk
\int\! k^2\,
\Big[\langle\BiiT_{{\bf k}\alpha}(t)\rangle\BoT^*_{{\bf k}\alpha}+{\rm c.c.}\Big]
\,d^2{\bf\k}
\quad
\nonumber\\
&&\quad{} = {\left(\frac{L}{2\pi}\right)}^{\!3}
\Big[{\cal T}_\diamond-{\cal T}'_\diamond-{\cal T}''_\diamond+{\cal T}'''_\diamond
+{\cal T}^{\rm iv}_\diamond\Big],
\qquad
\label{F_B_INTEGRAL}
\eeq
where
\beq
{\cal T}_\diamond \!\!&=&\!\!
\int\limits_{-\infty}^\infty\int\limits_{-\infty}^\infty
\mu^2 k^2 F(k)\,|\BoT_{\bf k'}|^2 \,d^3{\bf k'}d^3{\bf k}
\int\limits_0^t\!\!\int\limits_0^t 
\langle\ViT_{{\bf k''}\alpha}(t')\ViT^*_{{\bf k''}\alpha}(t'')\rangle
\,dt' dt''
\nonumber\\
{}\!\!&=&\!\!
\int\limits_{-\infty}^\infty \!\!|\BoT_{\bf k'}|^2 d^3{\bf k'} \!
\int\limits_{-\infty}^\infty \!\!\mu''^2 k''^2 F(k)\,d^3{\bf k''} 
\int\limits_0^t\!\!\int\limits_0^t 
\langle\ViT_{{\bf k''}\alpha}(t')\ViT^*_{{\bf k''}\alpha}(t'')\rangle
\,dt' dt'',
\quad
\label{T_d}
\\
{\cal T}'_\diamond  \!\!&=&\!\!
\int\limits_{-\infty}^\infty\int\limits_{-\infty}^\infty
\mu k F(k)\,|\BoT_{\bf k'}|^2 \,d^3{\bf k'}d^3{\bf k} 
\int\limits_0^t\!\!\int\limits_0^t 
\bo_\alpha\langle\ViT_{{\bf k''}\alpha}(t')\ViT^*_{{\bf k''}\beta}(t'')\rangle k_\beta
\,dt' dt''
\nonumber\\
{}\!\!&=&\!\!
\int\limits_{-\infty}^\infty \!\!|\BoT_{\bf k'}|^2 d^3{\bf k'} \!
\int\limits_{-\infty}^\infty \!\!\mu''k'' F(k)\, d^3{\bf k''}
\int\limits_0^t\!\!\int\limits_0^t 
\bo_\alpha\langle\ViT_{{\bf k''}\alpha}(t')\ViT^*_{{\bf k''}\beta}(t'')\rangle k'_\beta
\,dt' dt'',
\qquad
\label{T'_d}
\\
{\cal T}''_\diamond  \!\!&=&\!\!  
{\cal T}'_\diamond,
\label{T''_d}
\\
{\cal T}'''_\diamond  \!\!&=&\!\!
\int\limits_{-\infty}^\infty\int\limits_{-\infty}^\infty
F(k)\,|\BoT_{\bf k'}|^2 \,d^3{\bf k'}d^3{\bf k}
\int\limits_0^t\!\!\int\limits_0^t 
k_\alpha\langle\ViT_{{\bf k''}\alpha}(t')\ViT^*_{{\bf k''}\beta}(t'')\rangle k_\beta
\,dt' dt''
\nonumber\\
{}\!\!&=&\!\!
\int\limits_{-\infty}^\infty \!\!|\BoT_{\bf k'}|^2 d^3{\bf k'} \!
\int\limits_{-\infty}^\infty \!\! F(k)\,d^3{\bf k''}
\int\limits_0^t\!\!\int\limits_0^t 
k'_\alpha\langle\ViT_{{\bf k''}\alpha}(t')\ViT^*_{{\bf k''}\beta}(t'')\rangle k'_\beta
\,dt' dt'',
\quad
\label{T'''_d}
\\
{\bf k} \!\!&=&\!\! {\bf k'}+{\bf k''},
\label{k_k'_k''}
\eeq
and
\beq
{\cal T}^{\rm iv}_\diamond \!\!&=&\!\!
{}-2 \!\int\limits_{-\infty}^\infty\!\!
F(k)\,|\BoT_{\bf k}|^2 \,d^3{\bf k}
\!\int\limits_{-\infty}^\infty\!d^3{\bf k''}
\int\limits_0^t\!dt'\!\int\limits_0^{t'}\!dt''\;
k_\alpha\langle\ViT_{-{\bf k''}\alpha}(t'')\,\ViT_{{\bf k''}\beta}(t')\rangle k_\beta
\nonumber\\
{}\!\!&=&\!\!
{}- \int\limits_{-\infty}^\infty\!\!
F(k)\,|\BoT_{\bf k}|^2 \,d^3{\bf k}
\!\int\limits_{-\infty}^\infty\!d^3{\bf k''}
\int\limits_0^t\!\!\int\limits_0^t 
k_\alpha\langle\ViT_{-{\bf k''}\alpha}(t'')\,\ViT_{{\bf k''}\beta}(t')\rangle k_\beta
\,dt' dt''.
\qquad
\label{Tiv_d}
\eeq
Here, to obtain the final results in equations~(\ref{T_d})--(\ref{T'''_d}), we change 
the integration over $\bf k$ in these equations to integration over $\bf k''$ by
making use of formula~(\ref{k_k'_k''}). We use the divergence free condition on 
the fluid velocities, $k''_\alpha\ViT_{{\bf k''}\alpha}$, and therefore,
$k_\alpha\ViT_{{\bf k''}\alpha}=k'_\alpha\ViT_{{\bf k''}\alpha}$. We also use
the obvious formula
\beq
\mu k=\bobf\cdot{\bf k}=\bobf\cdot{\bf k''}=k''(\bobf\cdot{\bf\k''})
=\mu''k''=k''\sin\theta\cos\varphi,
\label{k''_mu''}
\eeq
see Figure~\ref{FIGURE_k'_perp_bo_and_k''}. To obtain the final result in 
equation~(\ref{Tiv_d}), we make use of equation~(\ref{ViT_ViT_AVERAGED_NEW}) and of 
the first equality in equation~(\ref{H_F_INTEGRAL_TT}).

We calculate ${\cal T}_\diamond$, ${\cal T}'_\diamond$, ${\cal T}'''_\diamond$
and ${\cal T}^{\rm iv}_\diamond$ separately, up to the second order in $k''\ll k,k'$.

First, following Kulsrud and Anderson~\cite{KA_92}, in 
equations~(\ref{T_d})--(\ref{T'''_d}) we expand the slowly varying function $F(k)$ 
in $k''\ll k$ at point $k'$ up to the second order. We have, see 
Figure~\ref{FIGURE_k'_perp_bo_and_k''},
\beq
k \!\!&=&\!\! 
{\Big[k'^2+k''^2+2({\bf k'}\cdot{\bf k''})\Big]}^{1/2}
= k'{\left[1+\frac{2({\bf k'}\cdot{\bf k''})}{k'^2}+\frac{k''^2}{k'^2}\right]}^{1/2}
\nonumber\\
{}\!\!&=&\!\! k'+\frac{({\bf k'}\cdot{\bf k''})}{k'}
+\frac{1}{2}\left[\frac{k''^2}{k'}-\frac{({\bf k'}\cdot{\bf k''})^2}{k'^3}\right]
= k'+k''\cos\theta+\frac{k''^2}{2k'}\sin^2\theta\,,
\qquad
\label{K_EXPANSION}
\\
F(k) \!\!&=&\!\! 
F(k')+\frac{dF}{dk'}(k-k')+\frac{1}{2}\,\frac{d^2F}{dk'^2}(k-k')^2
\nonumber\\
{}\!\!&=&\!\! F(k') + \frac{dF}{dk'}k''\cos\theta 
+\frac{1}{2k'}\frac{dF}{dk'}k''^2\sin^2\theta
+\frac{1}{2}\frac{d^2F}{dk'^2}k''^2\cos^2\theta,
\label{F_EXPANSION}
\eeq

Second, we calculate ${\cal T}_\diamond$, given by equation~(\ref{T_d}).
Because $\mu''^2k''^2$ is of the second order in $k''$, we need to keep only the zero 
order term in expansion~(\ref{F_EXPANSION}) for $F(k)$. Thus, we have
\beq
{\cal T}_\diamond \!\!&=&\!\!
\int\limits_{-\infty}^\infty \!F(k')\,|\BoT_{\bf k'}|^2 \,d^3{\bf k'}
\int\limits_{-\infty}^\infty \!\mu''^2k''^2 \,d^3{\bf k''}
\int\limits_0^t\!\!\int\limits_0^t 
\langle\ViT_{{\bf k''}\alpha}(t')\ViT^*_{{\bf k''}\alpha}(t'')\rangle
\,dt' dt''
\nonumber\\
\!\!&=&\!\!
t \int\limits_0^\infty \!F(k')\,dk' \int\! k'^2\,|\BoT_{\bf k'}|^2 \,d^2{\bf\k'}
\int\limits_0^\infty k''^4 J_{0k''}\,dk''
\int\limits_0^\pi\! d\theta\, \sin^3\theta
\nonumber\\
\!\!&&\!\!{}\times
\int\limits_0^{2\pi}\! d\varphi\, \cos^2\varphi
\left(1+\frac{{\bar\Omega}''}{\Omega''_{\rm rd}}\right)^{\!2}
\Bigg[1+\left(1+\frac{2\Omega''}{\Omega''_{\rm rd}}\right)^{\!-2}\Bigg],
\label{T_d_RESULT}
\eeq
where we use 
\beq
d^3{\bf k''} &=& k''^2\,dk''\,\sin\theta\,d\theta\,d\varphi,
\label{dk''_FORMULA}
\eeq
and equations~(\ref{FIRST_DOUBLE_TIME_INTEGRAL}) and~(\ref{k''_mu''}), see 
Figure~\ref{FIGURE_k'_perp_bo_and_k''}. Here and below in this section, functions 
${\bar\Omega}''$, $\Omega''$ and $\Omega_{\rm rd}''$ depend on $k''$ and on 
$\mu''^2=\sin^2\theta\cos^2\varphi$, see equations~(\ref{OMEGA_AVERAGED}), (\ref{OMEGA}) 
and~(\ref{OMEGA_DAMP}).

Third, we calculate ${\cal T}'_\diamond$, given by equation~(\ref{T'_d}). 
Because $\mu''k''$ is of the first order in $k''$, we need to keep only the zero and
the first order terms in expansion~(\ref{F_EXPANSION}) for $F(k)$. We have
\beq
{\cal T}'_\diamond \!\!&=&\!\!
\int\limits_{-\infty}^\infty \!\!F(k')
|\BoT_{\bf k'}|^2 d^3{\bf k'} \!
\int\limits_{-\infty}^\infty \!\!\mu''k'' d^3{\bf k''}
\int\limits_0^t\!\!\int\limits_0^t 
\bo_\alpha\langle\ViT_{{\bf k''}\alpha}(t')\ViT^*_{{\bf k''}\beta}(t'')\rangle k'_\beta
\,dt' dt''
\nonumber\\
{}\!\!&+&\!\!
\int\limits_{-\infty}^\infty \!\frac{dF}{dk'}|\BoT_{\bf k'}|^2 d^3{\bf k'}\!\!
\int\limits_{-\infty}^\infty \!\!\!\mu''k''^2 \cos\theta\, d^3{\bf k''}\!\!
\int\limits_0^t\!\!\int\limits_0^t \!
\bo_\alpha\langle\ViT_{{\bf k''}\alpha}(t')\ViT^*_{{\bf k''}\beta}(t'')\rangle k'_\beta
\,dt' dt''.
\qquad
\label{T'_d_SECOND}
\eeq
Using equations~(\ref{ViT_ViT_AVERAGED_NEW}),~(\ref{H_F_INTEGRAL_TT}),~(\ref{k''_mu''})
and Figure~\ref{FIGURE_k'_perp_bo_and_k''}, we obtain
\beq
&&\int\limits_0^t\!\!\int\limits_0^t 
\bo_\alpha\langle\ViT_{{\bf k''}\alpha}(t')\ViT^*_{{\bf k''}\beta}(t'')\rangle k'_\beta
\,dt' dt'' = {}
\nonumber\\
&&{}=-tk'J_{0k''}\sin\theta\cos\theta\cos\varphi\,
\left(1+\frac{{\bar\Omega}''}{\Omega''_{\rm rd}}\right)^{\!2}
\left(1+\frac{2\Omega''}{\Omega''_{\rm rd}}\right)^{\!-2}.
\eeq
Now, we substitute this equation and equation~(\ref{dk''_FORMULA}) into 
equation~(\ref{T'_d_SECOND}). The first term in equation~(\ref{T'_d_SECOND}) vanishes 
after the integration over $\theta$ because the integrand is an odd function of 
$\cos\theta$.~\footnote{
This is because the Braginskii viscosity is invariant under reflection 
${\bf\b}\rightarrow{}-{\bf\b}$. See also the footnote~\ref{BRAGINSKII_IS_EVEN_IN_b} 
on page~\pageref{BRAGINSKII_IS_EVEN_IN_b}.
}
The second term is nonzero. As a result, we have
\beq
{\cal T}'_\diamond \!\!&=&\!\!
-t \int\limits_0^\infty k'\,\frac{dF}{dk'}\,dk'\int\!k'^2\,|\BoT_{\bf k'}|^2\,d^2{\bf\k'}
\int\limits_0^\infty k''^4J_{0k''}\,dk''
\int\limits_0^\pi d\theta\, \sin^3\theta\cos^2\theta
\nonumber\\
\!\!&&\!\!\quad{}\times
\int\limits_0^{2\pi} d\varphi\, \cos^2\varphi
\left(1+\frac{{\bar\Omega}''}{\Omega''_{\rm rd}}\right)^{\!2}
\left(1+\frac{2\Omega''}{\Omega''_{\rm rd}}\right)^{\!-2}.
\label{T'_d_RESULT}
\eeq

Fourth, we calculate ${\cal T}'''_\diamond$, given by equation~(\ref{T'''_d}). 
We need to keep all terms in expansion~(\ref{F_EXPANSION}) for $F(k)$. We have
\beq
{\cal T}'''_\diamond  \!\!&=&\!\!
\int\limits_{-\infty}^\infty \!\!F(k')
|\BoT_{\bf k'}|^2 d^3{\bf k'} \!
\int\limits_{-\infty}^\infty \!\! d^3{\bf k''}
\int\limits_0^t\!\!\int\limits_0^t 
k'_\alpha\langle\ViT_{{\bf k''}\alpha}(t')\ViT^*_{{\bf k''}\beta}(t'')\rangle k'_\beta
\,dt' dt''
\nonumber\\
{}\!\!&+&\!\!
\int\limits_{-\infty}^\infty \frac{dF}{dk'}
|\BoT_{\bf k'}|^2 d^3{\bf k'} \!
\int\limits_{-\infty}^\infty \!\! k''\cos\theta\,d^3{\bf k''}
\int\limits_0^t\!\!\int\limits_0^t 
k'_\alpha\langle\ViT_{{\bf k''}\alpha}(t')\ViT^*_{{\bf k''}\beta}(t'')\rangle k'_\beta
\,dt' dt''
\nonumber\\
{}\!\!&+&\!\!
\int\limits_{-\infty}^\infty \frac{1}{2k'}\frac{dF}{dk'}
|\BoT_{\bf k'}|^2 d^3{\bf k'} \!
\int\limits_{-\infty}^\infty \!\! k''^2\sin^2\theta\,d^3{\bf k''}
\int\limits_0^t\!\!\int\limits_0^t 
k'_\alpha\langle\ViT_{{\bf k''}\alpha}(t')\ViT^*_{{\bf k''}\beta}(t'')\rangle k'_\beta
\,dt' dt''
\nonumber\\
{}\!\!&+&\!\!\!
\int\limits_{-\infty}^\infty \!\frac{1}{2}\frac{d^2F}{dk'^2}
|\BoT_{\bf k'}|^2 d^3{\bf k'} \!\!
\int\limits_{-\infty}^\infty \!\! k''^2\cos^2\theta\,d^3{\bf k''}\!
\int\limits_0^t\!\!\int\limits_0^t \!
k'_\alpha\langle\ViT_{{\bf k''}\alpha}(t')\ViT^*_{{\bf k''}\beta}(t'')\rangle k'_\beta
\,dt' dt''\!.
\qquad\quad
\label{T'''_d_SECOND}
\eeq
The first term in this equation is equal to minus ${\cal T}^{\rm iv}_\diamond$, 
which is given by equation~(\ref{Tiv_d}), [to see this, we simply change the notation 
in equation~(\ref{Tiv_d}), ${\bf k}\rightarrow{\bf k'}\,$]. To calculate the last three
terms in equation~(\ref{T'''_d_SECOND}), we again use
equations~(\ref{ViT_ViT_AVERAGED_NEW}),~(\ref{H_F_INTEGRAL_TT}),~(\ref{k''_mu''})
and Figure~\ref{FIGURE_k'_perp_bo_and_k''} to obtain
\beq
\!\!\!\!&&\!\!\!\int\limits_0^t\!\!\int\limits_0^t \!
k'_\alpha\langle\ViT_{{\bf k''}\alpha}(t')\ViT^*_{{\bf k''}\beta}(t'')\rangle k'_\beta
\,dt' dt''
\nonumber\\
\!\!\!\!\!&&\!\!\!{}=tk'^2 J_{0k''}\sin^2\theta\!
\left(1+\frac{{\bar\Omega}''}{\Omega''_{\rm rd}}\right)^{\!2}\!
\left\{ 
1-\Bigg[1-\left(1+\frac{2\Omega''}{\Omega''_{\rm rd}}\right)^{\!-2}\Bigg] 
\frac{\cos^2\theta\cos^2\varphi}{\cos^2\theta\cos^2\varphi+\sin^2\varphi}
\right\}\!.
\qquad\quad
\eeq
[Compare this equation with equation~(\ref{TT'_INTEGRAL_FOR_BiiT}) and 
Figure~\ref{FIGURE_k'_perp_bo_and_k''} with Figure~\ref{FIGURE_k_perp_bo_and_k''}]. Now, 
we substitute this equation and equation~(\ref{dk''_FORMULA}) into the last three terms 
of equation~(\ref{T'''_d_SECOND}). The second term of equation~(\ref{T'''_d_SECOND})
vanishes after the integration over $\theta$ because the integrand is an odd 
function of $\cos\theta$. The last two terms of equation~(\ref{T'''_d_SECOND}) 
are nonzero. As a result, we have
\beq
{\cal T}'''_\diamond  \!\!&=&\!\! {}-{\cal T}^{\rm iv}_\diamond
\nonumber\\
\!\!&&\!\! {}+\,
\frac{t}{2}\int\limits_0^\infty\! k'\,\frac{dF}{dk'}\,dk'
\int\! k'^2\, |\BoT_{\bf k'}|^2 d^2{\bf\k'}\!
\int\limits_0^\infty k''^4J_{0k''}\,dk''
\int\limits_0^\pi d\theta\, \sin^5\theta
\nonumber\\
\!\!&&\!\! \;{}\times
\int\limits_0^{2\pi} d\varphi\, 
\left(1+\frac{{\bar\Omega}''}{\Omega''_{\rm rd}}\right)^{\!2}\!
\left\{ 
1-\Bigg[1-\left(1+\frac{2\Omega''}{\Omega''_{\rm rd}}\right)^{\!-2}\Bigg] 
\frac{\cos^2\theta\cos^2\varphi}{\cos^2\theta\cos^2\varphi+\sin^2\varphi}
\right\}
\nonumber\\
\!\!&&\!\! {}+\,
\frac{t}{2}\int\limits_0^\infty\! k'^2\,\frac{d^2F}{dk'^2}\,dk'
\int\! k'^2\,|\BoT_{\bf k'}|^2 d^2{\bf\k'}\!
\int\limits_0^\infty k''^4J_{0k''}\,dk''
\int\limits_0^\pi d\theta\, \sin^3\theta\cos^2\theta
\nonumber\\
\!\!&&\!\! \;{}\times
\int\limits_0^{2\pi} d\varphi\, 
\left(1+\frac{{\bar\Omega}''}{\Omega''_{\rm rd}}\right)^{\!2}\!
\left\{ 
1-\Bigg[1-\left(1+\frac{2\Omega''}{\Omega''_{\rm rd}}\right)^{\!-2}\Bigg] 
\frac{\cos^2\theta\cos^2\varphi}{\cos^2\theta\cos^2\varphi+\sin^2\varphi}
\right\}\!.
\qquad\quad
\label{T'''_d_RESULT}
\eeq

Now, we substitute equations~(\ref{T''_d}),~(\ref{T_d_RESULT}),~(\ref{T'_d_RESULT})
and (\ref{T'''_d_RESULT}) into equation~(\ref{F_B_INTEGRAL}). The first term of 
equation~(\ref{T'''_d_RESULT}) cancels the ${\cal T}^{\rm iv}_\diamond$ term
in equation~(\ref{F_B_INTEGRAL}). Then, we substitute the result that we get in 
equation~(\ref{F_B_INTEGRAL}) into formula~(\ref{F_M_INTEGRAL}), and using
equation~(\ref{M_K}) for $M(0,k')$, we obtain
\beq
\int\limits_0^\infty\!F(k){\left.\frac{\partial M}{\partial t}\right|}_{t=0} dk 
\!\!&=&\!\!
{\left(\frac{L}{2\pi}\right)}^{\!3}\!
\int\limits_0^\infty \left[
\lambda_0F(k')+\lambda_1k'\frac{dF}{dk'} +\lambda_2k'^2\frac{d^2F}{dk'^2}
\right] M(0,k')\,dk',
\qquad\quad
\label{INTEGRAL_SMALL_SCALES_FIRST}
\eeq
where
\beq
\lambda_0 \!\!&=&\!\!
\int\limits_0^\infty k''^4 J_{0k''}\,dk''
\int\limits_0^\pi\! d\theta\, \sin^3\theta
\int\limits_0^{2\pi}\! d\varphi\, \cos^2\varphi
\left(1+\frac{{\bar\Omega}''}{\Omega''_{\rm rd}}\right)^{\!2}
\Bigg[1+\left(1+\frac{2\Omega''}{\Omega''_{\rm rd}}\right)^{\!-2}\Bigg],
\qquad
\label{lambda_0}
\\
\lambda_1 \!\!&=&\!\!
\int\limits_0^\infty k''^4 J_{0k''}\,dk''
\int\limits_0^\pi\! d\theta\, \sin^3\theta
\int\limits_0^{2\pi}\! d\varphi 
\left(1+\frac{{\bar\Omega}''}{\Omega''_{\rm rd}}\right)^{\!2}
\left\{ 2\cos^2\theta\cos^2\varphi+\frac{1}{2}\sin^2\theta
\vphantom{\Bigg[1-\left(1+\frac{2\Omega''}{\Omega''_{\rm rd}}\right)^{\!-2}\Bigg]}
\right.
\nonumber\\
\!\!&&\!\!\left.
{}-\Bigg[1-\left(1+\frac{2\Omega''}{\Omega''_{\rm rd}}\right)^{\!-2}\Bigg]
\left(2+\frac{1}{2}\,\frac{\sin^2\theta}{\cos^2\theta\cos^2\varphi+\sin^2\varphi}\right)
\cos^2\theta\cos^2\varphi\right\}\!,
\label{lambda_1}
\\
\lambda_2 \!\!&=&\!\!
\frac{1}{2}\int\limits_0^\infty k''^4J_{0k''}\,dk''
\int\limits_0^\pi d\theta\, \sin^3\theta\cos^2\theta
\nonumber\\
\!\!&&\!\! {}\times
\int\limits_0^{2\pi} d\varphi\, 
\left(1+\frac{{\bar\Omega}''}{\Omega''_{\rm rd}}\right)^{\!2}\!
\left\{ 
1-\Bigg[1-\left(1+\frac{2\Omega''}{\Omega''_{\rm rd}}\right)^{\!-2}\Bigg] 
\frac{\cos^2\theta\cos^2\varphi}{\cos^2\theta\cos^2\varphi+\sin^2\varphi}
\right\}\!.
\quad
\label{lambda_2}
\eeq
Next, we integrate the right-hand-side of equation~(\ref{INTEGRAL_SMALL_SCALES_FIRST}) 
by parts over some extent in $k'$ and choose $F(k')$, so that it and its derivative 
$dF/dk'$ vanish at the end points. We have
\beq
\int\limits_0^\infty\!F\frac{\partial M}{\partial t}dk =\!
{\left(\frac{L}{2\pi}\right)}^{\!3}\!
\int\limits_0^\infty \!\Big[
(\lambda_0-\lambda_1+2\lambda_2)M+(4\lambda_2-\lambda_1)k\frac{\partial M}{\partial k} 
+\lambda_2k^2\frac{\partial^2M}{\partial k^2}\Big] Fdk.
\quad\;
\eeq
This equation is valid for an arbitrary function $F$. As a result, the integrands
on the left- and right-hand-side of this equation should be equal, and we finally 
obtain the mode-coupling equation for the magnetic energy spectrum $M(t,k)$ on small 
(subviscous) scales
\beq
\frac{\partial M}{\partial t} = 
\frac{\Gamma}{5} \left[k^2\frac{\partial^2M}{\partial k^2}
-(\Lambda_1-1)k\frac{\partial M}{\partial k} + \Lambda_0 M\right],
\label{SMALL_SCALES_MODE_COUPLING}
\eeq
where
\beq
\Gamma \!\!&=&\!\!
\frac{5}{2}{\left(\frac{L}{2\pi}\right)}^{\!3}
\int\limits_0^\infty k^4J_{0k}\,dk
\int\limits_0^\pi d\theta\, \sin^3\theta\cos^2\theta
\nonumber\\
\!\!&&\!\! {}\times
\int\limits_0^{2\pi} d\varphi\, 
\left(1+\frac{{\bar\Omega}}{\Omega_{\rm rd}}\right)^{\!2}\!
\left\{ 
1-\Bigg[1-\left(1+\frac{2\Omega}{\Omega_{\rm rd}}\right)^{\!-2}\Bigg] 
\frac{\cos^2\theta\cos^2\varphi}{\cos^2\theta\cos^2\varphi+\sin^2\varphi}
\vphantom{\Bigg[1-\left(1+\frac{2\Omega''}{\Omega''_{\rm rd}}\right)^{\!-2}\Bigg]}
\right\}\!,
\qquad\quad
\label{CAPITAL_GAMMA}
\\
\Lambda_1 \!\!&=&\!\! {}-3 \:+\:
\frac{5}{\Gamma}{\left(\frac{L}{2\pi}\right)}^{\!3}
\int\limits_0^\infty k^4 J_{0k}\,dk
\int\limits_0^\pi\! d\theta\, \sin^3\theta
\int\limits_0^{2\pi}\! d\varphi 
\left(1+\frac{{\bar\Omega}}{\Omega_{\rm rd}}\right)^{\!2}
\nonumber\\
\!\!&&\!\!\quad{}\times
\left\{ 2\cos^2\theta\cos^2\varphi+\frac{1}{2}\sin^2\theta
-\Bigg[1-\left(1+\frac{2\Omega}{\Omega_{\rm rd}}\right)^{\!-2}\Bigg]
\vphantom{\Bigg[1-\left(1+\frac{2\Omega''}{\Omega''_{\rm rd}}\right)^{\!-2}\Bigg]}
\right. 
\nonumber\\
\!\!&&\!\!\qquad\quad{}\times \left.
\left(2+\frac{1}{2}\,\frac{\sin^2\theta}{\cos^2\theta\cos^2\varphi+\sin^2\varphi}\right)
\cos^2\theta\cos^2\varphi
\vphantom{\Bigg[1-\left(1+\frac{2\Omega''}{\Omega''_{\rm rd}}\right)^{\!-2}\Bigg]}
\right\}\!,
\label{LAMBDA_1}
\\
\Lambda_0 \!\!&=&\!\! 2 \:+\:
\frac{5}{\Gamma}{\left(\frac{L}{2\pi}\right)}^{\!3}
\int\limits_0^\infty k^4J_{0k}\,dk
\int\limits_0^\pi d\theta\, \sin^3\theta
\int\limits_0^{2\pi} d\varphi\, 
\left(1+\frac{{\bar\Omega}}{\Omega_{\rm rd}}\right)^{\!2}\!
\nonumber\\
\!\!&&\!\!\quad{}\times
\left\{2\sin^2\theta\cos^2\varphi-\frac{1}{2}\sin^2\theta
+\Bigg[1-\left(1+\frac{2\Omega}{\Omega_{\rm rd}}\right)^{\!-2}\Bigg]
\vphantom{\Bigg[1-\left(1+\frac{2\Omega''}{\Omega''_{\rm rd}}\right)^{\!-2}\Bigg]}
\right. 
\nonumber\\
\!\!&&\!\!\qquad\quad{}\times \left.
\left(\cos^2\theta-\sin^2\theta+\frac{1}{2}\,
\frac{\sin^2\theta\cos^2\theta}{\cos^2\theta\cos^2\varphi+\sin^2\varphi}
\right)\cos^2\varphi
\vphantom{\Bigg[1-\left(1+\frac{2\Omega''}{\Omega''_{\rm rd}}\right)^{\!-2}\Bigg]}
\right\}\!,
\label{LAMBDA_0}
\\
\frac{{\bar\Omega}}{\Omega_{\rm rd}} \!\!&=&\!\!
6\,\frac{k^2}{k_\nu^2}\,,
\label{OMEGA_BAR_DAMP_RATIO_THIRD}
\\
\frac{2\Omega}{\Omega_{\rm rd}} \!\!&=&\!\!
90\,\frac{k^2}{k_\nu^2}\,\sin^2\theta\cos^2\varphi\,(1-\sin^2\theta\cos^2\varphi)\,,
\label{OMEGA_DAMP_RATIO_THIRD}
\eeq
and function $J_{0k}$ is given by equation~(\ref{J_ZERO_K}).
Here, we drop double primes, $''$, and use equations~(\ref{lambda_0})--(\ref{lambda_2}) 
and equations~(\ref{OMEGA_AVERAGED}), (\ref{OMEGA}), (\ref{OMEGA_DAMP}).
We calculate $\Gamma$ and dimensionless number $\Lambda_0$ and $\Lambda_1$ numerically
in Appendix~(\ref{CALCULATION_SMALL_SCALES}), and obtain
\beq
\Gamma &\approx& 
100\:\left(\frac{U_0L}{\nu}\right)^{1/2}\frac{U_0}{L}
\label{CAPITAL_GAMMA_RESULT}
\\
\Lambda_1 &\approx& 2,
\label{LAMBDA_1_RESULT}
\\
\Lambda_0 &\approx& 5.
\label{LAMBDA_0_RESULT}
\eeq
Therefore,
\beq
\frac{\partial M}{\partial t} =
20\:\left(\frac{U_0L}{\nu}\right)^{1/2}\frac{U_0}{L} 
\left[k^2\frac{\partial^2M}{\partial k^2}
-k\frac{\partial M}{\partial k} + 5M\right].
\label{SMALL_SCALES_MODE_COUPLING_RESULT}
\eeq
It is interesting to compare it with equation~(\ref{SMALL_SCALES_MODE_COUPLING_KA}), 
obtained by Kulsrud and Anderson in the kinematic dynamo case~\cite{KA_92}.

Now, assume that $M(t,k_{\rm ref})$ is known as a function of time at some reference 
wave number $k=k_{\rm ref}$, then the solution of~(\ref{SMALL_SCALES_MODE_COUPLING}) is
\beq
M(t,k)=\int_{-\infty}^t M(t',k_{\rm ref}) G(k/k_{\rm ref},t-t')\,dt',
\label{SMALL_SCALES_SOLUTION}
\eeq
where the Green's function $G(k,t)$ is
\beq
G(k,t) &=& {\left(\frac{5}{4\pi}\right)}^{\!1/2}\,
\frac{k^{\Lambda_1/2}\ln{k}}{\Gamma^{1/2}t^{3/2}}
\,e^{(\Gamma/5)(\Lambda_0-\Lambda_1^2/4)t}
\,e^{\left.-5\ln^2{k}\right/4\Gamma t}
\nonumber\\
{}&=& 
{\left(\frac{5}{4\pi}\right)}^{\!1/2}\,
\frac{k\ln{k}}{\Gamma^{1/2}t^{3/2}}
\,e^{(4\Gamma/5)t}\,e^{\left.-5\ln^2{k}\right/4\Gamma t}\,.
\label{GREENS_FUNCTION}
\eeq
We derive these two equations in Appendix~\ref{GREENS_FUNCTION_SMALL_SCALES},
and use equations~(\ref{LAMBDA_1_RESULT}) and~(\ref{LAMBDA_0_RESULT})
for $\Lambda_1$ and $\Lambda_0$. 
We see that a ``signal'' $M(t,k_{\rm ref})$, at zero time, will increase 
exponentially as $e^{(4/5)\Gamma t}$ and will extend down to the scale 
$k_{\rm peak}\approx e^{(4/5)\Gamma t}\,k_{\rm ref}$, where $k_{\rm peak}$ is 
the peak of function $kG(k,t)$, (of course, the field scale can not become
less than the resistivity scale). As a result, in the magnetized dynamo theory
the magnetic energy tends to quickly propagate to very small subviscous scales,
the same way as it does in the kinematic dynamo theory, (this propagation is 
checked by the resistivity). However, the tail of the magnetic energy spectrum on 
$k_{\rm ref}\simlt k\simlt k_{\rm peak}$ scales increases with the wavenumber 
as ${}\propto {\it k}$ instead of ${}\propto {\it k}^{\rm 3/2}$ in the kinematic 
theory. Note, that according to equations~(\ref{GAMMA_RESULT}) 
and~(\ref{CAPITAL_GAMMA_RESULT}), the growth rate of the Green's function, 
$(4/5)\Gamma$, is approximately equal to a half of the growth rate of the 
total magnetic energy, $2\gamma$.

In the end of this section, let us consider the kinematic turbulent dynamo. In this 
case we need to take the limit ${\bar\Omega}\rightarrow 0$, $\Omega\rightarrow 0$, 
or alternatively, the limit $\Omega_{\rm rd}\rightarrow \infty$ in 
equations~(\ref{SMALL_SCALES_MODE_COUPLING})--(\ref{LAMBDA_0}) [see the 
footnote~\ref{KA_LIMIT_FOOTNOTE} on page~\pageref{KA_LIMIT_FOOTNOTE}]. After 
integrating over $\varphi$ and $\theta$, we have $\Gamma=\gamma_{\rm o}$, where 
$\gamma_{\rm o}$ is the Kulsrud-Anderson magnetic energy growth rate, given by 
equation~(\ref{GAMMA_0}), $\Lambda_1=3$ and $\Lambda_0=6$. As a result, as one might 
expect, in the kinematic dynamo case equation~(\ref{SMALL_SCALES_MODE_COUPLING}) 
reduces to equation~(\ref{SMALL_SCALES_MODE_COUPLING_KA}), and the Green's 
function~(\ref{GREENS_FUNCTION}) reduces to formula~(\ref{GREENS_FUNCTION_KA}).


%% file: chapter_5.tex
\chapter
[Discussion and Conclusions]
{Discussion and Conclusions}
\label{DISCUSSION}


As we discussed in the introductory section, the origin of galactic and 
extragalactic magnetic fields is one of the key questions in astrophysics.
This thesis is devoted to this question. There are two prevailing theories 
for the origin of cosmic magnetic fields. First, the galactic turbulent 
dynamo theory, which states that the magnetic fields have been primarily 
amplified in differentially rotating galactic disks after the galaxies had 
been formed. Second, the primordial turbulent dynamo theory, which states 
that the fields have primarily been produced in protogalaxies, undergoing 
gravitational collapse. It seems that both observational and theoretical 
results favor the second, the primordial dynamo theory. 

In calculations of the magnetic field evolution the previous dynamo theories 
assumed the regular {\it isotropic viscosity} for the turbulent plasma motions, 
and this is justified because of neutrals. However, in protogalaxies the 
temperature is so high, that there are no neutrals, and the viscosity 
is dominated by ions. Therefore, in a protogalaxy, as the magnetic field 
strength grows in time because of the dynamo inductive action, starting from 
its initial seed value, the plasma quickly becomes strongly magnetized, the 
viscosity becomes {\it the Braginskii tensor viscosity}, and the turbulent 
motions on the viscous scales become strongly altered from the isotropic case. 
As a result, the turbulent dynamo becomes {\it the magnetized turbulent dynamo}.

In this thesis we have developed a theoretical basis for the magnetized 
turbulent dynamo, which operates in protogalaxies. The results of the 
kinematic dynamo theory already seem to support the primordial 
(protogalactic) dynamo origin of cosmic magnetic fields~\cite{KCOR_97}. The 
results that we have obtained for the magnetized dynamo, further support this 
primordial origin theory. This is because the number of e-foldings of 
the total magnetic energy during the collapse of a protogalaxy, given 
by equation~(\ref{GAMMA_COLLAPSE_TIME}), is as much as ten time larger 
than that in the kinematic dynamo theory. Therefore, the number of e-foldings 
in the magnetized dynamo is more than large enough for the magnetic fields 
in protogalaxies to grow from their seed value, provided by the Biermann 
battery, up to the field-turbulence energy equipartition value. The number 
of e-foldings of the magnetic energy on the viscous scale, which is equal 
to the growth rate $(4/5)\Gamma$ of the Green's 
function~(\ref{GREENS_FUNCTION}), is less by one half, but it is still 
sufficiently large~\footnote{
Of course, our results~(\ref{GAMMA})--(\ref{GAMMA_RATIO}) for the magnetic 
energy growth rate are sensitive to the value of the physical parameter 
$\Omega_{\rm rd}$, which is estimated in equation~(\ref{OMEGA_DAMP}). We 
also left out the finite time correlation effects (see discussion below).
Therefore, our result for the number of magnetic energy e-foldings should 
be viewed as an estimate, valid within a factor of order two. However, it 
is important that the number of e-foldings that we found in the magnetized 
dynamo theory is large and is clearly larger than that in the kinematic 
dynamo theory.
}. 

Another of our predictions is that the tail of the magnetic energy 
spectrum on the small subviscous scales increases with the wavenumber as 
${}\propto {\it k}$ [see the Green's function~(\ref{GREENS_FUNCTION})], 
instead of ${}\propto {\it k}^{\rm 3/2}$ in the kinematic theory. Therefore, 
in the magnetized dynamo the magnetic energy is slightly shifted to 
larger scales as compared to the kinematic dynamo case.

Although the important results of this thesis are convincing, we left out 
several issues in our theory. Therefore, let us itemize and briefly discuss 
the possibilities of further research on the magnetized turbulent dynamos.
\begin{itemize}
\item 
{\bf Calculations of magnetic field curvature.}\\
In our calculations in this thesis we assumed that as far as we are interested
in the evolution of magnetic energy, we can consider the magnetic field lines 
to be initially straight. This our first working hypothesis. It relies on the 
assumption that in the case of the magnetized turbulent dynamo the magnetic 
field has a folding structure similar to the one that exists in the case of the 
kinematic turbulent dynamo~\cite{M_01,SCMM_01}. Unfortunately, the 
calculation methods employed in this thesis are not adequate to justify 
this our hypothesis because of complications that arise when one 
calculates the statistics of the field curvature. It would be interesting 
either to expand our theory to the case of curved magnetic field lines, 
or to carry out numerical simulations in order to check the 
field folding structure in the magnetized dynamo theory.
\item 
{\bf Effective rotational damping of turbulent velocities.}\\
Another hypothesis that we made, but did not proved, in this thesis was the 
inclusion of the effective rotational damping into our equations as a linear 
damping term. Basically, the rotational damping rate $\Omega_{\rm rd}$, which 
is in front of this linear damping term [see equation~(\ref{v_EVOLUTION_NEW})], 
enters our final results as a physical parameter. We obtained only an estimate 
of this parameter, so our final numerical results depend on its value. As we 
said above, the rotational damping is associated with non-linear coupling of 
velocity modes and is very important. It restricts the unlimited growth of 
velocity modes which are perpendicular to the magnetic field lines and are 
undamped by the Braginskii viscous forces. It is important to further 
theoretically study the mechanisms which stop the unlimited growth of the 
undamped velocities, and to carry out MHD numerical simulations, including 
the Braginskii viscosity.
\item 
{\bf Finite time correlation effects.}\\
In this thesis we used the quasi-linear (up to the second order) expansion 
procedure in time to solve the MHD equations. We assumed that we can choose 
our time expansion parameter $t$ small enough for the quasi-linear expansion 
to be valid, but large enough for it to be much larger than the turbulent eddy 
decorrelation time. This is equivalent to the assumption that the turbulent 
velocities are $\delta$-correlated in time. It is known from the kinematic 
turbulent dynamo theory that the effects of finite velocity correlation time 
can decrease the magnetic energy growth rate by a factor of order 
two~\cite{SK_01}. Thus, including the finite time correlation effects into the 
magnetized dynamo theory is important but not vital, since our calculation
predicts a very large number of e-foldings of the magnetic field strength.
\item 
{\bf Energy equipartition and the inverse cascade.}\\
Finally, the Green's function solution of the mode-coupling equation for the 
magnetic energy spectrum on subviscous scales indicates that in the 
magnetized dynamo theory the magnetic energy tends to quickly propagate to 
very small subviscous scales, similar to the kinematic dynamo case. On the 
other hand, the observed cosmic fields have rather large correlation lengths. 
Therefore, the magnetic field lines must be unwrapped on small scales by 
the Lorentz tension forces, while the field energy is transferred and 
amplified on larger scales during the inverse cascade stage. This stage happens
when the magnetic field energy is comparable to the turbulent kinetic energy, 
i.~e.~when there is the energy equipartition between the field and the turbulence. 
Thus, calculations and/or numerical simulations of the inverse cascade with 
the Braginskii viscosity are extremely important for the theory of the origin of 
cosmic magnetic fields. In the end of this chapter let us discuss the 
importance of the Braginskii viscosity for the inverse cascade in more details.
\end{itemize}

As we said in the first introductory chapter of this thesis, the theory of the 
inverse cascade in a plasma with the regular isotropic viscosity has 
a difficulty of unwrapping of the small-scale magnetic field lines.
This difficulty can be understood as follows~\cite{CKS_01}. The equation on 
the turbulent velocities $\bf V$ in the plasma with the isotropic viscosity,
including the Lorentz forces, is
\beq
\partial_t V_\alpha &=&
-P_{,\alpha}+f_\alpha+\nu\triangle V_\alpha
+\frac{1}{4\pi\rho}({\bf B}\cdot{\bf\nabla}) B_\alpha
-({\bf V}\cdot{\bf\nabla})V_\alpha.
\label{EQUATION_FOR_U_LORENTS_FORCES}
\eeq
[Compare this equation to equation~(\ref{EQUATION_FOR_U})]. 
The $(1/4\pi\rho)({\bf B}\cdot{\bf\nabla})B_\alpha$ term is the magnetic tension 
force, normalized to the plasma density $\rho$. The magnetic pressure term is 
included into the hydrodynamic pressure term, $-P_{,\alpha}$. 
The $\nu\triangle V_\alpha$ term is the viscous term. We can estimate 
the velocity of the magnetic field lines unwrapping, $V_{\rm unwrap}$,
by Fourier transforming equation~(\ref{EQUATION_FOR_U_LORENTS_FORCES}) in space,
${\bf r}\rightarrow{\bf k}$, and then balancing the viscous and the inertial 
forces against the magnetic tension force. For the isotropic case, the viscosity 
dominates on small scales, and we have
\beq
\nu k^2 V_{\rm unwrap} &\sim& \frac{1}{4\pi\rho}k_\parallel B^2,
\\
V_{\rm unwrap} &\sim& \frac{k_\parallel}{k}\,\frac{k_\nu}{k}\,
\frac{V_{\rm A}^2}{\nu k_\nu} \ll V_{\rm A},
\label{UNWRAPPING_VELOCITY}
\eeq
where $V_{\rm A}$ is the Alfven speed. The Alfven speed $V_{\rm A}\sim\nu k_\nu$ 
at the time of the field-turbulence energy equipartition on the viscous scale, 
and $V_{\rm A}<\nu k_\nu$ before the equipartition. If the field lines 
have a folding pattern (and they do in the kinematic dynamo theory), then 
$k\gg k_\parallel\sim k_\nu$, see Figure~\ref{FIGURE_FOLDING}, and therefore,  
the unwrapping velocity~(\ref{UNWRAPPING_VELOCITY}) is small compared to the 
Alfven speed, even at the equipartition. In other words, since $V_{\rm A}\sim V$ 
at the energy equipartition, then $k_\parallel V_{\rm A}\sim\gamma$ ($\gamma$ is 
the magnetic energy growth rate), and the unwrapping rate 
$k_\parallel V_{\rm unwrap}$ is much smaller than $\gamma$. This means that the 
field continues to grow on the viscous and subviscous scales even at the 
equipartition. 

However, in the case of the magnetized turbulent dynamo, the viscosity term
in the above equation is modified and the anti-unwrapping argument does not 
apply. Indeed, the field unwrapping velocity is parallel and varies 
perpendicularly to the magnetic field lines. The large velocity gradient
perpendicular to the magnetic field lines, which leads to a large perpendicular
stress in the isotropic viscosity case, is irrelevant in the case of the 
Braginskii viscous forces (because the transfer of the ion momentum in the 
perpendicular direction is inhibited). Therefore, in the magnetized dynamo theory
$k_\parallel V_{\rm unwrap}\sim k_\parallel V_{\rm A}\sim \gamma$ at the 
equipartition, and the magnetic field strength saturates on the viscous and 
subviscous scales. As a result, the Braginskii viscosity makes the inverse 
cascade of the magnetic energy more likely, because the larger turbulent eddies 
can not deliver their energy to the magnetic field on the viscous and subviscous 
scales~\cite{K_00}.


%% file: component_appendix_a.tex
\chapter
[Fourier and Laplace Transformations]
{Fourier and Laplace Transformations}
\label{TRANSFORMS}


Let us consider an arbitrary continuous function $f(t,{\bf r})$, where $t$ is time 
and $\bf r$ is position in space. We assume that $f$ is defined inside a spatial 
box of length $L$ and, in the rest of space, is periodic with the box size. 
Then $f(t,{\bf r})$ can be expressed in terms of its Fourier components 
${\tilde f}(t,{\bf k})$ as
\beq
f(t,{\bf r})=\sum_{\bf k} {\tilde f}_{\bf k}(t)\,e^{i{\bf kr}}.
\label{FOURIER_SUM_A}
\eeq
Here the summation is done over discrete values of vector $\bf k$, so that
the $x$-, $y$- and $z$- components of $\bf k$ are discrete and are given by
equations
\beq
k_x(n_x) &=& \frac{2\pi}{L}\,n_x,
\\
k_y(n_y) &=& \frac{2\pi}{L}\,n_y,
\\
k_z(n_z) &=& \frac{2\pi}{L}\,n_z,
\eeq
where $n_x$, $n_y$ and $n_z$ are arbitrary integer numbers. The Fourier coefficients 
${\tilde f}_{\bf k}(t)$ are given by the standard discrete Fourier transformation 
formula
\beq
{\tilde f}_{\bf k}(t)=\frac{1}{L^3}\int\limits_{-L/2}^{L/2}
f(t,{\bf r})\,e^{-i{\bf kr}}\,d^3{\bf r}.
\label{FOURIER_COEFFICIENTS_A}
\eeq
Now let us introduce a function ${\tilde f}(t,{\bf k'})$, which is a continuous
function of the wave number ${\bf k'}$,
\beq
{\tilde f}(t,{\bf k'}) \define {\tilde f}_{\bf k}(t) 
\qquad\mbox{if}\qquad 
\begin{array}{l}
k_x(n_x)-\pi L^{-1} \le k'_x < k_x(n_x)+\pi L^{-1},\\
k_y(n_y)-\pi L^{-1} \le k'_y < k_x(n_y)+\pi L^{-1},\\
k_z(n_z)-\pi L^{-1} \le k'_z < k_x(n_z)+\pi L^{-1}.
\end{array}
\eeq
This continuous function is constant over small three-dimensional cubic volume 
elements $(2\pi/L)^3$ in $\bf k$-space, and it is equal to the values of the 
discrete function ${\tilde f}_{\bf k}(t)$ at the center points of these volume 
elements~\footnote{
Note, that the important property $\tilde f(t,-{\bf k})= \tilde f^*(t,{\bf k})$ 
is an exact consequence of $\tilde f_{-\bf k}=\tilde f^*_{\bf k}$ everywhere 
in k-space but on the faces of the cubic volume elements. This is not a problem 
though, because the faces occupy an infinitely small three-dimensional volume.
}. 
As a result, the Fourier series~(\ref{FOURIER_SUM_A}) can be rewritten,
using the integration of this appropriately defined continuous function
${\tilde f}(t,{\bf k})$ over $\bf k$,
as
\beq
f(t,{\bf r})=\sum_{\bf k} {\tilde f}_{\bf k}(t)\,e^{i{\bf kr}}
={\left(\frac{L}{2\pi}\right)}^{\!3}
\int\limits_{-\infty}^\infty {\tilde f}(t,{\bf k})\,e^{i{\bf kr}}\,d^3{\bf k}.
\eeq
In this thesis we frequently use the appropriately defined functions that are 
continuous in $\bf k$, instead of discrete Fourier components.

We can further Fourier transform functions ${\tilde f}_{\bf k}(t)$ and 
${\tilde f}(t,{\bf k})$ in time, by making use of the standard continuous Fourier 
transformation $t\rightarrow\omega$. For example, for function ${\tilde f}_{\bf k}(t)$
we have 
\beq
{\tilde f}_{\bf k}(\omega) &=& \frac{1}{\sqrt{2\pi}}\int\limits_{-\infty}^\infty 
{\tilde f}_{\bf k}(t)\,e^{i\omega t}\,dt,
\\
{\tilde f}_{\bf k}(t) &=& \frac{1}{\sqrt{2\pi}}\int\limits_{-\infty}^\infty 
{\tilde f}_{\bf k}(\omega)\,e^{-i\omega t}\,d\omega.
\label{TIME_FOURIER_A}
\eeq
We can also form the continuous Laplace transformation in time, $t\rightarrow s$,
\beq
{\tilde f}_{\bf k}(s) &=& \int\limits_{0}^\infty 
{\tilde f}_{\bf k}(t)\,e^{-st}\,dt,
\\
\widetilde{\frac{\partial{\tilde f}_{\bf k}}{\partial t}}(s) &=& s{\tilde f}_{\bf k}(s)
-{\tilde f}_{\bf k}(t=0),
\label{LAPLACE_OF_DERIVATIVE}
\\
{\tilde f}_{\bf k}(t) &=& \frac{1}{2\pi i}\int\limits_{c-i\infty}^{c+i\infty} 
{\tilde f}_{\bf k}(s)\,e^{ts}\,ds.
\label{TIME_LAPLACE_A}
\eeq
In the last formula all poles of function ${\tilde f}_{\bf k}(s)$ should be to 
the left of the integration contour chosen in the complex plane.


%% file: component_appendix_b.tex
\chapter
[The Ensemble Average of the Second Order Velocities, 
{\boldmath${\langle\ViiT_{{\bf k}\alpha}(t)\rangle}$}]
{The Ensemble Average of the Second Order Velocities, 
{\boldmath${\langle\ViiT_{{\bf k}\alpha}(t)\rangle}$}}
\label{Vii_ENSEMBLE_AVERAGE}


To derive equation~(\ref{Vii_AVERAGE}) for the ensemble averaged second order velocity
in our case of an initially straight field $\bbo_{\alpha\beta}={\rm const}$, we proceed 
as follows. Let choose a system of coordinates in which the initial field $\bobf$
is along the $x$-direction. In this case 
$\bbo_{\alpha\beta}=\delta_{\alpha x}\delta_{\beta x}$, and we rewrite 
equation~(\ref{v_EVOLUTION_ii}) as
\beq
\partial_t\,\vii_\alpha &=& -\:\Pii_{,\alpha} 
+ 3\nu\delta_{\alpha x}\vii_{x,xx}
+ 3\nu{[\bbi_{\alpha\beta}(\vi_{x,x}+U_{x,x})]}_{,\beta}
\nonumber\\
{}&+& 3\nu\delta_{\alpha x}{[\bbi_{\mu\nu}(\vi_{\mu,\nu}+U_{\mu,\nu})]}_{,x}
- {[\vi_\alpha U_\beta-U_\alpha\vi_\beta-\vi_\alpha\vi_\beta]}_{,\beta}\;,
\label{vii_EQUATION_B}
\eeq
Here, we use formula $\bbi_{\alpha\beta\mu\nu}= 
\bbi_{\alpha\beta}\bbo_{\mu\nu} + \bbo_{\alpha\beta}\bbi_{\mu\nu}$, see 
definitions~(\ref{bbbb})--(\ref{bb}). 
Now, first, we Fourier transform this equation in space, ${\bf r}\rightarrow{\bf k}$.
Second, we multiply the transformed equation on the left by the tensor
$\delta^\perp_{\gamma\alpha}=\delta_{\gamma\alpha}-\k_\gamma\k_\alpha$
to eliminate the pressure $\Pii$, and we also use the fluid incompressibility
condition $k_\alpha\viiT_\alpha=0$. Third, we ensemble average the resulting
equation. As a result, we obtain 
\beq
&&\partial_t\,\langle\viiT_{{\bf k}\gamma}\rangle 
+ 3\nu k_x^2\delta^\perp_{\gamma x} \langle\viiT_{{\bf k}x}\rangle 
= 3\nu\delta^\perp_{\gamma\alpha}ik_\beta
\sum_{{\bf k'}\atop{{\bf k}''=\,{\bf k}-{\bf k}'}} ik''_x
\Bigl(\langle\bbiT_{{\bf k'}\alpha\beta}\viT_{{\bf k''}x}\rangle
+\langle\bbiT_{{\bf k'}\alpha\beta}{\tilde U}_{{\bf k''}x}\rangle\Bigr)
\nonumber\\
&&\qquad\qquad{} + 3\nu\delta^\perp_{\gamma x}ik_x
\sum_{{\bf k'}\atop{{\bf k}''=\,{\bf k}-{\bf k}'}} ik''_\nu
\Bigl(\langle\bbiT_{{\bf k'}\mu\nu}\viT_{{\bf k''}\mu}\rangle
+\langle\bbiT_{{\bf k'}\mu\nu}{\tilde U}_{{\bf k''}\mu}\rangle\Bigr)
\nonumber\\
&&\qquad\qquad{} - \delta^\perp_{\gamma\alpha}ik_\beta 
\sum_{{\bf k'}\atop{{\bf k}''=\,{\bf k}-{\bf k}'}}
\Bigl[\langle{\tilde U}_{{\bf k'}\beta}\viT_{{\bf k''}\alpha}\rangle +
 \langle{\tilde U}_{{\bf k'}\alpha}\viT_{{\bf k''}\beta}\rangle +
 \langle\viT_{{\bf k'}\alpha}\viT_{{\bf k''}\beta}\rangle\Bigr]\;.
\label{vii_T_EQUATION_B}
\eeq

Next, in order to calculate the first order quantities $\bbiT_{{\bf k}\alpha\beta}(t)$ 
and $\viT_{{\bf k}\alpha}(t)$, which enter equation~(\ref{vii_T_EQUATION_B}), we proceed
as follows. First, we Fourier transform equation~(\ref{v_EVOLUTION_i}) in space, 
and we multiply the transformed equation on the left by tensor
$\delta^\perp_{\gamma\alpha}$. We have
\beq
\partial_t\,\viT_{{\bf k}\gamma} 
+ 3\nu k_x^2\delta^\perp_{\gamma x}\viT_{{\bf k}x}
= {}-3\nu k_x^2\delta^\perp_{\gamma x}{\tilde U}_{{\bf k}x}
+\frac{1}{5}\nu k^2 {\tilde U}_{{\bf k}\gamma}\;.
\label{v_i_EVOLUTION_B}
\eeq
Integrating this equation with zero initial condition, ${\vibf|}_{t=0}=0$, we find
\beq
\viT_{{\bf k}x}(t) &=& ({\bar\Omega}-2\Omega)\int\limits_0^t
{\tilde U}_{{\bf k}x}(t')\,e^{-2\Omega(t-t')}\,dt'\;,
\label{viT_x_B}
\\
\viT_{{\bf k}\gamma_{\perp_x}}(t) &=& 2\Omega\,\frac{\k_x\k_{\gamma_{\perp_x}}}{1-\mu^2}
\int\limits_0^t \Bigl[{\tilde U}_{{\bf k}x}(t')+\viT_{{\bf k}x}(t')\Bigr]dt'
+{\bar\Omega}\int\limits_0^t {\tilde U}_{{\bf k}\gamma_{\perp_x}}(t')\,dt'\;,
\label{viT_perp_B}
\eeq
where index $\gamma_{\perp_x}$ is equal to $y$ or $z$, and the viscous frequencies 
$\bar\Omega$ and $\Omega$ are given by equations~(\ref{OMEGA_AVERAGED}) 
and~(\ref{OMEGA}).
Second, we use equation~(\ref{b_EVOLUTION_i}) to obtain
\beq
\partial_t\,\bbi_{\alpha\beta} \!\!&=&\!\! 
\partial_t\,(\bi_{\alpha}\bo_{\beta}+\bo_{\alpha}\bi_{\beta})
= \Vi_{\alpha,\eta}\bbo_{\eta\beta} 
+ \Vi_{\beta,\eta}\bbo_{\eta\alpha}
- 2\Vi_{\tau,\eta}\bbo_{\alpha\beta\tau\eta}
- \Vi_\tau\bbo_{\alpha\beta,\tau}
\nonumber\\
\!\!&=&\!\!  (\delta_{\alpha\gamma}\delta_{\beta x}
+ \delta_{\beta\gamma}\delta_{\alpha x}
- 2\delta_{\alpha x}\delta_{\beta x}\delta_{\gamma x})\Vi_{\gamma,x}\;.
\eeq
Now, we Fourier transform this equation and integrate the resulting equation. 
We have 
\beq
\bbiT_{{\bf k}\alpha\beta}(t) = ik_x
(\delta_{\alpha\gamma}\delta_{\beta x} + \delta_{\beta\gamma}\delta_{\alpha x}
- 2\delta_{\alpha x}\delta_{\beta x}\delta_{\gamma x})
\int\limits_0^t \Bigl[{\tilde U}_{{\bf k}\gamma}(t')+\viT_{{\bf k}\gamma}(t')\Bigr]dt'.
\label{bbiT_B}
\eeq

Finally, we substitute formulas~(\ref{viT_x_B}),~(\ref{viT_perp_B}) and~(\ref{bbiT_B}) 
into equation~(\ref{vii_T_EQUATION_B}), and carry out the ensemble averagings. Because 
${\bf{\tilde U}}_{\bf k'}{\bf{\tilde U}}_{\bf k''}\propto\delta_{{\bf k''},-{\bf k'}}$,
we find that ${\bf k}={\bf k'}+{\bf k''}$ is equal to zero, ${\bf k}=0$.
Thus, all terms in the right-hand-side of equation~(\ref{vii_T_EQUATION_B}) vanish.
Because $\langle\viiT_{{\bf k}\gamma}\rangle$ is initially (at $t=0$) is zero, it 
stays zero in time, $\langle\viiT_{{\bf k}\gamma}(t)\rangle=0$, and using 
equation~(\ref{Vii}), we immediately obtain formula~(\ref{Vii_AVERAGE}).


%% file: component_appendix_c.tex
\chapter
[The Calculation of the Magnetic Energy Growth Rate, {\boldmath$\gamma$}]
{The Calculation of the Magnetic\\ 
Energy Growth Rate, {\boldmath$\gamma$}}
\label{CALCULATION_OF_GAMMA}


It is convenient to split the right-hand-side of equation~(\ref{GAMMA}) into two terms:
\beq
\gamma &=& \gamma'+\gamma'',
\label{GAMMA_SPLITED_C}
\\
\gamma' &=& \pi{\left(\frac{L}{2\pi}\right)}^{\!3}
\int\limits_0^\infty k^4 J_{0k}\,dk
\int\limits_{-1}^1 \mu^2
\left(1+\frac{{\bar\Omega}}{\Omega_{\rm rd}}\right)^2 d\mu\,,
\label{GAMMA'_C}
\\
\gamma'' &=& \pi{\left(\frac{L}{2\pi}\right)}^{\!3}
\int\limits_0^\infty k^4 J_{0k}\,dk
\int\limits_{-1}^1 \mu^2
\left(1+\frac{{\bar\Omega}}{\Omega_{\rm rd}}\right)^2\,
\left(1+\frac{2\Omega}{\Omega_{\rm rd}}\right)^{-2} d\mu\,.
\label{GAMMA''_C}
\eeq
Using equations~(\ref{OMEGA_AVERAGED}),~(\ref{OMEGA}) and~(\ref{OMEGA_DAMP}), we have
\beq
\frac{{\bar\Omega}}{\Omega_{\rm rd}}=\frac{6k^2}{k_\nu^2},
\qquad
\frac{2\Omega}{\Omega_{\rm rd}}=\frac{90k^2}{k_\nu^2}\,\mu^2(1-\mu^2).
\eeq
First, we calculate the first term, $\gamma'$, given by equation~(\ref{GAMMA'_C}).
Integrating over $\mu$, and using equation~(\ref{J_ZERO_K}) for $J_{0k}$, we obtain
\beq
\gamma' &=& \frac{2\pi}{3}{\left(\frac{L}{2\pi}\right)}^{\!3}\int\limits_0^\infty 
k^4 J_{0k} \left(1+\frac{6k^2}{k_\nu^2}\right)^2 dk
\nonumber\\
{}&=& \frac{2\pi}{3}k_0^{-3}\int\limits_{k_0}^{k_\nu}
k^4 \:\frac{U_0}{2k_0}{(k/k_0)}^{\!-13/3} 
\left(1+\frac{6k^2}{k_\nu^2}\right)^2 dk
\nonumber\\
{}&=& \frac{\pi}{3}U_0k_0\left(\frac{k_\nu}{k_0}\right)^{2/3}
\int\limits_{k_0/k_\nu}^1 x^{-1/3}\left(1+6x^2\right)^2 dx
\nonumber\\
{}&\approx& \frac{32\pi}{7}\,U_0k_0\left(\frac{k_\nu}{k_0}\right)^{2/3}
= \frac{64\pi^2}{7}\left(\frac{5U_0L}{2\pi\nu}\right)^{1/2}\frac{U_0}{L}.
\label{GAMMA'_RESULT_C}
\eeq
Here, we substitute $x=k/k_\nu$ and use the fact that the integral is dominated by the 
upper limit. We also use formulas $k_0=2\pi/L$ ($L$ is the system size) and 
$k_\nu/k_0=R^{3/4}$, where $R=U_0/k_0\nu_{\rm eff}=5U_0/k_0\nu$ is the Reynolds number, 
and $\nu_{\rm eff}=(1/5)\nu$ is the effective viscosity (see 
Section~\ref{MAGNETIZED_DYNAMO}).

Second, we calculate $\gamma''$, given by equation~(\ref{GAMMA''_C}), in a similar way.
We have
\beq
\gamma'' \!\!&=&\!\! \pi{\left(\frac{L}{2\pi}\right)}^{\!3} 
\int\limits_0^\infty k^4 J_{0k} \left(1+\frac{6k^2}{k_\nu^2}\right)^2 \!dk\,
\int\limits_{-1}^1
\frac{\mu^2}{{[1+90(k/k_\nu)^2\mu^2(1-\mu^2)]}^2}\,d\mu
\nonumber\\
{}\!\!&=&\!\! 2\pi k_0^{-3}\int\limits_{k_0}^{k_\nu}
k^4 \:\frac{U_0}{2k_0}{(k/k_0)}^{\!-13/3} 
\left(1+\frac{6k^2}{k_\nu^2}\right)^2 \!dk
\int\limits_0^1\frac{\mu^2}{{[1+90(k/k_\nu)^2\mu^2(1-\mu^2)]}^2}\;d\mu
\nonumber\\
{}\!\!&\approx&\!\! \pi U_0k_0\left(\frac{k_\nu}{k_0}\right)^{2/3}
\int\limits_0^1 x^{-1/3}\left(1+6x^2\right)^2 dx
\int\limits_0^1\frac{\mu^2}{{[1+90x^2\mu^2(1-\mu^2)]}^2}\;d\mu
\nonumber\\
{}\!\!&\approx&\!\! 0.36\pi\, U_0k_0\left(\frac{k_\nu}{k_0}\right)^{2/3}
\,\ll\, \gamma'.
\label{GAMMA''_RESULT_C}
\eeq
Here, we again use equation~(\ref{J_ZERO_K}) for $J_{0k}$, substitute $x=k/k_\nu$, 
and carry out the integration numerically (it clear that $\gamma''\ll\gamma'$
because $90$ is a large number). Combining 
equations~(\ref{GAMMA'_RESULT_C}),~(\ref{GAMMA''_RESULT_C}) 
and~(\ref{GAMMA_SPLITED_C}), we obtain equation~(\ref{GAMMA_RESULT}).

Third, we calculate $\gamma_{\rm o}$, given by equation~(\ref{GAMMA_0}), in order 
to estimate the ratio $\gamma/\gamma_{\rm o}$. We have
\beq
\gamma_{\rm o} &=& \frac{1}{3}{\left(\frac{L}{2\pi}\right)}^{\!3}
\int\limits_{-\infty}^\infty k^2 J_{0k} \,d^3{\bf k}
=\frac{1}{3}k_0^{-3}\int\limits_{k_0}^{k_\nu} 4\pi k^4 
\frac{U_0}{2k_0}{(k/k_0)}^{\!-13/3} \,dk
\nonumber\\
{}&\approx& \frac{2\pi}{3} U_0k_0\left(\frac{k_\nu}{k_0}\right)^{2/3}
\int\limits_0^1 x^{-1/3}dx
= \pi U_0k_0\left(\frac{k_\nu}{k_0}\right)^{2/3}.
\eeq
In the case of the kinematic dynamo, the Reynolds number is $
R=U_0/k_0\nu=U_0L/2\pi\nu$. Therefore, using $k_\nu/k_0=R^{3/4}$, we obtain
\beq
\gamma_{\rm o} &=& \pi(2\pi)^{1/2}\left(\frac{U_0L}{\nu}\right)^{1/2}\frac{U_0}{L}.
\eeq
Dividing equation~(\ref{GAMMA_RESULT}) by this equation, we obtain 
formula~(\ref{GAMMA_RATIO}).


%% file: component_appendix_d.tex
\chapter
[The Calculation of Coefficients {\boldmath$\Gamma$}, {\boldmath$\Lambda_1$}
and {\boldmath$\Lambda_2$} in Equation~(\ref{SMALL_SCALES_MODE_COUPLING})]
{The Calculation of Coefficients\\ 
{\boldmath$\Gamma$}, {\boldmath$\Lambda_1$} and {\boldmath$\Lambda_2$} in 
Equation~(\ref{SMALL_SCALES_MODE_COUPLING})}
\label{CALCULATION_SMALL_SCALES}


Before we numerically calculate $\Gamma$, $\Lambda_0$ and $\Lambda_1$, given
by equations~(\ref{CAPITAL_GAMMA})--(\ref{OMEGA_DAMP_RATIO_THIRD}), let us
first check equation~(\ref{SMALL_SCALES_MODE_COUPLING}). To do this, we use 
equation~(\ref{TOTAL_MAGNETIC_ENERGY}) and equation~(\ref{SMALL_SCALES_MODE_COUPLING}) 
to obtain the change of the total magnetic energy $\cal E$ in time,
\beq
\frac{\partial{\cal E}}{\partial t} \!\!&=&\!\!
\frac{1}{2}\int\limits_0^\infty \frac{\partial M}{\partial t}\,dk
= \frac{1}{2}\,\frac{\Gamma}{5}\int\limits_0^\infty 
\left[k^2\frac{\partial^2M}{\partial k^2}
-(\Lambda_1-1)k\frac{\partial M}{\partial k} + \Lambda_0 M\right]dk
\nonumber\\
{}\!\!&=&\!\!
\frac{1}{2}\,\frac{\Gamma}{5}\,\Big[2+(\Lambda_1-1)+\Lambda_0\Big]
\int\limits_0^\infty M\,dk
=\frac{\Gamma}{5}\left(\Lambda_1+\Lambda_0+1\right){\cal E}
\nonumber\\
{}\!\!&=&\!\!
2\gamma{\cal E},
\label{ENERGY_GROWTH_D}
\eeq
where the growth rate $\gamma$ is given by equation~(\ref{GAMMA}). To obtain the 
second line of this equation, we integrate the first line by parts. To obtain
the last (third) line of this equation, we use equations~(\ref{LAMBDA_1}) 
and~(\ref{LAMBDA_0}). Equation~(\ref{ENERGY_GROWTH_D}) coincides with 
equation~(\ref{ENERGY_GROWTH}), as one might expect.

Now, let us refer to equations~(\ref{CAPITAL_GAMMA})--(\ref{OMEGA_DAMP_RATIO_THIRD}).
Using equation~(\ref{J_ZERO_K}) for $J_{0k}$, and changing the integration variables, 
$k\rightarrow x=k/k_\nu$, we obtain
\beq
\frac{{\bar\Omega}}{\Omega_{\rm rd}} \!\!&=&\!\!6 x^2
\\
\frac{2\Omega}{\Omega_{\rm rd}} \!\!&=&\!\!
90x^2\sin^2\theta\cos^2\varphi\,(1-\sin^2\theta\cos^2\varphi)\,,
\\
\Gamma \!\!&=&\!\! 
\left(\frac{U_0L}{\nu}\right)^{1/2}\frac{U_0}{L}\:\Gamma'\,,
\label{CAPITAL_GAMMA_D}
\\
\Gamma' \!\!&=&\!\! 
\frac{5}{2}{\left(\frac{5\pi}{2}\right)}^{\!1/2}
\int\limits_0^1 x^{-1/3}\,dx
\int\limits_0^\pi d\theta\, \sin^3\theta\cos^2\theta
\nonumber\\
\!\!&&\!\! {}\times
\int\limits_0^{2\pi} d\varphi\, 
\left(1+\frac{{\bar\Omega}}{\Omega_{\rm rd}}\right)^{\!2}\!
\left\{ 
1-\Bigg[1-\left(1+\frac{2\Omega}{\Omega_{\rm rd}}\right)^{\!-2}\Bigg] 
\frac{\cos^2\theta\cos^2\varphi}{\cos^2\theta\cos^2\varphi+\sin^2\varphi}
\vphantom{\Bigg[1-\left(1+\frac{2\Omega''}{\Omega''_{\rm rd}}\right)^{\!-2}\Bigg]}
\right\}\!,
\qquad\quad
\label{CAPITAL_GAMMA'_D}
\\
\Lambda_1 \!\!&=&\!\! {}-3 \:+\:
\frac{5}{\Gamma'}{\left(\frac{5\pi}{2}\right)}^{\!1/2}
\int\limits_0^1 x^{-1/3}\,dx
\int\limits_0^\pi\! d\theta\, \sin^3\theta
\int\limits_0^{2\pi}\! d\varphi 
\left(1+\frac{{\bar\Omega}}{\Omega_{\rm rd}}\right)^{\!2}
\nonumber\\
\!\!&&\!\!\quad{}\times
\left\{ 2\cos^2\theta\cos^2\varphi+\frac{1}{2}\sin^2\theta
-\Bigg[1-\left(1+\frac{2\Omega}{\Omega_{\rm rd}}\right)^{\!-2}\Bigg]
\vphantom{\Bigg[1-\left(1+\frac{2\Omega''}{\Omega''_{\rm rd}}\right)^{\!-2}\Bigg]}
\right. 
\nonumber\\
\!\!&&\!\!\qquad\quad{}\times \left.
\left(2+\frac{1}{2}\,\frac{\sin^2\theta}{\cos^2\theta\cos^2\varphi+\sin^2\varphi}\right)
\cos^2\theta\cos^2\varphi
\vphantom{\Bigg[1-\left(1+\frac{2\Omega''}{\Omega''_{\rm rd}}\right)^{\!-2}\Bigg]}
\right\}\!,
\label{LAMBDA_1_D}
\\
\Lambda_0 \!\!&=&\!\! 2 \:+\:
\frac{5}{\Gamma'}{\left(\frac{5\pi}{2}\right)}^{\!1/2}
\int\limits_0^1 x^{-1/3}\,dx
\int\limits_0^\pi d\theta\, \sin^3\theta
\int\limits_0^{2\pi} d\varphi\, 
\left(1+\frac{{\bar\Omega}}{\Omega_{\rm rd}}\right)^{\!2}\!
\nonumber\\
\!\!&&\!\!\quad{}\times
\left\{2\sin^2\theta\cos^2\varphi-\frac{1}{2}\sin^2\theta
+\Bigg[1-\left(1+\frac{2\Omega}{\Omega_{\rm rd}}\right)^{\!-2}\Bigg]
\vphantom{\Bigg[1-\left(1+\frac{2\Omega''}{\Omega''_{\rm rd}}\right)^{\!-2}\Bigg]}
\right. 
\nonumber\\
\!\!&&\!\!\qquad\quad{}\times \left.
\left(\cos^2\theta-\sin^2\theta+\frac{1}{2}\,
\frac{\sin^2\theta\cos^2\theta}{\cos^2\theta\cos^2\varphi+\sin^2\varphi}
\right)\cos^2\varphi
\vphantom{\Bigg[1-\left(1+\frac{2\Omega''}{\Omega''_{\rm rd}}\right)^{\!-2}\Bigg]}
\right\}\!.
\label{LAMBDA_0_D}
\eeq
Here we use formulas $k_0=2\pi/L$ ($L$ is the system size) and 
$k_\nu/k_0=R^{3/4}$, where $R=U_0/k_0\nu_{\rm eff}=5U_0/k_0\nu$ is the Reynolds number, 
and $\nu_{\rm eff}=(1/5)\nu$ is the effective viscosity (see 
Section~\ref{MAGNETIZED_DYNAMO}).
We calculate the triple integrals in equations~(\ref{CAPITAL_GAMMA'_D})--
(\ref{LAMBDA_0_D}) numerically. We obtain
\beq
\Gamma' \approx 104,
\qquad
\Lambda_1 \approx 2.13,
\qquad
\Lambda_0 \approx 5.21,
\label{RESULT_D}
\eeq
which, immediately give us equations~(\ref{CAPITAL_GAMMA_RESULT})--
(\ref{LAMBDA_0_RESULT}).

It is interesting that if $\Omega_{\rm rd}=\Omega_{\rm rd}(k)$ [it depends only on $k$, 
not on $\mu^2$], then in the limit $\Omega_{\rm rd}\Big/\nu_{\rm eff}k_\nu^2\rightarrow 0$ 
(weak rotational damping) we have $\Lambda_1=2$ and $\Lambda_0=5$. The results given 
by equation~(\ref{RESULT_D}) are close to these results.


%% file: component_appendix_e.tex
\chapter
[The Derivation of the Green's Function Solution~(\ref{GREENS_FUNCTION})]
{The Derivation of the Green's\\
Function Solution~(\ref{GREENS_FUNCTION})}
\label{GREENS_FUNCTION_SMALL_SCALES}


To solve equation~(\ref{SMALL_SCALES_MODE_COUPLING}), we Laplace transform it in time,
$t\rightarrow s$ and $M(t,k)\rightarrow {\tilde M}(s,k)$. We assume that $M(t,k)$ is 
zero at $t=0$ and that we know $M(t,k_{\rm ref})$ at some reference wave number 
$k=k_{\rm ref}$ at $t>0$. To simplify our notations we set $k_{\rm ref}=1$, so that
$M(t,k_{\rm ref})=M(t,1)$. As a result, we have
\beq
\frac{5}{\Gamma}s{\tilde M} = 
k^2\frac{\partial^2{\tilde M}}{\partial k^2}
- (\Lambda_1-1)k\frac{\partial {\tilde M}}{\partial k} 
+ \Lambda_0{\tilde M}.
\eeq
We consider the following solution of this equation:
\beq
{\tilde M}(s,k)={\tilde M}(s,1)\,
k^{(\Lambda_1/2)\pm\sqrt{5s/\Gamma-(\Lambda_0-\Lambda_1^2/4)}},
\label{LAPLACE_SOLUTION_E}
\eeq
where
\beq
{\tilde M}(s,1)=\int\limits_0^\infty e^{-st'}M(t',1)\,dt'
\label{LAPLACE_SOLUTION_REF_E}
\eeq
is the Laplace coefficient of function $M(t,1)$. This solution satisfies 
${\tilde M}|_{k=1}={\tilde M}(s,1)$. Now, the minus sign in the exponent in 
equation~(\ref{LAPLACE_SOLUTION_E}) should be chosen to satisfy the boundary 
condition at infinity in $k$, ${\tilde M}(s,\infty)=0$,~\cite{KA_92}. The Laplace 
inversion of ${\tilde M}(s,k)$, given by equations~(\ref{LAPLACE_SOLUTION_E}) 
and~(\ref{LAPLACE_SOLUTION_REF_E}), is
\beq
M(t,k) &=&
\frac{1}{2\pi i}\int\limits_{c-i\infty}^{c+i\infty} e^{st}{\tilde M}(s,k)\,ds
\nonumber\\
{}&=& 
\frac{1}{2\pi i}\int\limits_0^\infty dt'
\int\limits_{c-i\infty}^{c+i\infty} M(t',1)\,e^{s(t-t')}\,
k^{\Lambda_1/2}\,k^{-\sqrt{5s/\Gamma-(\Lambda_0-\Lambda_1^2/4)}}\,ds.
\quad
\eeq
Changing the integration variables
\beq
t'\rightarrow\tau=t-t',
\qquad\quad
s\rightarrow z=\sqrt{5s/\Gamma-(\Lambda_0-\Lambda_1^2/4)},
\eeq
we obtain
\beq
M(t,k) &=&
\frac{\Gamma}{5\pi i}\,k^{\Lambda_1/2}
\int\limits_{-\infty}^t  M(t-\tau,1)\,e^{(\Gamma/5)(\Lambda_0-\Lambda_1^2/4)\tau}\:d\tau
\int\limits_{-i\infty}^{i\infty} z\,e^{(\Gamma\tau/5)z^2-z\ln{k}}\,dz.
\qquad
\eeq
Here, the $z$ integration can be taken as the imaginary vertical axis in the complex 
$z$-plane. Carrying out the $z$ integration, we obtain
\beq
M(t,k) &=&
{\left(\frac{5}{4\pi}\right)}^{\!1/2}
\frac{k^{\Lambda_1/2}\ln{k}}{\Gamma^{1/2}}
\int\limits_{-\infty}^t  M(t-\tau,1)\,
\frac{e^{(\Gamma/5)(\Lambda_0-\Lambda_1^2/4)\tau-5\ln^2{k}/4\Gamma\tau}}{\tau^{3/2}}
\,d\tau.
\qquad
\eeq
This function satisfies equation~(\ref{SMALL_SCALES_MODE_COUPLING}), but does not
satisfy the initial condition equation, $M(0,k)=0$. It is clear that function
\beq
M(t,k) &=&
{\left(\frac{5}{4\pi}\right)}^{\!1/2}
\frac{k^{\Lambda_1/2}\ln{k}}{\Gamma^{1/2}}
\int\limits_0^t M(t-\tau,1)\,
\frac{e^{(\Gamma/5)(\Lambda_0-\Lambda_1^2/4)\tau-5\ln^2{k}/4\Gamma\tau}}{\tau^{3/2}}
\,d\tau
\qquad
\label{SMALL_SCALES_SOLUTION_E}
\eeq
satisfies both equations. Because $M(t-\tau,1)$ is zero if $\tau>t$, we can extend
the upper integration limit in equation~(\ref{SMALL_SCALES_SOLUTION_E}) to infinity.
Therefore, this equation is equivalent to equations~(\ref{SMALL_SCALES_SOLUTION}) 
and~(\ref{GREENS_FUNCTION}), after we replace $k$ by $k/k_{\rm ref}$ 
and $t-\tau$ by $t'$.


%% file: component_bibliography.tex


%% file: thesis.bbl
\begin{thebibliography}{99}
\vspace{-0.18cm}
\addcontentsline{toc}{chapter}{Bibliography}


\bibitem{A_43}
Alfv$\acute{\rm e}$n, H., 1943,
Arkiv f$\ddot{\rm o}$r Matematik, Astronomi och Fysik, {\bf 29B}, No.~2


\bibitem{A_92}
Anderson, S. W., 1992,
{\em Limits on galactic dynamo theory due to magnetic fluctuations},
Ph.~D.~Thesis, Princeton University (January 1992)


\bibitem{B_67}
Barnes, A., 1967,
{\em Stochastic electron heating and hydromagnetic wave damping},
The Physics of Fluids, {\bf 10}, 2427


\bibitem{BBMSS_96}
Beck, R., Branderburg, A., Moss, D., Shukurov, A., \& Sokoloff, D., 1996,
{\em Galactic magnetism: recent developments and perspectives},
Annual Review of Astronomy and Astrophysics, {\bf 34}, 155


\bibitem{BPSS_94}
Beck, R., Poezd, A. D., Shukurov, A., \& Sokoloff, D. D., 1994,
{\em Dynamos in evolving galaxies},
Astronomy and Astrophysics, {\bf 289}, 94


\bibitem{B_50}
Biermann, L., 1950,
{\em $\ddot {\rm U}$ber den ursprung der magnetfelder auf sternen und im 
interstellaren raum},
Zeitschrift fur Naturforschung, {\bf 5a}, 65


\bibitem{B_65}
Braginskii, S. I., 1965,
{\em Transport processes in a plasma},
Reviews of Plasma Physics, {\bf 1}, 205


\bibitem{B_01}
Brandenburg, A., 2001,
{\em The inverse cascade and nonlinear alpha-effect in simulations of isotropic 
helical hydromagnetic turbulence},
The Astrophysical Journal, {\bf 550}, 824


\bibitem{BS_00}
Boldyrev, S. A., \& Schekochihin, 2000,
{\em Geometric properties of passive random advection},
Physical Review E, {\bf 62}, 545


\bibitem{CV_91}
Cattaneo, F., \& Vainshtein, S. I., 1991,
{\em Suppression of turbulent transport by a weak magnetic field},
The Astrophysical Journal Letters, {\bf 376}, L21


\bibitem{C_99}
Cowley, S., 1999, unpublished


\bibitem{CKS_01}
Cowley, S., Kulsrud, R. M., \& Schekochihin, A. A., 2001, private communication


\bibitem{DW_00}
Davies, G., \& Widrow, L. M., 2000,
{\em A possible mechanism for generating galactic magnetic fields},
The Astronomical Journal, {\bf 540}, 755


\bibitem{E_99} 
Eilek, J., 1999, 
{\em Magnetic fields in clusters: theory vs. observations},
Proceedings of the 1999 Ringberg Workshop, Germany, April 19-23, 1999, 71;
{astro-ph/9906485}\-


\bibitem{F_49}
Fermi, E., 1949,
{\em On the origin of cosmic radiation},
Physical Review, {\bf 75}, 1169


\bibitem{FFFGGMMS_99}
Fusco-Femiano, R., dal Fiume, D., Feretti, L., Giovannini, G., Grandi, P., Matt, G., 
Molendi, S., \& Santangelo, A., 1999,
{\em Hard X-ray radiation in the Coma cluster spectrum},
The Astrophysical Journal Letters, {\bf 1999}, L21


\bibitem{HM_49}
Hall, J. S., \& Mikesell, A. H., 1949,
{\em Observations of polarized light from stars},
The Astronomical Journal, {\bf 54}, 187 


\bibitem{GW_66}
Gardner, F. F., \& Whiteoak, J. B., 1966,
{\em The polarization of cosmic radio waves},
Annual Review of Astronomy and Astrophysics, {\bf 4}, 245


\bibitem{GD_94}
Gruzinov, A. V., \& Diamond, P. H., 1994,
{\em Self-consistent theory of mean-field electrodynamics},
Physical Review Letters, {\bf 72}, 1651


\bibitem{H_49}
Hiltner, W. A., 1949,
{\em On the presence of polarization in the continuous radiation of stars. II},
The Astrophysical Journal, {\bf 109}, 471


\bibitem{HK_97}
Howard, A. M., \& Kulsrud, R. M., 1997,
{\em The evolution of a primordial galactic magnetic field},
The Astrophysical Journal, {\bf 483}, 648


\bibitem{K_68}
Kazantsev, A. P., 1968,
{\em Enhancement of a magnetic field by a conducting fluid},
Soviet Physics JETP, {\bf 26}, 1031


\bibitem{KN_67}
Kraichnan, R. H., \& Nagarajan, S., 1967,
{\em Growth of turbulent magnetic fields},
The Physics of Fluids, {\bf 10}, 859


\bibitem{K_94} 
Kronberg, P. P., 1994, 
{\em Extragalactic magnetic fields},
Reports on Progress in Physics, {\bf 57}, 325


\bibitem{K_99}
Kulsrud, R. M., 1999,
{\em A critical review of galactic dynamos},
Annual Review of Astronomy and Astrophysics, {\bf 37}, 37


\bibitem{K_00}
Kulsrud, R. M., 2000,
{\em The origin of galactic magnetic fields},
Proceedings of the International School of Physics ``Enrico Fermi'' Course CXLII, 
ed.~Coppi, B., Ferrari, A., and 
Sindoni, E., IOS Press, Amsterdam 2000, page 107


\bibitem{KA_92}
Kulsrud, R. M., \& Anderson, S. W., 1992,
{\em The spectrum of random magnetic fields in the mean field dynamo theory of 
the galactic magnetic field},
The Astrophysical Journal, {\bf 396}, 606


\bibitem{KCOR_97}
Kulsrud, R. M., Cen, R., Ostriker, J. P., \& Ryu, D., 1997,
{\em The protogalactic origin for cosmic magnetic fields},
The Astrophysical Journal, {\bf 480}, 481


\bibitem{LL_84}
Landau, L. D., \& Lifshitz, E. M., 1984, 
{\it Electrodynamics of continuous media}, Oxford; New York: Pergamon, page 225


\bibitem{LSTC_97}
Lemoine, M., Schramm, D. N., Truran, J. W., \& Copi, C. J., 1997,
{\em On the significance of population II~ $^6$Li abundances},
The Astrophysical Journal, {\bf 478}, 554


\bibitem{M_01} 
Malyshkin, L., 2001,
{\em Evolution of magnetic field curvature in the Kulsrud-Anderson dynamo theory},
The Astrophysical Journal, in press;\\
{astro-ph/0103191}\-


\bibitem{M_74}
Manchester, R. N., 1974,
{\em Structure of the local galactic magnetic field},
The Astrophysical Journal, {\bf 188}, 637


\bibitem{MF_70}
Mathewson, D. S., \& Ford, V. L., 1970,
{\em The magnetic-field structure of the Magellanic clouds},
The Astrophysical Journal Letters, {\bf 190}, L43


\bibitem{NHBBKK_93}
Neininger, N., Horellou, C., Beck, R., Berkhuijsen, E. M., Krause, M., \& Klein, U., 1993,
{\em The Magnetic Field of M 51}, 
The cosmic dynamo: proceedings of the 157th Symposium of the International Astronomical 
Union; Edited by F. Krause, K. H. Radler, and Gunther Rudiger.; Kluwer Acad.~Publ.; 
Dordrecht, page~313
{http://www.mpifr-bonn.mpg.de/staff/wsherwood/mag-fields.html}\-


\bibitem{P_70}
Parker, E. N., 1970,
{\em The generation of magnetic fields in astrophysical bodies. I. The dynamo equations},
The Astrophysical Journal, {\bf 162}, 665


\bibitem{P_71}
Parker, E. N., 1971,
{\em The generation of magnetic fields in astrophysical bodies. II. The galactic field},
The Astrophysical Journal, {\bf 163}, 255


\bibitem{P_79}
Parker, E. N., 1979,
{\em Cosmical magnetic fields},
Clarendon Press: Oxford, 1979


\bibitem{P_94}
Perry, J. J., 1994,
{\em Magnetic fields at high redshift},
In Cosmical Magnetism. Contributed Papers in Honor of Professor L.~Mestel, 
ed. D.~Lyndon-Bell, page 144. Cambridge: Inst. Astron.


\bibitem{PFL_76}
Pouquet, A., Frisch, U., \& L$\acute {\rm e}$orat, J., 1976,
{\em Strong MHD helical turbulence and the nonlinear dynamo effect},
Journal of Fluid Mechanics, {\bf 77}, 321 


\bibitem{PS_89}
Pudritz, R. E., \& Silk, J., 1989,
{\em The origin of magnetic fields and primordial stars in protogalaxies},
The Astrophysical Journal, {\bf 342}, 650


\bibitem{RK_00}
Rafikov, R. R., \& Kulsrud, R. M., 2000,
{\em Magnetic flux expulsion in powerful superbubble explosions and the 
$\alpha$--$\Omega$ dynamo},
Monthly Notices of the Royal Astronomical Society, {\bf 314}, 839


\bibitem{RK_89}
Rand, R. J., \& Kulkarni, S. R., 1989,
{\em The local galactic magnetic field},
The Astrophysical Journal, {\bf 343}, 760


\bibitem{RD_89}
Rosner, R., \& Deluca, E., 1989,
{\em On the galactic dynamo},
The Center of the Galaxy, IAU Symp.~136, ed. Morris, M., page 319


\bibitem{SK_00}
Sarazin, C. L., \& Kempner, J. C., 2000,
{\em Nonthermal bremsstrahlung and hard X-ray emission from clusters of galaxies},
The Astrophysical Journal, {\bf 533}, 73


\bibitem{SBK_01}
Schekochihin, A. A., Boldyrev, S. A., \& Kulsrud, R. M., 2001,
{\em Spectra and Growth Rates of Fluctuating Magnetic Fields in the Kinematic Dynamo
Theory with Large Magnetic Prandtl Numbers},
submitted to The Astrophysical Journal,
astro-ph/0103333


\bibitem{SCMM_01}
Schekochihin, A., Cowley, S., Maron, J., \& Malyshkin, L., 2001,
{\em Structure of small-scale magnetic fields in the kinematic dynamo theory},
submitted to Physical Review E; {astro-ph/0105322}\-


\bibitem{SK_01}
Schekochihin, A. A., \& Kulsrud, R. M., 2001,
{\em Finite-correlation-time effects in the kinematic dynamo problem},
astro-ph/0002175


\bibitem{SMOC_01}
Schekochihin, A. A., Maron, J., Opher, M., \& Cowley, S., 2001, 
{\em Magnetic-Field Structure and Saturation in the Small-Scale Dynamo Theory},
American Astronomical Society Meeting 198, \#90.01


\bibitem{S_62} 
Spitzer Jr, L., 1962, 
{\it Physics of fully ionized gases}, 
New York: Wiley


\bibitem{V_70}
Vainshtein, S. I., 1970,
{\em The generation of a large-scale magnetic field by a turbulent fluid},
Soviet Physics JETP, {\bf 31}, 87


\bibitem{V_82}
Vainshtein, S. I., 1982,
{\em Theory of small-scale magnetic fields},
Soviet Physics JETP, {\bf 56}, 86


\bibitem{VC_92}
Vainshtein, S. I., \& Cattaneo, F., 1992,
{\em Nonlinear restrictions on dynamo action},
The Astrophysical Journal, {\bf 393}, 165


\bibitem{VR_72a}
Vainshtein, S. I., \& Ruzmaikin, A. A., 1972,
{\em Generation of the large-scale galactic magnetic field},
Soviet Astronomy, {\bf 15}, 714


\bibitem{VR_72b}
Vainshtein, S. I., \& Ruzmaikin, A. A., 1972,
{\em Generation of the large-scale galactic magnetic field. II},
Soviet Astronomy, {\bf 16}, 365


\bibitem{VZ_72}
Vainshtein, S. I., \& Zel'dovich, Ya. B., 1972,
{\em Origin of magnetic fields in Astrophysics},
Soviet Physics Uspekhi, {\bf 15}, 159


\bibitem{WLO_92}
Wolfe, A. M., Lanzetta, K. M., \& Oren, A. L., 1992,
{\em Magnetic fields in damped Ly$\alpha$ systems},
The Astrophysical Journal, {\bf 388}, 17


\bibitem{V_69}
Verschuur, G. L., 1969,
{\em Measurements of magnetic fields in interstellar clouds of neutral hydrogen},
The Astrophysical Journal, {\bf 156}, 861


\bibitem{ZH_97}
Zweibel, E. G., \& Heiles, C., 1997,
{\em Magnetic fields in galaxies and beyond},
Nature, {\bf 385}, 131


\end{thebibliography}
